\documentclass[a4paper,12pt]{article}

\usepackage{jheppub}
\usepackage{hyperref}

\usepackage{bm, bbm, bbold}
\usepackage{latexsym}
\usepackage{dcolumn}
\usepackage[T1]{fontenc}
\usepackage{comment}
\usepackage{graphicx,epsfig}
\usepackage{environ}
\usepackage{mathtools}
\usepackage{braket}
\usepackage{graphicx,epstopdf}
\usepackage{tikz}
\usepackage{float}
\usepackage{framed}
\usepackage{soul}
\usetikzlibrary{positioning,decorations.markings, decorations.pathmorphing, calc}
\tikzset{snake it/.style={decorate, decoration=snake}}

\usepackage{stmaryrd}
\usepackage{slashed}
\usepackage{array,multirow}

\usepackage[framemethod=default]{mdframed}
\newmdenv[skipabove=7pt,
skipbelow=7pt,
rightline=false,
leftline=false,
topline=false,
bottomline=false,
backgroundcolor=gray!10,
linecolor=gray,
innerleftmargin=5pt,
innerrightmargin=5pt,
innertopmargin=5pt,
innerbottommargin=5pt,
leftmargin=0cm,
rightmargin=0cm,
linewidth=4pt]{eBox}

\def\s{\sigma}

\def\a{\alpha}

\def\e{\epsilon}

\def\s{\sigma}

\def\ph{\phi} 
\def\p{\partial}
\def\tt{\tilde{\tau}}

\newcommand{\la}{\langle}
\newcommand{\ra}{\rangle}
\newcommand{\be}{\begin{equation}}
\newcommand{\ee}{\end{equation}}
\newcommand{\bes}{\begin{equation*}}
\newcommand{\ees}{\end{equation*}}

\def\Mpl([#1],[#2]){\textrm{Li}_{#1}\left(#2\right)}

\NewEnviron{derivation}{
\begin{framed}
\begin{center}
{\bf-: Derivation :-}\\
\end{center}
  \BODY

\end{framed}
}

\NewEnviron{eqn}{
\begin{align}
\begin{split}
  \BODY
\end{split}
\end{align}
}

\NewEnviron{eqn*}{
\begin{align*}
\begin{split}
  \BODY
\end{split}
\end{align*}
}

\newcommand{\intinf}{\int_{-\infty}^{\infty}} 
\newcommand{\intsinf}{\int_{0}^{\infty}} 
\newcommand{\partiald}[2]{\frac{\partial #1}{\partial #2}}

\newcommand{\reference}[1]{{\fontfamily{lmtt}\selectfont #1}}

\newcommand{\red}[1]{{\color{red}#1}}
\newcommand{\blue}[1]{{\color{blue}#1}}

\usepackage{cancel}
\usepackage{tikz, pgf}
\usetikzlibrary{shapes.misc}
\usetikzlibrary{shapes}
\usetikzlibrary{fit}
\usetikzlibrary{arrows}

\usetikzlibrary{decorations.pathmorphing}	% For Feynman Diagrams
\usetikzlibrary{decorations.markings}
\tikzset{
    sugra/.style={decorate, decoration={snake}, draw=black},
    scalarphi/.style={dashed,draw=black, postaction={decorate},
        },
    hwbou/.style={draw=blue, postaction={decorate}, ultra thick
        },
    vector/.style={draw=blue,decorate, decoration={snake}, draw},
	provector/.style={decorate, decoration={snake,amplitude=2.5pt}, draw},
	antivector/.style={decorate, decoration={snake,amplitude=-2.5pt}, draw},
   	 fermion/.style={draw=cyan, postaction={decorate},
        decoration={markings,mark=at position .55 with {\arrow[draw=black]{>}}}},
    fermionbar/.style={draw=cyan, postaction={decorate},
        decoration={markings,mark=at position .55 with {\arrow[draw=black]{<}}}},
    fermionnoarrow/.style={draw=black},
    gluon/.style={decorate, draw=red,
        decoration={coil,amplitude=4pt, segment length=5pt}},
    scalar/.style={dashed,draw=black, postaction={decorate},
        decoration={markings,mark=at position .55 with {\arrow[draw=black]{>}}}},
    scalarbar/.style={dashed,draw=black, postaction={decorate},
        decoration={markings,mark=at position .55 with {\arrow[draw=black]{<}}}},
    electron/.style={draw=black, postaction={decorate},
        decoration={markings,mark=at position .55 with {\arrow[draw=black]{>}}}},
    scalarnoarrow/.style={dashed, draw=black},
    electron/.style={draw=black, postaction={decorate},
        decoration={markings, mark=at position .55 with {\arrow[draw=black]{>}}}},
	bigvector/.style={decorate, decoration={snake, amplitude=4pt}, draw},
    photon/.style={draw=red, decorate, decoration={zigzag}, draw},
    higgs/.style={dashed, draw=black, postaction={decorate},
        },	
        goldstone/.style={draw=brown, postaction={decorate},
        },    
          ghost/.style={dashed, draw=magenta, postaction={decorate},
        decoration={markings, mark=at position .55 with {\arrow[draw=black]{>}}}
        },  
          antighost/.style={dashed, draw=magenta, postaction={decorate},
        decoration={markings, mark=at position .55 with {\arrow[draw=black]{<}}}
        },  
          mphoton/.style={decorate, decoration={snake}, draw=violet},
            realscalar/.style={draw=black}, 
           mgluon/.style={decorate, draw=blue,
        decoration={coil,amplitude=4pt, segment length=5pt}},
         weylfermion/.style={draw=orange, postaction={decorate},
        decoration={markings,mark=at position .55 with {\arrow[draw=black]{>}}}},
         weylfermionbar/.style={draw=orange, postaction={decorate},
        decoration={markings,mark=at position .55 with {\arrow[draw=black]{<}}}}, 
   	wboson/.style={draw=blue,decorate, decoration={snake,amplitude=4pt}, draw},  
    zboson/.style={draw=violet, decorate, decoration={snake}, draw},   
    lepton/.style={draw=black, postaction={decorate},
        decoration={markings,mark=at position .55 with {\arrow[draw=black]{>}}}},
    leptonbar/.style={draw=black, postaction={decorate},
        decoration={markings,mark=at position .55 with {\arrow[draw=black]{<}}}}, 
        graviton/.style={draw=blue,decorate, decoration={snake,amplitude=4pt}, draw},  
        gravitino/.style={draw=red, postaction={decorate},
        decoration={snake, markings, mark=at position .55 with {\arrow[draw=black]{>}}}},
    gravitinobar/.style={draw=red, postaction={decorate},
        decoration={snake, markings,mark=at position .55 with {\arrow[draw=black]{<}}} },    
}

\newcommand{\HlineN}[3]{
\begin{tikzpicture}[baseline]
\draw[dashed] (0,0) -- (1.5,0);
\node at (0,0) {\textbullet};
\node at (1.5,0) {\textbullet};
\node at (0,-0.25) {$#1$};
\node at (1.5,-0.25) {$#2$};
\node at (0.75,0.25) {$#3$};
\end{tikzpicture}
}

\newcommand{\tadpoleD}[2]{
\begin{tikzpicture}[baseline]
\draw[very thick] (0,0.5) circle (0.5);
\draw[very thick] (0, 0) -- (-0.4, -0.4);
\draw[very thick] (0, 0) -- (0.4, -0.4);
\node at (-0.0, -0.35) {\footnotesize $#1$};
\node at (0, 1.25) {\footnotesize $#2$};
\end{tikzpicture}
}

\newcommand{\tadpoleN}[2]{
\begin{tikzpicture}[baseline]
\draw[dashed, very thick] (0,0.5) circle (0.5);
\draw[very thick] (0, 0) -- (-0.5, -0.5);
\draw[very thick] (0, 0) -- (0.5, -0.5);
\node at (-0.0, -0.35) {\footnotesize $#1$};
\node at (0, 1.25) {\footnotesize $#2$};
\end{tikzpicture}
}

\newcommand{\tadpoleDD}[4]{
\begin{tikzpicture}[baseline]
\draw[very thick] (0,0.5) circle (0.5);
\draw[very thick] (0,1.5) circle (0.5);
\node at (0, -0.45) {\footnotesize $#1$};
\node at (0, 0.75) {\footnotesize $#2$};
\node at (-0.65, 0.5) {\footnotesize $#3$};
\node at (0, 2.25) {\footnotesize $#4$};
\draw[black, very thick] (0, 0) -- (-0.5, -0.5);
\draw[black,very thick] (0, 0) -- (0.5, -0.5);
\node at (0.5, 1) {$x$};
\node at (0.5, 0) {$z$};
\end{tikzpicture}
}

\newcommand{\tadpoleDN}[4]{
\begin{tikzpicture}[baseline]
\draw[very thick] (0,0.5) circle (0.5);
\draw[dashed, very thick] (0,1.5) circle (0.5);
\node at (0, -0.45) {\footnotesize $#1$};
\node at (0, 0.75) {\footnotesize$#2$};
\node at (-0.65, 0.5) {\footnotesize$#3$};
\node at (0, 2.25) {\footnotesize$#4$};
\draw[very thick] (0, 0) -- (-0.5, -0.5);
\draw[very thick] (0, 0) -- (0.5, -0.5);
\draw[ black, very thick] (0, 0) -- (-0.5, -0.5);
\draw[black,very thick] (0, 0) -- (0.5, -0.5);
\node at (0.5, 1) {$x$};
\node at (0.5, 0) {$z$};
\end{tikzpicture}
}

\newcommand{\tadpoleND}[4]{
\begin{tikzpicture}[baseline]
\draw[dashed, very thick] (0,0.5) circle (0.5);
\draw[very thick] (0,1.5) circle (0.5);
\node at (0, -0.45) {\footnotesize$#1$};
\node at (0, 0.75) {\footnotesize$#2$};
\node at (-0.65, 0.5) {\footnotesize$#3$};
\node at (0, 2.25) {\footnotesize$#4$};
\draw[very thick] (0, 0) -- (-0.5, -0.5);
\draw[very thick] (0, 0) -- (0.5, -0.5);
\draw[ black, very thick] (0, 0) -- (-0.5, -0.5);
\draw[black,very thick] (0, 0) -- (0.5, -0.5);
\node at (0.5, 1) {$x$};
\node at (0.5, 0) {$z$};
\end{tikzpicture}
}

\newcommand{\tadpoleNN}[4]{
\begin{tikzpicture}[baseline]
\draw[dashed, very thick] (0,0.5) circle (0.5);
\draw[dashed, very thick] (0,1.5) circle (0.5);
\node at (0, -0.45) {\footnotesize$#1$};
\node at (0, 0.75) {\footnotesize$#2$};
\node at (-0.65, 0.5) {\footnotesize$#3$};
\node at (0, 2.25) {\footnotesize$#4$};
\draw[very thick] (0, 0) -- (-0.5, -0.5);
\draw[very thick] (0, 0) -- (0.5, -0.5);
\draw[ very thick] (0, 0) -- (-0.5, -0.5);
\draw[very thick] (0, 0) -- (0.5, -0.5);
\node at (0.5, 1) {$x$};
\node at (0.5, 0) {$z$};
\end{tikzpicture}
}

\newcommand{\HlineDD}[5]{
\begin{tikzpicture}[baseline]
\draw (-1,0) -- (0,0);
\draw (0,0) -- (1,0);
\node at (-1, 0) {\textbullet};
\node at (0, 0) {\textbullet};
\node at (1, 0) {\textbullet};
\node at (-1, -0.25) {$#1$};
\node at (0, -0.25) {$#2$};
\node at (1, -0.25) {$#3$};
\node at (-0.5, 0.25) {$#4$};
\node at (0.5, 0.25) {$#5$};
\end{tikzpicture}
}

\newcommand{\HlineNN}[5]{
\begin{tikzpicture}[baseline]
\draw[dashed] (-1,0) -- (0,0);
\draw[dashed] (0,0) -- (1,0);
\node at (-1, 0) {\textbullet};
\node at (0, 0) {\textbullet};
\node at (1, 0) {\textbullet};
\node at (-1, -0.25) {$#1$};
\node at (0, -0.25) {$#2$};
\node at (1, -0.25) {$#3$};
\node at (-0.5, 0.25) {$#4$};
\node at (0.5, 0.25) {$#5$};
\end{tikzpicture}
}

\newcommand{\vertex}[1]{\begin{tikzpicture}[baseline]
\node at (0,0) {\textbullet};
\node at (0,-0.25) {$#1$};
\end{tikzpicture}}

\newcommand{\eqs}[1]{\begin{equation}\begin{split}#1\end{split}\end{equation}}

\hypersetup{pagebackref,urlcolor=brown,citecolor=blue,linkcolor=purple}
\usepackage{amsthm,amsmath,latexsym,amssymb,amsfonts,amscd,soul}
\usepackage{mathrsfs}
\usepackage{mathtools,commath}
\usepackage{soul}
\usepackage{comment}
\usepackage{subcaption}
\definecolor{bubbles}{rgb}{0.91, 1.0, 1.0}
\definecolor{aquamarine}{rgb}{0.5, 1.0, 0.83}
\definecolor{bubblegum}{rgb}{0.99, 0.76, 0.8}
\definecolor{blackbell}{rgb}{0.64, 0.64, 0.82}
\definecolor{dollarbill}{rgb}{0.72, 0.93, 0.6}
 \preprint{\vbox{\hbox{LMU-ASC 37/23}}
 \vbox{\hbox{IPhT-t23/119}}}

\title{The Subtle Simplicity of Cosmological Correlators}

\author[a, b]{Chandramouli Chowdhury,}
\author[c]{Arthur Lipstein,} 
\author[c]{Jiajie Mei,}
\author[d]{Ivo Sachs,}
\author[e]{and Pierre Vanhove}

\affiliation[a]{Mathematical Sciences and STAG Research Centre, University of Southampton, Highfield, Southampton SO17 1BJ, United Kingdom}
\affiliation[b]{International Centre for Theoretical Sciences - TIFR, Bengaluru - 560089, India} 
\affiliation[c]{Department of Mathematical Sciences, Durham University, Stockton Road, DH1 3LE, Durham, United Kingdom}
\affiliation[d]{Arnold-Sommerfeld-Center for Theoretical Physics, Ludwig-Maximilians-Universit\"at M\"unchen, Theresienstr. 37, D-80333 Munich, Germany}
\affiliation[e]{Institut de Physique Th\'eorique, Universit\'e Paris-Saclay, CEA, CNRS, F-91191 Gif-sur-Yvette Cedex, France}
\abstract{We investigate cosmological correlators for conformally coupled $\phi^4$ theory in four-dimensional de Sitter space. These \textit{in-in} correlators differ from scattering amplitudes for massless particles in flat space due to the spacelike structure of future infinity in de Sitter. They also require a regularization which preserves de Sitter-invariance, which makes the flat space limit subtle to define at loop-level. Nevertheless we find that up to two loops, the \textit{in-in} correlators  are structurally simpler than the wave function and have the same transcendentality as flat space amplitudes. Moreover, we show that their loop integrands can be recast in terms of flat space integrands and can be derived from a novel recursion relation.  

}
\begin{document}

\maketitle
\flushbottom

\newpage

\section{Introduction}

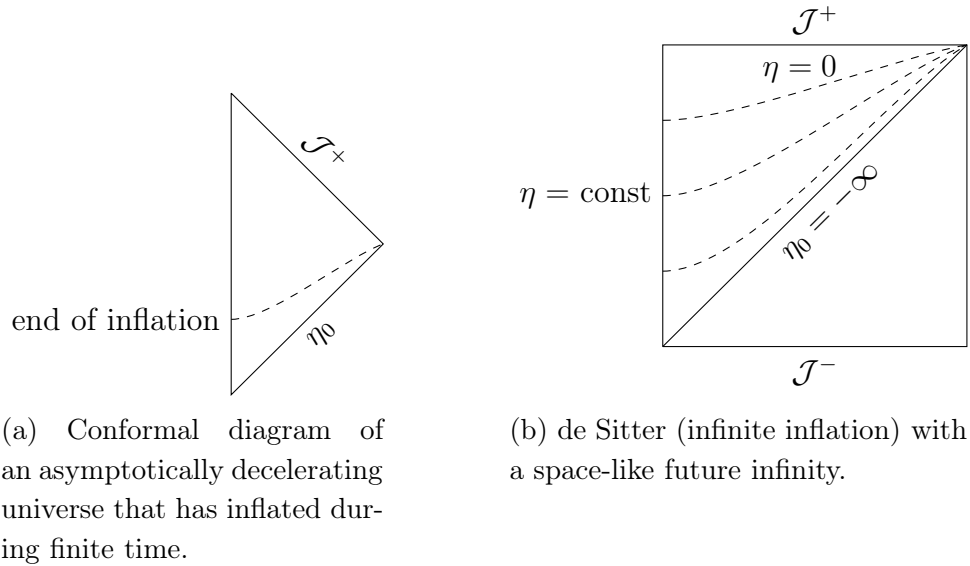
\begin{figure}[h]
	\centering
	\subcaptionbox{Conformal diagram of an asymptotically decelerating universe that has  inflated during finite time. \label{fig:finite}}{\begin{tikzpicture}
		%Global patch
		%\draw (-2,0) to 
		%node[midway, below] {$\mathcal{J}^-$}(2,0) 
		%to (2,4) to node[midway, above] {$\mathcal{J}^+$} node[pos=.55, below]{$\eta=0$} (-2,4) to  (-2,0);
		\draw (-1,4) to node[midway,above,sloped] {$\mathcal{J}^+$}(1,2)  to node[midway,below,sloped] {$\eta_0$} (-1,0) to  (-1,4) ;
		
		%Poincare patch
		%\draw[dashed] (-1,1.5) to [out=0, in=190,looseness=0.6](1,2);
		\draw[dashed] (-1,1) to [out=0, in=200,looseness=0.6] node[at start, left]{end of inflation}(1,2);
		%\draw[dashed] (-1,0.5) to [out=0, in=210,looseness=0.6](1,2);
	\end{tikzpicture}}\qquad\qquad
 \subcaptionbox{de Sitter (infinite inflation) with a space-like future infinity.\label{fig:infinite}}{\begin{tikzpicture}
		%Global patch
		\draw (-2,0) to 
		node[midway, below] {$\mathcal{J}^-$}(2,0) 
		to (2,4) to node[midway, above] {$\mathcal{J}^+$} node[pos=.55, below]{$\eta=0$} (-2,4) to  (-2,0);
		\draw (-2,0) to node[midway,below,sloped] {$\eta_0=-\infty$}(2,4);
		
		%Poincare patch
		\draw[dashed] (-2,3) to [out=0, in=190,looseness=0.6](2,4);
		\draw[dashed] (-2,2) to [out=0, in=200,looseness=0.6] node[at start, left]{$\eta=$ const}(2,4);
		\draw[dashed] (-2,1) to [out=0, in=210,looseness=0.6](2,4);
	\end{tikzpicture}}
 \caption{Conformal diagram where $\eta$ is the conformal time.}
 \label{fig:conformal_diagram_dS}
\end{figure}

Correlation functions of quantum fields at the end of inflation are important in cosmology as they provide the seeds for the formation of structure in the Universe~\cite{Mukhanov:1981xt} as sketched in Figure \ref{fig:finite}, where correlation functions are to be evaluated on the dashed line. This is demonstrated by the imprint they leave on the comic microwave background~\cite{Planck:2013oqw}.  In this paper we consider an idealization of a cosmological space-time that consists of an eternally inflating Universe as in Figure \ref{fig:infinite}.   
In such a space-time there is no structure formation, but the calculation of correlation function is 
still of theoretical interest, since the early universe was approximately described by a de Sitter geometry during  inflation, and it is possible to derive inflationary correlators from de Sitter correlators by giving a small mass to one of the legs of a four-point function in de Sitter space
(proportional to the slow-roll parameter) and then taking a soft limit~\cite{Creminelli:2003iq,Assassi:2012zq, Kundu:2014gxa,Kundu:2015xta}.
We consider equal-time correlators but, since there is no ``end of inflation,'' the natural location to evaluate them is at future infinity ($\eta=0$ in fig. \ref{fig:infinite}). As a consequence of the $SO(4,1)$-isometry of de Sitter, these correlation functions will transform in a representation of the 3-dimensional conformal group acting on $\mathcal{J}^+$. This is an instance of the dS/CFT duality~\cite{Strominger:2001pn, 
Maldacena:2002vr,McFadden:2009fg}. 

When computing correlation functions in de Sitter, being a non-static space-time, one furthermore has to specify the initial and final state of the system. There are two natural choices. One is the $\bra{\textrm{out}}\cdots\ket{\textrm{in}}$ expectation value, where $\ket{\textrm{in}}$ is the de Sitter-invariant Bunch-Davies initial state at $\eta_0$ (the beginning of inflation), and $\bra{\textrm{out}}$ is given by the boundary condition at future infinity. With this choice the (integrated) correlators at $\mathcal{J}^+$ compute the expansion coefficients of the wave function $\Psi[\phi]$ of the quantum field $\phi$~\cite{Hartle:1983ai,Maldacena:2002vr}. In recent years, many tools have been developed to compute the wavefunction perturbatively, inspired by the study of scattering amplitudes and AdS/CFT. These tools include geometric approaches~\cite{Arkani-Hamed:2017fdk,Bzowski:2020kfw}, methods based on locality and unitarity~\cite{Arkani-Hamed:2015bza,Arkani-Hamed:2018kmz,Meltzer:2021zin,Goodhew:2020hob,Jazayeri:2021fvk,Lee:2023kno}, the double copy~\cite{Farrow:2018yni,Armstrong:2020woi,Albayrak:2020fyp,Jain:2021qcl,Herderschee:2022ntr,Cheung:2022pdk,Armstrong:2023phb,Mei:2023jkb}, scattering equations~\cite{Gomez:2021qfd}, Mellin-Barnes representations~\cite{Sleight:2019hfp}, and holographic methods~\cite{Bzowski:2012ih,Bzowski:2013sza,Bzowski:2017poo,Bzowski:2018fql,Heckelbacher:2020nue,Heckelbacher:2022hbq}.

From the point of view of cosmology the natural quantity is the $\bra{\mbox{in}}\cdots\ket{\mbox{in}}$ correlator, as it only depends on the initial condition before inflation. It can be computed by squaring the wavefunction and computing an expectation value or using the Schwinger-Keldysh formalism~\cite{Maldacena:2002vr,Weinberg:2005vy}. The calculation of \textit{in-in} correlators can also be mapped to Euclidean anti-de Sitter~\cite{Sleight:2019mgd, Sleight:2020obc, Sleight:2021plv} using an effective action involving shadow fields~\cite{DiPietro:2021sjt}. The shadow fields have ghost-like kinetic terms, although this not an issue in Euclidean anti-de Sitter and the action is ultimately designed to compute unitary observables in de Sitter. While \textit{in-in} correlators \cite{Weinberg:2005vy} relate more directly to experimentally measurable observables, they naively appear to be more complicated objects than wavefunction coefficients. 

One of the main messages of this paper will be that the loop corrections to the \textit{in-in} correlators are actually much simpler than wavefunction coefficients. The simplicity arises from nontrivial cancellations in the loop integrands due to shadow ghost-like contributions which renders \textit{in-in} correlators closer in structure to flat space scattering amplitudes than the wavefunction coefficients. The resulting loop integrands can be written in terms of standard four-dimensional Feynman integrals in flat space although the presence of a boundary in the radial direction means that integrals over the radial loop momentum (which is not conserved) must be performed separately from the boundary loop momentum. We then obtain integrals over three-dimensional boundary loop momentum whose integrands can also be derived from a recursion relation analogous to the one developed for wavefunction coefficients in~\cite{Arkani-Hamed:2017fdk}. After performing the loop integration, the resulting analytic structure of \textit{in-in} correlators is remarkably similar to that of flat space scattering amplitudes. In particular, they have the same transcendentality and are significantly simpler than wavefunction coefficients. We demonstrate this up to two loops for the conformally coupled $\phi^4$ theory, but we expect this simplicity to extend to more general theories, as we discuss in the conclusion. In the process we establish renormalisability of the effective action of~\cite{DiPietro:2021sjt} up to two loops.

Since the de Sitter metric is locally conformally flat, one might naively expect that correlators of conformally coupled $\phi^4$ theory can be mapped into flat space scattering amplitudes after a conformal transformation. 
The story is not so simple, however. For one thing, energy conservation is absent. Also, because the asymptotic structure of de Sitter and Minkowski space are different. The conformal boundary of the latter is a null-infinity instead of the space-like infinity of de Sitter.\footnote{There is no globally defined transformation that maps the two into each other.} Consequently, the bulk to boundary propagators in the two space-times are different. More precisely, the mode functions in de Sitter for the given initial condition admit two fall-off behaviours at future infinity, corresponding to Neumann- and Dirichlet boundary conditions respectively. Only one linear combination of the two modes gives rise to a flat space ``external leg'' with tree-level integrands that are conformally related to the flat space result~\cite{Heckelbacher:2022fbx}. Another crucial difference between working in de Sitter and flat space concerns regularization and renormalisation of divergent loop integrals. Standard flat space regulators such as a cut-off, or dimensional regularization break the de Sitter-invariance~\cite{Senatore:2009cf}.
At 1-loop, conformal symmetry can then be restored by choosing non-minimal and non-invariant counterterms, although we do not have a systematic way to derive them using these regulators. This makes the flat space limit subtle to take, because in the flat space limit the isometry group of de Sitter space (which is the conformal group of the three-dimensional Euclidean boundary) gets broken to that of Minkowski space (which is the four dimensional Poincar\'e group), so we must break the conformal symmetry by introducing a dimensionful renormalisation scale. We will give examples of this in Section~\ref{sec:cut-off}.

A manifestly de Sitter-invariant regularization was found in postion space~\cite{Bertan:2018afl,Bertan:2018khc, Banados:2022nhj}, but in this paper we are  interested in the momentum space since this is the standard choice in cosmology and makes the relation to scattering amplitudes more manifest~\cite{Maldacena:2011nz,Raju:2011mp,Raju:2012zr}. Momentum space is also very convenient for the study of soft limits~\cite{Maldacena:2002vr,Assassi:2012zq,Hinterbichler:2013dpa,Armstrong:2022vgl} and higher-loop correlators~\cite{Albayrak:2020isk, Chowdhury:2023khl} since their functional form in momentum space is much simpler than in position space (as we will see later). As we will explain, there is a generalization of analytic regularization in momentum space which keeps de Sitter-invariance manifest. This will automatically produce correlators which obey the three-dimensional conformal Ward identities and will preserve much of the flat space structure of the loop integrands described above, although it does not seem to be compatible with the recursion for loop integrands. We will also briefly describe an alternative de Sitter-invariant regularization scheme in Section~\ref{sec:analytic-reg} for which the recursion rules do not hold. This is based on the dimensional regularisation scheme introduced in~\cite{Bzowski:2013sza}, where one shifts both the boundary dimension $d$ and the scaling dimensions of the dual operators $\Delta$ in order to preserve the spectral parameter $i\nu=\Delta-d/2$, although the resulting loop integrals are very challenging to evaluate.

The structure of this paper is as follows. In Section~\ref{sec:review}, we review the definition of wavefunction and \textit{in-in} correlators, as well the effective action for computing \textit{in-in} correlators via Witten diagrams in Euclidean anti-de Sitter and derive its Feynman rules in momentum space. In Section~\ref{sec:recursion}, we derive a recursion relation for the three-dimensional loop integrands of \textit{in-in} correlators. In section \ref{sec:cut-off}, we then compute various \textit{in-in} correlators up to two loops using a cut-off in the boundary loop momentum and describe how to renormalise them. We also compute an infinite class one-loop polygon diagrams which do not require regularisation because they are finite. Along the way, we will show that the loop integrands can be recast in terms of standard four dimensional flat space Feynman integrals and renormalised by introducing a dimensionful renormalisation scale. The resulting correlators have the correct flat space limit but break the three-dimensional conformal symmetry. Remarkably, conformal symmetry can then be restored by setting the renormalisation scale equal to the energy times a dimensionless renormalisation scale, although this makes the flat space limit more subtle to define. In Section~\ref{sec:analytic-reg}, we derive a de Sitter-invariant regulator and show that it preserves much of the flat space structure found when using a cut-off while giving correlators which automatically obey the conformal Ward identities. We carry out calculations up to two loops in this regularisation scheme and show that the correlators can be renormalised in a consistent way. We also briefly describe an alternative de-Sitter invariant regularisation scheme based on dimensional regularisation. Finally, we present our conclusions in Section~\ref{sec:conclusion}. We also include a number of Appendices containing further results and technical details.

\section{Effective action and Feynman rules}\label{sec:review}
%{\bf review definition of wavefunction vs \textit{in-in} correlator, Komatsu Lagrangian}

We will work in the Poincar\'e patch of dS$_4$ equipped with the metric 
\begin{equation}
{\rm d}s^{2}=\ell_{dS}^2{-{\rm d}\eta^{2}+{\rm d}\vec{x}^{2}\over \eta^{2}},
\label{metric}
\end{equation}
where $\ell_{dS}$ denotes the curvature radius, $-\infty<\eta<0$ is the conformal time and $\vec{x}$ denotes the Euclidean boundary directions, with individual components $x^{i}$, $i=1,2,3$. Cosmological correlators (or {\textit{in-in}} correlators) can be computed as follows:
\begin{equation}
\left\langle \phi(\vec{k}_{1})\cdots\phi(\vec{k}_{n})\right\rangle =\frac{\int\mathcal{D}\phi \, \phi(\vec{k}_{1})\cdots\phi(\vec{k}_{n})\left|\Psi\left[\phi\right]\right|^{2}}{\int\mathcal{D}\phi\left|\Psi\left[\phi\right]\right|^{2}},
\end{equation}
where $\phi$ represents the value of a generic bulk field in the future boundary Fourier transformed to momentum space, $\vec{k}_a$ are boundary momenta, and $\Psi\left[\phi\right]$ is the cosmological wavefunction~\cite{Hartle:1983ai}, which is a functional of $\phi$. For simplicity, we are considering a scalar field but in general, we should integrate over the boundary values of all the bulk fields, including the metric. We describe this approach in more detail in Appendix \ref{sec:wave}.

The wavefunction $\Psi\left[\phi\right]$ for a scalar field with action $S[\phi]$ can be perturbatively expanded as follows: (where $d$ denotes the number of spatial dimensions, which for us is $d =3$)
\begin{equation}
\ln\Psi\left[\phi\right]=-\sum_{n=2}^{\infty}\frac{1}{n!}\int\prod_{i=1}^{n}\frac{{\rm d}^{d}k_{i}}{(2\pi)^{d}}\psi_{n}\left(\vec{k}_{1},\dots,\vec{k}_{n}\right)\phi(\vec{k}_{1})\cdots\phi(\vec{k}_{n}),
\end{equation}
where the wavefunction coefficients $\psi_n$ can be expressed as
\begin{equation}
\psi_{n}(\vec{k}_{1}, \cdots, \vec{k}_{n})=\delta^{d}(\vec{k}_{1}+\cdots+\vec{k}_{n})\Big\langle\!\!\Big\langle \mathcal{O}\left(\vec{k}_{1}\right)\cdots\mathcal{O}\left(\vec{k}_{n}\right)\Big\rangle\!\!\Big\rangle ,
\label{psid}
\end{equation}
where %$\vec{k}_T=\vec{k}_{1}+\cdots+\vec{k}_{n}$ and 
the object in double brackets can be treated as a CFT correlator\footnote{In the sense that the correlators solve the conformal bootstrap equations. They do not arise from a $local$ conformal field theory, however.} in the future boundary~\cite{Maldacena:2011nz,Bzowski:2012ih,Bzowski:2013sza,Arkani-Hamed:2015bza,Bzowski:2017poo,Arkani-Hamed:2018kmz,Bzowski:2018fql,Baumann:2020dch}. A novel feature in de Sitter space is that while momentum is conserved along the boundary,  the total energy defined as 
\begin{equation}
E=\sum_{a=1}^n k_a\;,\quad k_a = |\vec{k}_a| 
\end{equation}
is not conserved in the "scattering". We also define the shorthand $k_{ij}=k_i+k_j$ and $k_{ijl}=k_i+k_j+k_l$.

Alternatively, the { \textit{in-in}} correlators can be computed using the Schwinger-Keldysh formalism~\cite{Keldysh:1964ud} or, equivalently, via analytic continuation to Witten diagrams in Euclidean  anti-de Sitter~\cite{Sleight:2019mgd, Sleight:2020obc, Sleight:2021plv}. For scalar theories with polynomial interactions the resulting Feynman rules are conveniently encoded in the effective Lagrangian given below~\cite{DiPietro:2021sjt}:
\begin{multline}
    \label{eq:EAdS_action_gen_pot}
	iS_c
=\int\limits_0^{\infty}\frac{{\rm d} z {\rm d}^dx}{z^{d+1}}\left[\sin\left(\pi(\Delta_+ \!-\!\frac{d}{2})\right)\left((\partial\phi_+)^2\!-\!m^2{\phi_+}^2\right)\right.\cr+\sin\left(\pi(\Delta_- \!-\!\frac{d}{2})\right)\left((\partial\phi_-)^2\!-\!m^2{\phi_-}^2\right)\cr
	\left.+e^{i\pi\frac{d-1}{2}}V\left(e^{-i\frac{\pi}{2}\Delta_+}\phi_+ +e^{-i\frac{\pi}{2}\Delta_-}\phi_-\right)
	+e^{-i\pi\frac{d-1}{2}}V\left(e^{i\frac{\pi}{2}\Delta_+}\phi_+ +e^{i\frac{\pi}{2}\Delta_-}\phi_-\right)\right],
\end{multline}
where $\Delta_{\pm}$ are the scaling dimensions of the dual CFT operators which are related to the mass of the scalar fields $\phi_{\pm}$ via $\Delta_{\pm}=\frac{d}{2}\pm\sqrt{\left(\frac{d}{2}\right)^{2}-(\ell_{dS}m)^{2}}$. To keep the discussion self-contained, we briefly review the analytical continuation of the fields and refer the reader to~\cite{Gorbenko:2019rza, Heckelbacher:2022hbq} for more details. Using the standard representation of the Schwinger-Keldysh or \textit{in-in} formalism, the field in de Sitter is first split into two fields $\phi_L$ and $\phi_R$. These are then analytically continued to anti-de Sitter via   $\phi_L(\eta = z e^{-i \frac\pi2}) = \phi_T, \ \phi_R(\eta = z e^{i \frac\pi2}) = \phi_A$. The fields $\phi_\pm$ appearing in the action above~\eqref{eq:EAdS_action_gen_pot} are expressed as a linear combination in terms of these via
\begin{equation}
\phi_A = e^{-i \frac\pi2 \Delta_+} \phi_+ + e^{-i \frac\pi2 \Delta_-} \phi_- , \qquad 
\phi_T = e^{i \frac\pi2 \Delta_+} \phi_+ + e^{i \frac\pi2 \Delta_-} \phi_- .
\end{equation}
Note that $\phi_\pm$ have kinetic terms with the opposite sign for all values of $\Delta$, so $\phi_+$ can be thought of as a ghost field. However this is not a concern for unitarity as the fields are not viewed as an analytical continuation of the fields in Lorentzian  anti-de Sitter. In this work we consider the case of the conformally coupled scalar with $\Delta_+={d+1\over2}$ and $\Delta_-={d-1\over2}$ with $d=3$. The action~\eqref{eq:EAdS_action_gen_pot} then becomes \footnote{This shows the simplicity in the number of vertices as compared to the Schwinger-Keldysh action, where we have 5 vertices for $\phi^4$ theory with no apriori simplification for any particular mass. }
\begin{multline}	\label{eq:EAdS_action}
	iS_c
	=\int\limits_0^{\infty}\frac{{\rm d} z {\rm d}^3x}{z^4}\left[\frac{1}{2}\left((\partial\phi^+)^2-m^2{\phi^+}^2\right)-\frac{1}{2}\left((\partial\phi^-)^2-m^2{\phi^-}^2\right)\right.\cr
	\left.- \frac{1}{2}{\lambda\over 4!} \left({\phi^+}^4-6{\phi^+}^2{\phi^-}^2+{\phi^-}^4\right)\right].
\end{multline}
Note that $\phi_+$ and $\phi_-$ satisfy Dirichlet and Neumann boundary conditions, respectively \footnote{The mass of the field falls within the unitarity bound and quantization with both Dirichlet and Newmann boundary conditions are possible~\cite{Klebanov:1999tb}.}. 

The equations above describe how the fields in dS are expressed in terms of the fields in EAdS. This can be then used to study cosmological correlation functions. At four-points we have the following contributions at the leading and subleading orders as $\eta \to 0$  (see \cite{Heckelbacher:2022fbx} for further details):
\begin{eqn}\label{eq:dStoAdS}
&\left\langle \phi(\vec{k}_{1})\phi(\vec{k}_{2}) \phi(\vec{k}_{3})\phi(\vec{k}_{4})\right\rangle_{dS} \\
 &= \eta^{4\Delta_-} \left\langle \phi_-(\vec{k}_{1})\phi_-(\vec{k}_{2}) \phi_-(\vec{k}_{3})\phi_-(\vec{k}_{4}) \right\rangle_{AdS} 
 + \eta^{4\Delta_+} \left\langle \phi_+(\vec{k}_{1})\phi_+(\vec{k}_{2}) \phi_+(\vec{k}_{3})\phi_+(\vec{k}_{4}) \right\rangle_{AdS} \\
 &+ \eta^{2(\Delta_- + \Delta_+)} \Big( \left\langle \phi_+(\vec{k}_{1})\phi_+(\vec{k}_{2}) \phi_-(\vec{k}_{3})\phi_-(\vec{k}_{4})\right\rangle_{AdS} 
 + 
 \left\langle \phi_+(\vec{k}_{1})\phi_-(\vec{k}_{2}) \phi_-(\vec{k}_{3})\phi_-(\vec{k}_{4})\right\rangle_{AdS}\\
 &\qquad\qquad \qquad + 
 \left\langle \phi_+(\vec{k}_{1})\phi_-(\vec{k}_{2}) \phi_-(\vec{k}_{3})\phi_+(\vec{k}_{4})\right\rangle_{AdS}\Big), 
\end{eqn}
where the left-hand-side corresponds to the in-in correlator in de Sitter while the right hand-side is expressed in terms of EAdS correlators. Note that the all-$\phi_-$ correlator dominates at late time ($\eta \to 0$), so this will be our primary focus, but the other correlators are also of interest for studying the dS/CFT correspondence \cite{Heckelbacher:2022fbx}. 

\begin{comment}
For the action given above the leading order contribution to the cosmological correlator arises from the first term in equation \eqref{eq:dStoAdS}, 
\begin{eqn}
\left\langle \phi(\vec{k}_{1})\phi(\vec{k}_{2}) \phi(\vec{k}_{3})\phi(\vec{k}_{4})\right\rangle_{dS} 
 &= \eta^{4\Delta_-} \left\langle \phi^-(\vec{k}_{1})\phi^-(\vec{k}_{2}) \phi^-(\vec{k}_{3})\phi^-(\vec{k}_{4}) \right\rangle_{AdS} ~.
 \end{eqn}
 Hence the Witten diagrams relevant for the cosmological correlator have the $\phi^-$ fields in their external legs. Due to the $\phi_{\pm} \to \phi_{\mp}$ symmetry of the potential in \eqref{eq:EAdS_action}, some diagrams with external $\phi^+$ are proportional to diagrams with external $\phi^-$ legs, but beyond 1-loop we find exmamples (see sections \ref{sec:cut-off} and \ref{sec:analytic-reg}) where there is a difference and we shall highlight them where appropriate.
\end{comment}

\subsection{Propagators in dS momentum space}

The momentum space decomposition of the bulk-to-bulk Feynman propagator in de Sitter with Dirichlet boundary conditions at future infinity, corresponding to $\Delta=2$, is (here $\eta,\eta'<0$)
\begin{align}\label{eq:ds_bulk}
G_D(\vec x, \eta, \vec x', \eta')&=\frac{\eta\eta'}{\pi}\int d^3 k\int\limits_0^\infty d\omega\;\frac{\sin(\omega \eta)\sin(\omega \eta')}{-\omega^2+\vec{k^2}-i\varepsilon}e^{i\vec{k}\cdot(\vec{x}-\vec{x}')},\cr
&=\frac{\eta\eta'
}{2\pi i}\int d^3 k\;\frac{1}{k}e^{i\vec{k}\cdot(\vec{x}-\vec{x}')}\left(e^{i(k-i\varepsilon)(\eta-\eta')}-e^{i(k-i\varepsilon)(\eta+\eta')}\right)\, .
\end{align}
Then, performing the remaining integral over ${\rm d}^3 k$ gives 
\begin{align}
G_D(\vec x, \eta, \vec x', \eta')&=\frac{i}{\pi}\left(\frac{2\eta\eta' e^{\varepsilon(\eta-\eta')}
}{| \vec{x}-\vec{x}'|^2-(\eta-\eta')^2}- \frac{2\eta\eta'e^{\varepsilon(\eta+\eta')}
}{| \vec{x}-\vec{x}'|^2-(\eta+\eta')^2}\right)\\
&=\frac{i}{\pi}\left(\frac{Ke^{\varepsilon(\eta-\eta')}}{1+K}-\frac{Ke^{\varepsilon(\eta+\eta')}}{1-K}\right),\nonumber
\end{align}
where
\begin{align}
K=\frac{2\eta\eta'
}{| \vec{x}-\vec{x}'|^2-\eta^2-\eta'^2}
\end{align}
is a function of the  anti-de Sitter-invariant distance, with $K\to -1$ at coincident points. % (i.e. for $X\to X'$). 
Note that the  Feynman $i\varepsilon$ does not deal with this light cone singularity.  

The bulk-to-bulk propagator with Neumann boundary conditions at future infinity, corresponding to  $\Delta=1$, is given by a similar expression with the replacement of $\sin$ functions by $\cos$ functions
\begin{equation}\label{eq:GN_1}
G_N(\vec x, \eta, \vec x', \eta')=-\frac{\eta\eta'}{\pi}\int d^3 k\int\limits_0^\infty d\omega\;\frac{\cos(\omega \eta)\cos(\omega \eta')}{-\omega^2+k^2-i\varepsilon}e^{i\vec{k}\cdot(\vec{x}-\vec{x}')},
\end{equation}
which evaluates to 
\begin{align}
G_N(\vec x, \eta, \vec x', \eta')&=-\frac{i}{\pi}\left(\frac{2\eta\eta' e^{\varepsilon(\eta-\eta')}
}{| \vec{x}-\vec{x}'|^2-(\eta-\eta')^2}+\frac{2\eta\eta'e^{\varepsilon(\eta+\eta')}
}{| \vec{x}+\vec{x}'|^2-(\eta+\eta')^2}\right)\\
&=-\frac{i}{\pi}\left(\frac{Ke^{\varepsilon(\eta-\eta')}}{1+K}+\frac{Ke^{\varepsilon(\eta+\eta')}}{1-K}\right).\nonumber
\end{align}
The minus sign in~\eqref{eq:GN_1}  reflects the fact that $\phi_-$ is ghost-like. 

Finally, for the bulk-boundary boundary propagator in de Sitter space we obtain a momentum space representation by letting $\eta'\to 0$ and divide \eqref{eq:ds_bulk} by $\eta'$ giving
\begin{align}\label{eq:ds_bulk_boun}
G(\vec x,\eta;\vec x')&=\frac{\eta}{\pi}\int d^3 k\int\limits_0^\infty d\omega\;\frac{\omega\,\sin(\omega \eta')}{-\omega^2+\vec{k^2}-i\epsilon}e^{i\vec{k}\cdot(\vec{x}-\vec{x}')},\\
&=\frac{\eta}{\pi}\int d^3 k\ e^{i\vec{k}\cdot(\vec{x}-\vec{x}')}e^{i(|\vec{k}|-i\epsilon)\eta}\,.\nonumber
\end{align}
Note that the effective action~\eqref{eq:EAdS_action} is defined in Euclidean
anti-de Sitter which is obtained from de Sitter by a double Wick
rotation
\begin{equation}
    \eta \to -iz,\qquad \ell_{dS}\to i \ell_{AdS}, \qquad \omega\to i p,
\end{equation}
see Section~2.1 of~\cite{Heckelbacher:2022hbq} for some details. Hence, We may apply these Wick rotations to the propagators computed above to obtain propagators in EAdS. 

\subsection{Feynman rules in EAdS}
For a conformally coupled scalar in EAdS, it is convenient to make the conformal mapping to half of $\mathbb{R}^{4}$ with a boundary at $z=0$ through 
\begin{align}
    g_{\mu\nu}\to \frac{1}{z^2}g_{\mu\nu}\,,\quad\phi_\pm\to z^{\frac{d-1}{2}}\phi_\pm\,,
\end{align}
giving the action 
\begin{align}\label{eq:action-conf1}
S[\phi_+,\phi_-]&=\int dz d^{d}x\left(- \frac{1}{2}(\p \phi_+)^2  +\frac{1}{2}(\p \phi_-)^2 
    - V(\phi_+,\phi_-)\right),
\end{align}
where $V(\phi_+, \phi_-) = \frac{1}{2}\frac{\lambda}{4!} \big( \phi_+^4 - 6 \phi_+^2 \phi_-^2 +  \phi_-^4 \big)$. The bulk-to-boundary propagators by the Lagrangian~\eqref{eq:action-conf1} are then as follows:\footnote{We suppress the boundary of AdS from all Witten diagrams.} 
\begin{eqn}\begin{gathered}
\begin{tikzpicture}[baseline]
\draw[very thick] (0, 1) -- (0.25, 0);
\node at (0, 1.25) {$\vec k$};
\end{tikzpicture}\end{gathered}
= e^{- k z}, \quad\quad\begin{gathered}
\begin{tikzpicture}[baseline]
\draw[dashed, very thick] (0, 1) -- (0.25, 0);
\node at (0, 1.25) {$\vec k$};
\end{tikzpicture}\end{gathered}
=- e^{- k z}~\,,
\end{eqn}
where $k\equiv |\vec{k}|$, solid lines describe $\phi_+$, and dotted lines describe $\phi_-$. In practice we will only consider diagrams with $\phi_-$ on external lines. 

The bulk-to-bulk propagator for $\phi_+$ is given as, 
\begin{eqn}\label{GD}\begin{gathered}
\begin{tikzpicture}[baseline]
\draw[very thick] (-1, 0) -- (1, 0);
\node at (-1, -0.25) {$z_1$};
\node at (1, -0.25) {$z_2$};
\node at (0, 0.25) {$\vec k$};    
\end{tikzpicture}    
\end{gathered}
&=G_D(z_1,z_2,k):= \frac{1}{\pi}\int_{-\infty}^\infty \frac{d p }{ p ^2 + \vec{k}^2} \sin( p  z_1) \sin( p  z_2) \\
&= \frac{1}{2k} \Big[\Theta(z_1 - z_2) e^{- k (z_1 - z_2)} + \Theta(z_2 - z_1) e^{- k (z_2 - z_1)} - e^{- k(z_1 + z_2)}  \Big] ~,  
\end{eqn}
and the bulk-to-bulk propagator for $\phi_-$ is given as, 
\begin{eqn}\label{GN}
\begin{tikzpicture}[baseline]
\draw[dashed, thick] (-1, 0) -- (1, 0);
\node at (-1, -0.25) {$z_1$};
\node at (1, -0.25) {$z_2$};
\node at (0, 0.25) {$\vec k$};    
\end{tikzpicture}
&=G_N(z_1,z_2,k):= -\frac{1}{\pi} \int_{-\infty}^\infty \frac{d p }{ p ^2 + k^2} \cos( p  z_1) \cos( p  z_2)\\
&= - \frac{1}{2k} \Big[ e^{- k(z_1 + z_2)} + \Theta(z_1 - z_2) e^{- k(z_1 - z_2)} + \Theta(z_2 - z_1) e^{- k(z_2 - z_1)} \Big]\,.
\end{eqn}
The appearance of the $\sin$'s and $\cos$'s in the above formulas encode the boundary conditions at $z=0$. One can also perform the integral over $\omega$ via as we did in eq.~\eqref{eq:ds_bulk}, and we will derive a recursion relation for the integrated form in the next section. However, in order to construct a de Sitter-invariant regularization of loops the unintegrated form is more convenient as shown in section \ref{sec:analytic-reg}. For the $\phi^4$ contact interaction in~\eqref{eq:action-conf1} we have (where an integration $\int\frac{dz}{z^4}$ is understood)
\begin{eqn}
\begin{tikzpicture}[baseline]
\draw[very thick] (-1, 0.5) -- (1, -0.5);
\draw[very thick] (1, 0.5) -- (-1, -0.5);
\end{tikzpicture} = \frac{\lambda}{2} 
, \qquad 
\begin{tikzpicture}[baseline]
\draw[dashed, very thick] (-1, 0.5) -- (1, -0.5);
\draw[very thick] (1, 0.5) -- (-1, -0.5);
\end{tikzpicture} = -3\lambda
, \qquad
\begin{tikzpicture}[baseline]
\draw[dashed, very thick] (-1, 0.5) -- (1, -0.5);
\draw[dashed, very thick] (1, 0.5) -- (-1, -0.5);
\end{tikzpicture} = \frac{\lambda}{2}\,.
\end{eqn}

%%%%%%%%%%%%%%%%%%%%%%%%%%%%%%%%%%%%%%%%%%%%%%%%%%%%%%%%%%%%%%%%%%%%%%%%
\section{Recursion relation for integrands}\label{sec:recursion}
In this section we derive a set of recursion relations that are useful for simplifying the integrand for cosmological correlation functions.\footnote{We provide a  Mathematica notebook with the submission that implements the recursion relations of this section for all the diagrams discussed in this paper.} A precursor of these recursion relations was introduced in~\cite{Arkani-Hamed:2017fdk} where the authors gave a recursive formula for obtaining the wavefunction of the universe for fields satisfying Dirichlet boundary conditions. This was based on using integration by parts and made use of the vanishing of the integrand at the boundary and does not rely on a specific interaction vertex. We shall first review their construction and then explain how that can be generalized to cosmological correlation functions. 

\subsection{Exchange of $\phi_+$}
Let us now illustrate the idea of the recursion relations using the simplest tree level example. Consider the expression for the diagram shown below 
\begin{eqn}
\begin{gathered}\begin{tikzpicture}
\draw[very thick] (-1, 0) -- (1, 0);
\draw[very thick, dashed] (-1, 0) -- (-1.25, 0.5);
\draw[very thick, dashed] (-1, 0) -- (-1, 0.5);
\draw[very thick, dashed] (-1, 0) -- (-0.75, 0.5);
\draw[very thick, dashed] (1, 0) -- (1.25, 0.5);
\draw[very thick, dashed] (1, 0) -- (1, 0.5);
\draw[very thick, dashed] (1, 0) -- (0.75, 0.5);
\node at (-1, -0.25) {$x_1$ };
\node at (1, -0.25) {$x_2 $ };
\node at (0, -0.25) {$k$ };
\end{tikzpicture}\end{gathered}
= \intsinf dz_1 dz_2 e^{- x_1 z_1}  e^{- x_2 z_2} G_D(z_1, z_2, k) ,
\label{6pt-1}\end{eqn}
where $x_1 = k_{123}$, $x_2 = k_{456}$. From the expression for $G_D(z_1, z_2, k)$ given in~\eqref{GD} it is clear that $G_D(0, z_2, k)= G_D(z_1, 0, k) = 0$~, i.e, it satisfies Dirichlet boundary conditions. We now insert the $z$-translation operator
\begin{eqn}\label{e:Delta2def}
\hat \Delta_2 :=  \partiald{}{z_1}  + \partiald{}{z_2},
\end{eqn}
in the integrand and consider the following integral 
\begin{eqn}\label{e:totalderivative}
\intsinf dz_1 dz_2 \hat \Delta_2 \Big[ e^{- x_1 z_1}  e^{- x_2 z_2} G_D(z_1, z_2, k) \Big] = 0,
\end{eqn}
where the vanishing of integral  follows from the boundary condition satisfied by $G_D(z_1, z_2, k)$. The action of the differential operator on $e^{- x_1 z_1}e^{- x_2 z_2}$ is trivial
\begin{eqn}\label{e:Delta2exp}
\hat \Delta_2 \left(e^{- x_1 z_1}  e^{- x_2 z_2}\right) = - (x_1 + x_2) e^{- x_1 z_1}  e^{- x_2 z_2},
\end{eqn}
and leaves us with the total energy factor $E=x_1+x_2=k_1+\cdots +k_6$.

By acting with the differential operator on $G_D(z_1, z_2, y)$, that the terms containing the $\Theta$-functions in~\eqref{GD} are annihilated because they are functions of $z_1 - z_2$ and hence we are left with the following:
\begin{eqn}\label{e:Delta2GD}
\hat \Delta_2 G_D(z_1, z_2, k) = - e^{- k z_1} e^{- k z_2}.
\end{eqn}
This is a crucial step in the computation as it converts the problem of integrating functions to a set of diagrammatic rules that can be interpreted as {clipping rules}. We shall give a detailed prescription to express these below. 

Expanding the total derivative~\eqref{e:totalderivative}, using the results of~\eqref{e:Delta2exp} and~\eqref{e:Delta2GD}, we can express the six-point correlator~\eqref{6pt-1} as
\begin{equation}
(x_1+x_2)~\intsinf dz_1 dz_2 e^{- x_1 z_1}  e^{- x_2 z_2} G_D(z_1, z_2, k)= -\intsinf dz_1 dz_2 e^{-(k+ x_1) z_1}  e^{-(k+ x_2) z_2},
\end{equation}
which can be diagrammatically expressed as 
\begin{equation}
(x_1 + x_2)\begin{tikzpicture}[baseline]
\draw[very thick] (-1, 0) -- (1, 0);
\draw[very thick, dashed] (-1, 0) -- (-1.25, 0.5);
\draw[very thick, dashed] (-1, 0) -- (-1, 0.5);
\draw[very thick, dashed] (-1, 0) -- (-0.75, 0.5);
\draw[very thick, dashed] (1, 0) -- (1.25, 0.5);
\draw[very thick, dashed] (1, 0) -- (1, 0.5);
\draw[very thick, dashed] (1, 0) -- (0.75, 0.5);
\node at (-1, -0.25) {$x_1$ };
\node at (1, -0.25) {$x_2 $ };
\node at (0, -0.25) {$k$ };
\end{tikzpicture}
= \begin{tikzpicture}[baseline]
\draw[very thick, dashed] (-1, 0) -- (-1.25, 0.5);
\draw[very thick, dashed] (-1, 0) -- (-1, 0.5);
\draw[very thick, dashed] (-1, 0) -- (-0.75, 0.5);
\draw[very thick, dashed] (1, 0) -- (1.25, 0.5);
\draw[very thick, dashed] (1, 0) -- (1, 0.5);
\draw[very thick, dashed] (1, 0) -- (0.75, 0.5);
\node at (-1, -0.25) {$x_1 + k$ };
\node at (1, -0.25) {$x_2 +k $ };
\end{tikzpicture}.
\end{equation}
Performing the integration over $z_1$ and $z_2$ on the right-hand-side then gives an expression for the six-point correlator in terms of a product of three-point correlators:
\begin{equation}
 \begin{tikzpicture}[baseline]
\draw[very thick] (-1, 0) -- (1, 0);
\draw[very thick, dashed] (-1, 0) -- (-1.25, 0.5);
\draw[very thick, dashed] (-1, 0) -- (-1, 0.5);
\draw[very thick, dashed] (-1, 0) -- (-0.75, 0.5);
\draw[very thick, dashed] (1, 0) -- (1.25, 0.5);
\draw[very thick, dashed] (1, 0) -- (1, 0.5);
\draw[very thick, dashed] (1, 0) -- (0.75, 0.5);
\node at (-1, -0.25) {$x_1$ };
\node at (1, -0.25) {$x_2 $ };
\node at (0, -0.25) {$k$ };
\end{tikzpicture} = \frac{1}{(x_1 + x_2)(x_1 + k)(x_2 + k)} .
\end{equation}
Proceeding in a similar manner, one can derive a set of recursion relations for higher-point graphs that can be summarized diagrammatically as,
\begin{eqn}\label{recursionD}
E
\ \begin{tikzpicture}[baseline]
\draw[color=gray!60, fill=gray!5, very thick] (-1.2,0) circle (0.5);
\draw[very thick] (-0.7, 0) -- (0.7, 0);
\node at (-0.7, 0) {\textbullet};
\node at (-0.5, -0.25) {\footnotesize$x_1$};
\node at (0.7, -0.25) {\footnotesize$x_2$};
\node at (0, 0.25) {\footnotesize$k$};
\draw[very thick, dashed] (0.7, 0) -- (0.9, 0.5);
\draw[very thick, dashed] (0.7, 0) -- (0.5, 0.5);
\draw[very thick, dashed] (0.7, 0) -- (0.7, 0.5);
\end{tikzpicture}
= 
\begin{tikzpicture}[baseline]
\draw[color=gray!60, fill=gray!5, very thick] (-1.2,0) circle (0.5);
\node at (-0.7, 0) {\textbullet};
\node at (-0.5, 0.25) {\footnotesize $x_1+k$};
\node at (0.5, -0.25) {\footnotesize $x_2+k$};
\draw[very thick, dashed] (0.5, 0) -- (0.75, 0.5);
\draw[very thick, dashed] (0.5, 0) -- (0.5, 0.5);
\draw[very thick, dashed] (0.5, 0) -- (0.25, 0.5);
\end{tikzpicture} 
\end{eqn}
where $E$ denotes the total energy of the external legs. These recursion relations can be used to derive the wave function in de Sitter~\cite{Arkani-Hamed:2017fdk} or transition amplitudes in anti-de Sitter.\footnote{For an extension of these recursion relations to spinning correlators, see~\cite{Albayrak:2019asr}~.}

\subsection{Exchange of the shadow $\phi_-$}
To compute \textit{in-in} correlators in de Sitter, we need to use the effective Lagrangian~\eqref{eq:EAdS_action} and therefore will have also diagrams with $\phi_-$ exchange. For the purpose of deriving the recursion relations for these, it suffices to consider the following $\phi_-$ exchange diagram using the bulk-bulk Green function given in~\eqref{GN}:
\begin{eqn}\label{GNexchange}
\begin{tikzpicture}[baseline]
\draw[dashed, very thick] (-1, 0) -- (1, 0);
\draw[very thick, dashed] (-1, 0) -- (-1.25, 0.5);
\draw[very thick, dashed] (-1, 0) -- (-1, 0.5);
\draw[very thick, dashed] (-1, 0) -- (-0.75, 0.5);
\draw[very thick, dashed] (1, 0) -- (1.25, 0.5);
\draw[very thick, dashed] (1, 0) -- (1, 0.5);
\draw[very thick, dashed] (1, 0) -- (0.75, 0.5);
\node at (-1, -0.25) {$x_1$ };
\node at (1, -0.25) {$x_2 $ };
\node at (0, -0.25) {$k$ };
\end{tikzpicture} =  \intsinf dz_1 dz_2 e^{- x_1 z_1} e^{- x_2 z_2} G_N(z_1, z_2, k).
\end{eqn}
The Green function~\eqref{GN} satisfies the Neumann boundary conditions and
\begin{eqn}
G_N(0, z, k) = G_N(z, 0, k) = - \frac{1}{k} e^{- k z}.
\end{eqn}
This implies that the integrand in~\eqref{GNexchange} does not go to zero as $z_i \to 0$. By inserting the $z$-translation operator of eq.~\eqref{e:Delta2def} inside the integrand we now also need to keep track of contributions arising from $z_i \to 0$: 
\begin{eqn}\label{boundaryterm}
&\intsinf dz_1 dz_2 \hat \Delta_2 \Big[ e^{- x_1 z_1}  e^{- x_2 z_2} G_N(z_1, z_2, k)  \Big]\\
&=  \intsinf dz_2 e^{- x_2 z_2} G_N(0, z_2, y) + \intsinf dz_1 e^{- x_1 z_1} G_N(z_1, 0, k)\\
&=  \frac{1}{k} \Big[ \frac{1}{x_2 +k} + \frac{1}{x_1 +k}\Big], 
\end{eqn}
which gets contributions from the two boundaries, $z_1=0$ and $z_2=0$. 
Thus the boundary term splits into a product of terms that can be generated recursively.
Using that 
\begin{equation}
    \hat \Delta_2 G_N(z_1,z_2,k)= - e^{-k(z_1+z_2)},
\end{equation}
and following similar steps as before we get
\begin{eqn}\label{bulkterm}
&\intsinf dz_1 dz_2 \hat \Delta_2 \Big[ e^{- x_1 z_1}  e^{- x_2 z_2} G_N(z_1, z_2, k) \Big] \\
&=- (x_1 + x_2) \begin{tikzpicture}[baseline]
\draw[dashed, very thick] (-1, 0) -- (1, 0);
\draw[very thick, dashed] (-1, 0) -- (-1.25, 0.5);
\draw[very thick, dashed] (-1, 0) -- (-1, 0.5);
\draw[very thick, dashed] (-1, 0) -- (-0.75, 0.5);
\draw[very thick, dashed] (1, 0) -- (1.25, 0.5);
\draw[very thick, dashed] (1, 0) -- (1, 0.5);
\draw[very thick, dashed] (1, 0) -- (0.75, 0.5);
\node at (-1, -0.25) {$x_1$ };
\node at (1, -0.25) {$x_2$ };
\node at (0, -0.25) {$k$ };
\end{tikzpicture}
+\begin{tikzpicture}[baseline]
\draw[very thick, dashed] (-1, 0) -- (-1.25, 0.5);
\draw[very thick, dashed] (-1, 0) -- (-1, 0.5);
\draw[very thick, dashed] (-1, 0) -- (-0.75, 0.5);
\draw[very thick, dashed] (1, 0) -- (1.25, 0.5);
\draw[very thick, dashed] (1, 0) -- (1, 0.5);
\draw[very thick, dashed] (1, 0) -- (0.75, 0.5);
\node at (-1, -0.25) {$x_1 + k$ };
\node at (1, -0.25) {$x_2 +k $ };
\end{tikzpicture}
\end{eqn}
Equating the contribution of the bulk term~\eqref{bulkterm} with the boundary term~\eqref{boundaryterm} we obtain 
\begin{eqn}
\begin{tikzpicture}[baseline]
\draw[dashed, very thick] (-1, 0) -- (1, 0);
\draw[very thick, dashed] (-1, 0) -- (-1.25, 0.5);
\draw[very thick, dashed] (-1, 0) -- (-1, 0.5);
\draw[very thick, dashed] (-1, 0) -- (-0.75, 0.5);
\draw[very thick, dashed] (1, 0) -- (1.25, 0.5);
\draw[very thick, dashed] (1, 0) -- (1, 0.5);
\draw[very thick, dashed] (1, 0) -- (0.75, 0.5);
\node at (-1, -0.25) {$x_1$ };
\node at (1, -0.25) {$x_2$ };
\node at (0, -0.25) {$k$ };
\end{tikzpicture} =  -\frac{1}{x_1 + x_2} \Bigg[ \underbrace{ 
\begin{tikzpicture}[baseline]
\draw[very thick, dashed] (1, 0) -- (1.25, 0.5);
\draw[very thick, dashed] (1, 0) -- (1, 0.5);
\draw[very thick, dashed] (1, 0) -- (0.75, 0.5);
\node at (-0.75, -0) {$\frac{1}{k}$ };
\node at (1, -0.25) {$x_2 +k $ };
\end{tikzpicture}
+  
\begin{tikzpicture}[baseline]
\draw[very thick, dashed] (1, 0) -- (1.25, 0.5);
\draw[very thick, dashed] (1, 0) -- (1, 0.5);
\draw[very thick, dashed] (1, 0) -- (0.75, 0.5);
\node at (-0.75, -0) {$\frac{1}{k}$ };
\node at (1, -0.25) {$x_1 +k $ };
\end{tikzpicture}
}_{\mbox{boundary}~\p z} 
- \underbrace{
\begin{tikzpicture}[baseline]
\draw[very thick, dashed] (-1, 0) -- (-1.25, 0.5);
\draw[very thick, dashed] (-1, 0) -- (-1, 0.5);
\draw[very thick, dashed] (-1, 0) -- (-0.75, 0.5);
\draw[very thick, dashed] (1, 0) -- (1.25, 0.5);
\draw[very thick, dashed] (1, 0) -- (1, 0.5);
\draw[very thick, dashed] (1, 0) -- (0.75, 0.5);
\node at (-1, -0.25) {$x_1 + k$ };
\node at (1, -0.25) {$x_2 +k $ };
\end{tikzpicture}
}_{\mbox{bulk }z}  \Bigg]
\end{eqn}
We therefore have a recursion relation obtained by snipping dashed lines, \footnote{Since these recursion relations only rely on snipping the propagators they can also be used for other interaction vertices such as $\phi_+^3 \phi_-$, etc.}
\begin{eqn}\label{recursionN}
E
\begin{tikzpicture}[baseline]
\draw[color=gray!60, fill=gray!5, very thick] (-1.2,0) circle (0.5);
\draw[dashed, very thick] (-0.7, 0) -- (0.7, 0);
\node at (-0.7, 0) {\textbullet};
\node at (-0.5, -0.25) {\footnotesize$x_1$};
\node at (0.7, -0.25) {\footnotesize$x_2$};
\draw[very thick, dashed] (0.7, 0) -- (0.95, 0.5);
\draw[very thick, dashed] (0.7, 0) -- (0.7, 0.5);
\draw[very thick, dashed] (0.7, 0) -- (0.45, 0.5);
\node at (0, 0.25) {\footnotesize$k$};
\end{tikzpicture}
= 
- \begin{tikzpicture}[baseline]
\draw[color=gray!60, fill=gray!5, very thick] (-1.2,0) circle (0.5);
\node at (-0.7, 0) {\textbullet};
\node at (-0.5, 0.25) {\footnotesize $x_1+k$};
\node at (0.5, -0.25) {\footnotesize $x_2+k$};
\draw[very thick, dashed] (0.5, 0) -- (0.75, 0.5);
\draw[very thick, dashed] (0.5, 0) -- (0.5, 0.5);
\draw[very thick, dashed] (0.5, 0) -- (0.25, 0.5);
\end{tikzpicture} 
- \quad 
\begin{tikzpicture}[baseline]
\draw[color=gray!60, fill=gray!5, very thick] (-1.2,0) circle (0.5);
\node at (-0.7, 0) {\textbullet};
\node at (-0.5, -0.25) {\footnotesize $x_1+k$};
\node at (0.5, 0) {$\frac{1}{k}$};
\end{tikzpicture}
- \quad
\begin{tikzpicture}[baseline]
\draw[color=gray!60, fill=gray!5, very thick] (-1.2,0) circle (0.5);
\node at (-0.7, 0) {\textbullet};
\node at (-0.5, 0.25) {\footnotesize 0};
\node at (0.7, -0.25) {\footnotesize $x_2 + k$};
\node at (0, 0) {$\frac{1}{k}$};
\draw[very thick, dashed] (0.7, 0) -- (0.95, 0.5);
\draw[very thick, dashed] (0.7, 0) -- (0.7, 0.5);
\draw[very thick, dashed] (0.7, 0) -- (0.45, 0.5);
\end{tikzpicture}
\end{eqn}

\subsection{Cosmological correlators}
In the evaluation of cosmological correlators we have to sum the exchange of the $\phi_+$ field using the recursion~\eqref{recursionD} and the exchange of the $\phi_-$ field using the recursion~\eqref{recursionN}.
 When adding these  contributions, there is  a cancellation between two terms and the poles that remain in the final answer are simpler than the ones that were originally present for the wavefunction. This shows that the \textit{in-in} correlator has simpler pole structure than the corresponding wave function coefficient.\footnote{We find that the loop integrals that appear for the wavefunction compared to the correlator compared to the flat space amplitude are of the following schematic form 
$\int \frac{d^3 l}{|\bm l + k|} \mbox{vs} \int \frac{d^3 l}{|\bm l|} \mbox{vs} \int \frac{d^4 l}{l^2}$. This expression shows the simplicity of the singularities for the cosmological correlator as compared to the corresponding wavefunction and their resemblance with the scattering amplitude in flat space (a similar simplification was also noticed in~\cite{Lee:2023jby} for the one-loop tadpole).} We explicitly demonstrate the evaluation of the loop integrand for the triangle diagram using these recursion relations in appendix \ref{app:tria-int}.

The recursion relations allow one to express the correlation functions in terms of partial fractions and make the physical singularities manifest. To summarize, the order of performing the integrals is as follows. One first performs the auxiliary frequency integrals (the variable $\omega$ in~\eqref{GD} and~\eqref{GN}) to express the Green functions as a function of $(z_i, \vec k)$. The integrals over $z_i$ are then performed recursively as explained in this section.

Having derived the recursion relations it is now possible to avoid any reference to the bulk and use these as a tool to express the {\textit{in-in}} correlators in terms of functions depending on the boundary momenta. For diagrams that are finite so that the loop integration does not need to be regulated, the recursion relations~\eqref{recursionD} and~\eqref{recursionN} result in an important simplification of higher-order diagrams. This will be applied in the next section. If the loop integrals are infinite, the recursion relations can still be used to derive the integrand provided one uses a simple regulator like the cut-off while integrating. However, the recursions cannot be used in this form for the de Sitter-invariant regularization described in Section~\ref{sec:analytic-reg}.

%%%%%%%%%%%%%%%%%%%%%%%%%%%%%%%%%%%%%%%%%%%%%%%%%%%%%%%%%%%%%%%%%%%%%%%%%%%%%%%%%%%%%%%%%%%%%%%%%%%%%%%%%%%%%%%%%%

\section{Momentum cut-off regularisation}\label{sec:cut-off}

In this section, we compute cosmological correlators using a cut-off in the boundary loop momentum. In particular, we compute two and four-point correlators up to two loops and an infinite class of 1-loop correlators arising from polygon diagrams. These can be computed using the Feynman rules in section \ref{sec:review} and we show that due to non-trivial contributions from the ghosts they can be expressed in terms of flat space integrands. These integrands will have four-dimensional Lorentz invariance, but the four-dimensional covariance is actually broken by the boundary at $z=0$ so the integrals over the loop momentum in this direction are computed by taking residues in the upper half plane. The integrands of the resulting 3d loop integrals can be derived more directly using the recursion relations described in section \ref{sec:recursion} and are regulated using a hard momentum cut-off. Likewise, when performing dimensional regularisation only the three-dimensional loop momentum integration is regulated. A covariant four-dimensional analytic regularisation is described in section~\ref{sec:analytic-reg}.

We also perform renormalisation of two-point correlators up to two loops and four-point correlators at one-loop and show that the boundary conformal symmetry is generically broken by the cut-off, but can be restored by setting the renormalisation scale proportional to the energy. This result will be reproduced more systematically and generalised using analytic regularisation in the next section. An alternative cut-off prescription was considered for 2-point cosmological correlators in~\cite{Senatore:2009cf}, where it was pointed out that logarithms of the energy should not appear. We find a similar result after restoring conformal symmetry in the manner described above. See section \ref{sec:renormalization} for more details.

%----------------------------------------------------
\subsection{Two-point correlators}

In this section we evaluate the two-point correlation functions  of $\langle \phi_- \phi_-\rangle$ up to two-loop order.

\subsection*{One-loop tadpole}

The one-loop tadpole is given by the sum of $\phi_+$ and   $\phi_-$ fields running in the loop: 
\begin{align}\label{eq:tad_cut}
I^{(2)}_\circ(k)&:= \begin{gathered}
     \tadpoleD{2k}{l}\end{gathered}\quad +\quad \begin{gathered}\tadpoleN{2k}{l}
\end{gathered}\cr
&=\lambda\int\frac{d^{3} ld p }{(2\pi)^{4}}\frac{1}{ p ^{2}+{l}^{2}}\int_{0}^{\infty}dze^{-2kz}\left(\sin^{2}( p  z)+\cos^{2}\left( p  z\right)\right)\cr
&=\frac{\lambda}{ k}\int\frac{d^{4}L}{(2\pi)^{4}}\frac{1}{L^{2}}    \,,
\end{align}
where the subscript denotes the topology of the diagram and the superscript indicates the number of external legs. 
The tadpole integral is expressed using the  Euclidean four-dimensional loop momentum $L:=\big(\omega,\vec{l}\big)$ where $\vec l$ is integrated over $\mathbb R^3$ and $\omega$ over the real axis. Here (and everywhere in this section) the cut-off is understood to act only on the boundary components $\vec{l}$ of the loop integral.  
Hence, from \eqref{eq:tad_cut}, we see that the Dirichlet and Neumann boundary conditions of the two internal propagators combine to give an integrand which has 4d Lorentz invariance and we end up with the same one-loop integrand we would get
in flat space (without a boundary) times an overall energy pole $1/k$.
This simple example illustrates a general phenomenon for \textit{in-in} correlators
that we will see in more complicated examples.

To evaluate the integral, we first compute residue at $\omega=i| l|$
giving an integral over the boundary loop momentum (where the integrand could have been derived using the recursion relations given in the previous section)\footnote{We implement the hard-cutoff procedure as usually done in flat space. For alternative prescriptions where the cutoff changes with the radial coordinate, we refer the reader to \cite{Senatore:2009cf}. }:
\begin{equation}
I^{(2)}_\circ(k)=\frac{\lambda}{ k}\int_{| l|\leq \Lambda}\frac{d^{3} l}{(2\pi)^{3}}\frac{1}{2| l|}=\frac{\lambda}{k}\frac{\Lambda^{2}}{8\pi^{2}}.
\label{tadpole1}
\end{equation}
To evaluate the integral over the loop momentum, we went to polar coordinates and introduced the cutoff $\Lambda$ on
the magnitude of the loop momentum. Again, this simple example illustrates
the general method we will use for performing integration using a
cut-off in more complicated examples: first we compute the residues
of the radial momenta in the upper half-plane to yield an integral over
boundary loop momentum, and then we evaluate the remaining integrals by
going to polar coordinates and imposing a cutoff on the magnitude
of the loop momentum. The quadratic divergence needs the introduction of a counterterm. We will discuss the renormalisation in Section~\ref{sec:renormalization}.

%----------------------------------------------------

\subsection*{Two-loop tadpole}
Next we turn to the two-loop tadpole, which is given by the sum of four contributions (we label the radial coordinates of the vertex factors by $x$ and $z$)
\begin{equation}
I^{(2)}_{\circ\circ}(k):=\begin{gathered}
\tadpoleDD{2k}{}{l_1}{l_2}\end{gathered}+\begin{gathered}
    \tadpoleDN{2k}{}{l_1}{l_2}\end{gathered}+\begin{gathered}
    \tadpoleND{2k}{}{l_1}{l_2}\end{gathered}+ \begin{gathered}\tadpoleNN{2k}{}{l_1}{l_2},
\end{gathered}
\end{equation}
which reads
\begin{multline}
I_{\circ\circ}^{(2)}(k)=\frac{1}{2}\frac{\lambda^2}{(2\pi)^9}\int\frac{d^{3}l_{1}d^{3}l_{2}d p _{1}d p _{2}d p _{3}}{\left( p _{1}^{2}+l_{1}^{2}\right)\left( p _{2}^{2}+l_{2}^{2}\right)\left( p _{3}^{2}+l_{1}^{2}\right)}\cr
\times\int_0^\infty dzdxe^{-2kz}\left(\cos p _{1}x\cos p _{3}x\cos p _{1}z\cos p _{3}z+\cos\leftrightarrow\sin\right).
\end{multline}
The boundary loop momenta $l_i$ are integrated over the three-dimensional space $\mathbb R^3$ and the radial momenta $p_i$ over the real axis.
The integral over $x$ is oscillatory so it needs to be regulated at
$x\rightarrow\infty$ (this regularization was discussed in detail in appendix~A of~\cite{Chowdhury:2023khl}). For this we  introduce 
a damping factor

\begin{equation}
\int_{0}^{\infty}dx\cos p _{1}x\cos p _{3}x=\lim_{\epsilon\rightarrow0}\int_{0}^{\infty}dxe^{-\epsilon x}\cos p _{1}x\cos p _{3}x,
\end{equation}
where $\epsilon>0$. It is now straightforward to evaluate the integral
over $x$, with the help of the identity
\begin{equation}
\lim_{\epsilon\rightarrow0}\frac{\epsilon}{\epsilon^{2}+y^{2}}=\pi\delta(y)
\label{deltaidentity}
\end{equation}
to obtain
\begin{align}
\lim_{\epsilon\rightarrow0}\int_{0}^{\infty}dxe^{-\epsilon x}\cos p _{1}x\cos p _{3}x&=\lim_{\epsilon\rightarrow0}\left(\frac{\epsilon/2}{\epsilon^{2}+\left( p _{1}- p _{3}\right)^{2}}+\frac{\epsilon/2}{\epsilon^{2}+\left( p _{1}+ p _{3}\right)^{2}}\right)\nonumber\\
&=\frac{\pi}{2}\left(\delta( p _{1}- p _{3})+\delta( p _{1}+ p _{3})\right)\,.
\end{align}
Similarly, we regulate the integral 
\begin{align}
\lim_{\epsilon\rightarrow0}\int_{0}^{\infty}dxe^{-\epsilon x}\sin p _{1}x\sin p _{3}x&=\lim_{\epsilon\rightarrow0}\int_{0}^{\infty}dxe^{-\epsilon x}\sin p _{1}x\sin p _{3}x\cr
&={\pi\over 2}\left( \delta( p _1- p _3)-\delta( p _1+ p _3)\right).
\end{align}
Performing the integration over $x$ and $ p _{3}$ then gives
\begin{align}
I_{\circ\circ}^{(2)}(k)&=\frac{\pi^2}{2}\frac{\lambda^2}{(2\pi)^9}\int\frac{d^{3}l_{1}d p _{1}d^{3}l_{2}d p _{2}}{\left( p _{1}^{2}+l_{1}^{2}\right)^2\left( p _{2}^{2}+l_{2}^{2}\right)}\int_0^\infty dze^{-2kz}\left(\cos^{2} p _{1}z+\sin^{2} p _{1}z\right)\cr
&=\frac{\pi^2}{4k}\frac{\lambda^2}{(2\pi)^9}\int\frac{d^{4}L_{1}d^{4}L_{2}}{\left(L_{1}^{2}\right)^{2}L_{2}^{2}}\,,
\label{2pt2loopa}
\end{align}
where $L_{1}:=\left( p _{1},\vec{l}_{1}\right)$, $L_{2}:=\left( p _{2},\vec{l}_{2}\right)$.
Hence, we are once again left with a  flat space four-dimensional Feynman integral multiplied by an
 energy pole. 
Integrating out $ p _{1}$ and $ p _{2}$
via residues then produces the same integrand as the one obtained from the recursion presented in Section~\ref{sec:recursion}:
\begin{eqn}
I_{\circ\circ}^{(2)}(k) &\propto\frac{\lambda^2}{ k} \int {d^3 l_1 d^3 l_2\over l_1^3 l_2} \propto  \frac{\Lambda^2}{2k} \log \frac{\Lambda}{\Lambda_{IR}},
\label{2cactus}
\end{eqn}
where $\Lambda_{IR}$ denotes the lower limit of the $l_1$ integral. This is similar to the behavior in flat space and is zero in a scale invariant regularization.

Similarly the one-particle reducible  diagram with two one-loop tadpoles connected by a propagator vanishes as it is proportional to the square of the single tadpole:
\begin{eqn}
&\begin{tikzpicture}
\draw[very thick, dashed] (-1, 0) -- (1, 0);
\draw[very thick] (1, 0.5) circle  (0.5);
\draw[very thick] (-1, 0.5) circle  (0.5);
\draw[very thick, dashed] (-1, 0) -- (-1.7, 0);
\draw[very thick, dashed] (1, 0) -- (1+0.7, -0);
\end{tikzpicture}  
+ 
\begin{tikzpicture}
\draw[very thick, dashed] (-1, 0) -- (1, 0);
\draw[dashed, very thick] (-1, 0.5) circle  (0.5);
\draw[very thick] (1, 0.5) circle  (0.5);
\draw[very thick, dashed] (-1, 0) -- (-1.7, 0);
\draw[very thick, dashed] (1, 0) -- (1+0.7, -0);
\end{tikzpicture} 
+ 
\begin{tikzpicture}
\draw[very thick, dashed] (-1, 0) -- (1, 0);
\draw[very thick] (-1, 0.5) circle  (0.5);
\draw[dashed, very thick] (1, 0.5) circle  (0.5);
\draw[very thick, dashed] (-1, 0) -- (-1.7, 0);
\draw[very thick, dashed] (1, 0) -- (1+0.7, -0);
\end{tikzpicture} 
+ 
\begin{tikzpicture}
\draw[very thick, dashed] (-1, 0) -- (1, 0);
\draw[dashed, very thick] (-1, 0.5) circle  (0.5);
\draw[dashed, very thick] (1, 0.5) circle  (0.5);
\draw[very thick, dashed] (-1, 0) -- (-1.7, 0);
\draw[very thick, dashed] (1, 0) -- (1+0.7, -0);
\end{tikzpicture}\\
&\propto [\mbox{Single Tadpole}]^2 \int_0^\infty dz_1 dz_2 G_N(z_1, z_2, k) = 0. 
\end{eqn}

%----------------------------------------------
\subsection*{Two-loop sunset}

We turn the two-loop sunset diagram contributions with the loop momentum $y\equiv|\vec l_1 + \vec k - \vec l_2|$ in the diagrams below. There are two types of diagrams, one where all the propagators are the same and one where two of the propagators are shadow fields. The first comes with a symmetry factor of $\frac{1}{6}$ while the second type comes with a symmetry factor of $\frac{1}{2}$, so it is convenient to split this diagram into three contributions as illustrated in the figure below: 
\begin{multline}
 \label{sun-rec-integrand}
I^{(2)}_{\circleddash}(k):=
\begin{tikzpicture}[baseline]
\draw[very thick, dashed] (1,0) circle (1);
\draw[very thick, dashed] (0,0) -- (2,0);
\node at (-0.25, 0.25) {$k$};
\node at (1,1.25) {$l_1$};
\node at (1,0.25) {$y$};
\node at (1,-1.25) {$l_2$};
\draw[very thick, dashed] (2,0) -- (2.7, 0);
\draw[very thick, dashed] (0,0) -- (-0.7, 0);
\end{tikzpicture}
+
\begin{tikzpicture}[baseline]
\draw[ very thick] (1,0) circle (1);
\draw[very thick, dashed] (0,0) -- (2,0);
\node at (-0.25, 0.25) {$k$};
\node at (1,1.25) {$l_1$};
\node at (1,0.25) {$y$};
\node at (1,-1.25) {$l_2$};
\draw[very thick, dashed] (2,0) -- (2.7, 0);
\draw[very thick, dashed] (0,0) -- (-0.7, 0);
\end{tikzpicture}
\cr
+\begin{tikzpicture}[baseline]
\draw[ very thick] (1,0) circle (1);
\draw[very thick, dashed](0,0) -- (2,0);
\node at (-0.25, 0.25) {$k$};
\node at (1,1.25) {$y$};
\node at (1,0.25) {$l_1$};
\node at (1,-1.25) {$l_2$};
\draw[very thick, dashed] (2,0) -- (2.7, 0);
\draw[very thick, dashed] (0,0) -- (-0.7, 0);
\end{tikzpicture}
 + 
 \begin{tikzpicture}[baseline]
\draw[ very thick] (1,0) circle (1);
\draw[very thick, dashed] (0,0) -- (2,0);
\node at (-0.25, 0.25) {$k$};
\node at (1,1.25) {$l_1$};
\node at (1,0.25) {$l_2$};
\node at (1,-1.25) {$y$};
\draw[very thick, dashed] (2,0) -- (2.7, 0);
\draw[very thick, dashed] (0,0) -- (-0.7, 0);
\end{tikzpicture}.
\end{multline}

For example, the first diagram where all the propagators are the same
is given by
\begin{multline}\label{e:sunsetphi}
\begin{tikzpicture}[baseline]
\draw[very thick, dashed] (1,0) circle (1);
\draw[very thick, dashed] (0,0) -- (2,0);
\node at (-0.25, 0.25) {$k$};
\node at (1,1.25) {$l_1$};
\node at (1,0.25) {$y$};
\node at (1,-1.25) {$l_2$};
\draw[very thick, dashed] (2,0) -- (2.7, 0);
\draw[very thick, dashed] (0,0) -- (-0.7, 0);
\end{tikzpicture}=
{ \lambda^2\over (2\pi)^9}\int\frac{d^{3}l_{1}d^{3}l_{2}d p _{1}d p _{2}d p _{3}}{\left( p _{1}^{2}+l_{1}^{2}\right)\left( p _{2}^{2}+l_{2}^{2}\right)\left( p _{3}^{2}+\left(\vec{l}_{1}+\vec{l}_{2}+\vec{k}\right)^{2}\right)}\cr
\qquad \qquad \times\int_0^\infty dxdze^{-k(x+z)}\cos p _{1}x\cos p _{1}z\cos p _{2}x\cos p _{2}z\cos p _{3}x\cos p _{3}z    
\end{multline}
Performing integrals over $x$ and $z$ gives 
\begin{align}
\eqref{e:sunsetphi}
&=
{\pi^2 \lambda^2 \over (2\pi)^9}\int\frac{d^{3}l_{1}d^{3}l_{2}d p _{1}d p _{2}d p _{3}}{\left( p _{1}^{2}+l_{1}^{2}\right)\left( p _{2}^{2}+l_{2}^{2}\right)\left( p _{3}^{2}+\left(\vec{l}_{1}+\vec{l}_{2}+\vec{k}\right)^{2}\right)}
\nonumber\\
&\times \Bigg(\frac{ p _{1}+ p _{2}- p _{3}}{k^{2}+\left( p _{1}+ p _{2}- p _{3}\right)^{2}}+\frac{ p _{1}- p _{2}+ p _{3}}{k^{2}+\left( p _{1}- p _{2}+ p _{3}\right)^{2}} +\frac{- p _{1}+ p _{2}+ p _{3}}{k^{2}+\left(- p _{1}+ p _{2}+ p _{3}\right)^{2}}\nonumber\\
&\qquad -\frac{ p _{1}+ p _{2}+ p _{3}}{k^{2}+\left( p _{1}+ p _{2}+ p _{3}\right)^{2}}\Bigg)^{2}.    
\end{align}
We can similarly evaluate the other three diagrams. After performing a
change of variables, we obtain a very compact result for the sum over all the
sunset diagrams:
\begin{equation}
I^{(2)}_{\circleddash}(k)\propto \lambda^2\int\frac{d p  k^{2}}{( p ^{2}+k^{2})^2}\int\frac{d^{3}l_{1}d^{3}l_{2}d p _{1}d p _{2}}{\left( p _{1}^{2}+l_{1}^{2}\right)\left( p _{2}^{2}+l_{2}^{2}\right)\left(\left( p + p _{1}+ p _{2}\right)^{2}+\left(\vec{l}_{1}+\vec{l}_{2}+\vec{k}\right)^{2}\right)}.
\end{equation}
As before this expression can be expressed 
in terms of a four-dimensional flat space integrand by introducing the four-vectors  $L_{i}:=\left( p _{i},\vec{l}_{i}\right)$, $P:=\left( p ,{k}\right)$
\begin{equation}
I^{(2)}_{\circleddash}(k)\propto \lambda^2\int\frac{d p  k^{2}}{( p ^{2}+k^{2})^2}\int\frac{d^{4}L_{1}d^{4}L_{2}}{L_{1}^{2}L_{2}^{2}\left(L_{1}+L_{2}+P\right)^{2}}.  
\end{equation}
By taking the residues of $p_{1,2,3}$ in upper half-plane, we once again obtain an integrand which matches with the result from recursion: 
\begin{align}
I^{(2)}_{\circleddash}(k)&\propto \lambda^2 \frac{i}{16}\int\limits^\Lambda d^{3}l_{1}d^{3}l_{2}\frac{2k + l_{1}+l_{2}+\left|\vec{l}_{1}+\vec{l}_{2}+\vec{k}\right|}{l_{1}l_{2}\left|\vec{l}_{1}+\vec{l}_{2}+\vec{k}\right|\left(l_{1}+l_{2}+2k+\left|\vec{l}_{1}+\vec{l}_{2}+\vec{k}\right|\right)^{2}} \nonumber\\
&=\lambda^2 \frac{i}{16}\frac{k\pi^2}{12} \left[ -30 \log \left( \frac{2\Lambda}{3k} \right) + 12 \left(\Lambda\over k\right)^2  - 5 \right] 
\label{sunsetintegrated}
\end{align}
The renormalization of this diagram is discussed in section \ref{sec:analytic-reg}.

\subsection{Four-point correlators}

In this section we evaluate the four-point correlation functions $\phi_+$ fields up to two-loop order.
%------------------------------------------------

\subsection*{One-loop bubble}
At 1-loop, we need to consider bubble diagrams. The s-channel bubble diagrams are
\begin{equation}
 I^{(4)}_\circ=\frac{\lambda^2}{8} \left( \begin{gathered}\begin{tikzpicture}[baseline,scale=0.6]
  \filldraw [color = black, fill=none, very thick] (0,1) circle (1cm);
  \draw [black,very thick, dashed] (0.5,-1) to (0,0);
    \draw [black,very thick, dashed] (-0.5,-1) to (0,0);
 \draw [black,very thick, dashed] (0.5,3) to (0,2);
    \draw [black,very thick, dashed] (-0.5,3) to (0,2);
  \end{tikzpicture}\end{gathered}
   + \begin{gathered} 
  \begin{tikzpicture}[baseline,scale=0.6]
  \filldraw [color = black, fill=none, very thick,dashed] (0,1) circle (1cm);
  \draw [black,very thick, dashed] (0.5,-1) to (0,0);
    \draw [black,very thick, dashed] (-0.5,-1) to (0,0);
 \draw [black,very thick, dashed] (0.5,3) to (0,2);
    \draw [black,very thick, dashed] (-0.5,3) to (0,2);
\end{tikzpicture} \end{gathered}  \right)
\label{fig:bubble}
    \end{equation}
Adding the two contributions gives
\begin{align}
I^{(4)}_\circ&= \frac{\lambda^2}{2}\int\frac{d^{3}ld p _{1}d p _{2}}{(2\pi)^{5}\left( p _{1}^{2}+l^{2}\right)\left( p _{2}^{2}+\left(\vec{l}-\vec{k}_{12}\right)^{2}\right)}\\
&\quad \times\intsinf dz_{1}dz_{2}\left(\sin p _{1}z_{1}\sin p _{1}z_{2}\sin p _{2}z_{1}\sin p _{2}z_{2}+\sin\leftrightarrow\cos\right) e^{- k_{12} z_1} e^{- k_{34} z_2}\nonumber\\
&=\left(\lambda\over2\right)^2k_{12}k_{34}\int\frac{d p _{+}}{(2\pi)^5}\frac{1}{\left( p _{+}^{2}+k_{12}^{2}\right)\left( p _{+}^{2}+k_{34}^{2}\right)}\int\frac{d^{4}L}{L^{2}(L-P)^{2}},
\label{bubble}
\end{align}
where $ p _{+}= p _{1}+ p _{2}$, $L=\left( p _{1},\vec{l}\right)$, and
$P=\left( p _{+},\vec{k}_{12}\right)$ and $\vec k_{12} = \vec k_1 + \vec k_2$. 
After integrating $ p _{1}$
and $ p _{2}$ via residues in the upper half plane we obtain
\begin{equation}\label{bubble-rec}
I^{(4)}_\circ=\frac{\lambda^{2}}{16(k_{12}+k_{34})}\int\frac{d^{3}l}{(2\pi)^{3}}\frac{l+\left|\vec{l}-\vec{k}_{12}\right|+k_{12}+k_{34}}{l\left|\vec{l}-\vec{k}_{12}\right|\left(l+\left|\vec{l}-\vec{k}_{12}\right|+k_{12}\right)\left(l+\left|{l}-\vec{k}_{12}\right|+k_{34}\right)},
\end{equation}
in agreement with the result of recursion. Going to polar coordinates and integrating over the magnitude of the
loop momentum with cut-off $\Lambda$ gives
\begin{equation}
I^{(4)}_\circ=\frac{1}{(2\pi)^{2}}\frac{\lambda^{2}}{32(k_{12}+k_{34})}\left[\ln\left(\frac{(k_{12}+|\vec k_{12}|)(k_{34}+|\vec k_{12}|)}{4\Lambda^{2}}\right)+\frac{k_{12}+k_{34}}{k_{12}-k_{34}}\ln\left(\frac{k_{34}+|\vec k_{12}|}{k_{12}+|\vec k_{12}|}\right)\right]\,.
\label{1loop4ptcutoff}
\end{equation}
By comparing with the computation for the wave function coefficient for the bubble diagram at one-loop~\cite{Albayrak:2020isk}, we see that the cosmological correlator does indeed take a simpler form. In particular, it  has lower transcendentality.  

\subsection*{Two-loop necklace}

Next we will look at the necklace topology. Four diagrams contribute to the s-channel:
\begin{figure}[H]
\centering\begin{tikzpicture}[scale=0.6]
  \filldraw [color = black, fill=none, very thick] (0,1) circle (1cm);
  \filldraw [color = black, fill=none, very thick] (0,-1) circle
  (1cm);
  \draw [black,very thick, dashed] (0.5,-3) to (0,-2);
    \draw [black,very thick, dashed] (-0.5,-3) to (0,-2);
 \draw [black,very thick, dashed] (0.5,3) to (0,2);
    \draw [black,very thick, dashed] (-0.5,3) to (0,2);
    \node at (0.65, 2.25) {$z_1$};
\node at (1.15, 0) {$z_3$};
\node at (0.65, -2.25) {$z_2$};
  \end{tikzpicture}
  \qquad\begin{tikzpicture}[scale=0.6]
  \filldraw [color = black, fill=none, very thick,dashed] (0,1) circle (1cm);
  \filldraw [color = black, fill=none, very thick] (0,-1) circle
  (1cm);
  \draw [black,very thick, dashed] (0.5,-3) to (0,-2);
    \draw [black,very thick, dashed] (-0.5,-3) to (0,-2);
 \draw [black,very thick, dashed] (0.5,3) to (0,2);
    \draw [black,very thick, dashed] (-0.5,3) to (0,2);
    \node at (0.65, 2.25) {$z_1$};
\node at (1.15, 0) {$z_3$};
\node at (0.65, -2.25) {$z_2$};
\end{tikzpicture}\qquad\begin{tikzpicture}[scale=0.6]
  \filldraw [color = black, fill=none, very thick] (0,1) circle (1cm);
  \filldraw [color = black, fill=none, very thick,dashed] (0,-1) circle
  (1cm);
  \draw [black,very thick, dashed] (0.5,-3) to (0,-2);
    \draw [black,very thick, dashed] (-0.5,-3) to (0,-2);
 \draw [black,very thick, dashed] (0.5,3) to (0,2);
    \draw [black,very thick, dashed] (-0.5,3) to (0,2);
    \node at (0.65, 2.25) {$z_1$};
\node at (1.15, 0) {$z_3$};
\node at (0.65, -2.25) {$z_2$};
\end{tikzpicture}\qquad\begin{tikzpicture}[scale=0.6]
  \filldraw [color = black, fill=none, very thick,dashed] (0,1) circle (1cm);
  \filldraw [color = black, fill=none, very thick,dashed] (0,-1) circle
  (1cm);
  \draw [black,very thick, dashed] (0.5,-3) to (0,-2);
    \draw [black,very thick, dashed] (-0.5,-3) to (0,-2);
 \draw [black,very thick, dashed] (0.5,3) to (0,2);
    \draw [black,very thick, dashed] (-0.5,3) to (0,2);
\node at (0.65, 2.25) {$z_1$};
\node at (1.15, 0) {$z_3$};
\node at (0.65, -2.25) {$z_2$};
\end{tikzpicture}
\caption{Two-loop necklace. The vertices are labelled as $z_1, z_2, z_3$}\label{fig:necklace2loop}
\end{figure}
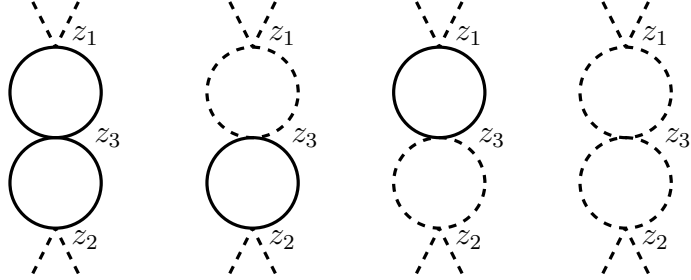

Adding up the $s$-channel diagrams we obtain
\begin{multline}
I^{(4)}_{\circ\circ}\propto\lambda^3\int_{0}^{\infty} dz_{1}dz_{2}dz_{3} \int \frac{d^{3}l_{1}d^{3}l_{2}d p _{1}d p _{2}d p _{3}d p _{4}\;\;e^{-k_{12}z_{1}-k_{34}z_{2}}}{( p _1^2 + l_1^2) ( p _2^2 + ( l_1 + \vec k_{12})^2)( p _3^2 + l_2^2) ( p _4^2 + ( l_2 + \vec k_{34})^2) } \cr
 \times\left(\cos p _{1}z_{1}\cos p _{1}z_{3}\cos p _{2}z_{1}\cos p _{2}z_{3}+\cos\leftrightarrow\sin\right)\cr
\times\left(\cos p _{3}z_{2}\cos p _{3}z_{3}\cos p _{4}z_{2}\cos p _{4}z_{3}+\cos\leftrightarrow\sin\right).
\end{multline}
Note that all four diagrams have the same symmetry factor, which includes the factor of 6 of the mixed four-point vertex. The integral over the middle vertex $z_{3}$ is oscillatory and therefore
needs to be regulated, just like case of two-loop tadpole. This can
be accomplished by inserting $e^{-\epsilon z_{3}}$, where $\epsilon>0$.
The integral over $z_{3}$ is then straightforward to evaluate. For
example, one of the four integrals is given by
\begin{equation}
\lim_{\epsilon\rightarrow0}\int_{0}^{\infty}dz_{3}\cos p _{1}z_{3}\cos p _{2}z_{3}\cos p _3 z_{3}\cos p _{4}z_{3}e^{-\epsilon z_{3}}=-\frac{\pi}{16}\sum_{\sigma_{i}=\pm1}\delta\left(\sum_{i=1}^4\left(-1\right)^{\sigma_{i}} p _{i}\right),
\end{equation}
where we used the identity in~\eqref{deltaidentity}. 
The other three integrals over $z_{3}$ can be evaluated in a similar
way. After changing integration variables, all the delta functions can be mapped to the energy conserving delta function $\delta\left( p _{1}+ p _{2}+ p _{3}+ p _{4}\right)$ up to an overall sign.

If we then integrate $ p _{4}$ against the delta function we end up with 
\begin{equation}
I^{(4)}_{\circ\circ}\propto \lambda^3k_{12}k_{34}\int\frac{d p _{+}}{\left( p _{+}^{2}+k_{12}^{2}\right)\left( p _{+}^{2}+k_{12}^{2}\right)}\int\frac{d^{4}L_{1}d^{4}L_{2}}{L_{1}^{2}L_{2}^{2}\left(L_{1}-P\right)^{2}\left(L_{2}+P\right)^{2}},
\end{equation}
where $L_{1}=\big( p _{1},\vec{l}_{1}\big)$, $L_{2}=\big( p _{2},\vec{l}_{2}\big)$, 
$P=\big( p _{+},\vec{k}_{12}\big)$, $ p _{+}=- p _{3}- p _{2}$.
Hence, we once again get a four dimensional flat space integrand with an auxiliary
integral over $ p _{+}$. Note that the auxiliary is exactly the
same as the one that appeared in the one-loop bubble in~\eqref{bubble}, while the four-dimensional integral
is simply the square of the one-loop one, as familiar from flat space
calculations. This suggests that the necklace diagrams can be
obtained to any order at the integrand level via exponentiation as in flat space. See~\cite{Sachs:2023eph} for a position space discussion.

Finally, let us note after integrating out $p_{1}$, $p_{2}$,
and $p_{+}$ using residues in the upper half-plane we obtain the same loop integrand derived using recursion relations\footnote{The companion Mathematica notebook contains this derivation.}:  

\begin{equation}\label{2loopnecklacerecursion}
I^{(4)}_{\circ\circ}\propto \lambda^3 \int   \frac{\left((\sigma_1 + \sigma_2) (E + \sigma_1)  (E + \sigma_2) + E x_1 x_2 \right) d^3 l_1 d^3 l_2 }{E y_1 y_2 y_3 y_4 (x_1 + \sigma_1)(x_1 + \sigma_2) (x_2 + \sigma_1) (x_2 + \sigma_2)\sigma_1 + \sigma_2 }
\end{equation}
where  $E = x_1 + x_3,\ y_1 = | \vec l_1|, \ y_2 = |\vec l_1 + \vec k_1 + \vec k_2|,\ y_3 = |\vec l_2|,\ y_4 = |\vec l_2 + \vec k_1 + \vec k_2|,\ x_1 = |\vec k_1| + |\vec k_2|,\ x_2 = |\vec k_3| + |\vec k_4|, \ \sigma_1 = y_1 + y_2, \ \sigma_2 = y_3 + y_4$.
The first term inside the parenthesis in the numerator encodes the square of the 1-loop bubble and the second term vanishes in the flat space limit, thus recovering the expected integrand in the flat space limit. In appendix \ref{app:necklace-hard} we evaluate the loop integrals using a hard cutoff. 
%------------------------------------------------

\subsection*{Two-loop Ice-cream}

The final topology of diagram to consider is the ice-cream cone diagram. There are three diagrams that contribute to the $s$-channel, two of which have symmetry factor $1/2$ and one of which has symmetry factor $1$ (including the factor of 6 in the mixed interaction vertex):
\begin{figure}[H]\centering
\begin{tikzpicture}[scale=0.6]
\node at (0, 2.5) {$p_4$};
\node at (0, 0.5) {$\vec \ell_4, p_2$};
\filldraw [fermion, color = black, fill=none, very thick,dashed] (0,0) circle (2cm);
\draw [fermionbar,dashed, black,very thick] (-2,0) to (2,0);
\filldraw [black] (2,0) circle (2pt);
\filldraw [black] (0,-2) circle (2pt);
\filldraw [black] (-2,0) circle (2pt);
\draw [fermionbar, black,very thick, dashed] (-2,0) to (-3,0);
\draw [fermion, black,very thick, dashed] (2,0) to (3,0);
\draw [fermion, black,very thick, dashed] (0,-2) to (0.5,-3);
\draw [fermion, black,very thick, dashed] (0,-2) to (-0.5,-3);
\node at (-3, 0.65) {$\vec k_3$};
\node at (3, 0.65) {$\vec k_4$};
%\node at (-0.75, -2.5) {$\vec k_2$};
%\node at (0.75, -2.5) {$\vec k_1$};

\node at (0, -1.75) {$z_1$};
\node at (-1.55, -0.25) {$z_2$};
\node at (1.55, -0.25) {$z_3$};

\node at (2.4, -1.5) {$\vec \ell_1, p_1$};
\node at (-2.25, -1.5) {$p_3$};
  
\node at (-0.5, -3) {$k_1$};
\node at (0.5, -3) {$k_2$};

\end{tikzpicture}
\quad
\begin{tikzpicture}[scale=0.6]
\draw[color = black, fill=none, very thick]  (-2,0) -- (2,0)
arc(0:180:2) --cycle;
\draw[color = black, fill=none, very thick,dashed] (2,0) arc(0:-180:2);
\filldraw [black] (2,0) circle (2pt);
\filldraw [black] (0,-2) circle (2pt);
\filldraw [black] (-2,0) circle (2pt);
\draw [fermionbar, black,very thick , dashed] (-2,0) to (-3,0);
\draw [fermion, black,very thick, dashed] (2,0) to (3,0);
\draw [fermion, black,very thick, dashed] (0,-2) to (0.5,-3);
\draw [fermion, black,very thick, dashed] (0,-2) to (-0.5,-3);
\node at (-2.5, 0.65) {$\vec k_3$};
\node at (2.5, 0.65) {$\vec k_4$};
\node at (-0.75, -2.5) {$\vec k_2$};
\node at (0.75, -2.5) {$\vec k_1$};

\node[align=center] at (-4, 0) {$+$};
\end{tikzpicture}
%\qquad
\begin{tikzpicture}[scale=0.6]
\draw[color = black, fill=none, very thick,dashed] (2,0) arc(0:180:2);
\draw[color = black, fill=none, very thick]  (-2,0) -- (2,0)
arc(0:-180:2)  --cycle;
\filldraw [black] (2,0) circle (2pt);
\filldraw [black] (0,-2) circle (2pt);
\filldraw [black] (-2,0) circle (2pt);
\draw [fermionbar, black,very thick, dashed] (-2,0) to (-3,0);
\draw [fermion, black,very thick, dashed] (2,0) to (3,0);
\draw [fermion, black,very thick, dashed] (0,-2) to (0.5,-3);
\draw [fermion, black,very thick, dashed] (0,-2) to (-0.5,-3);
\node at (-2.5, 0.65) {$\vec k_3$};
\node at (2.5, 0.65) {$\vec k_4$};
\node at (-0.75, -2.5) {$\vec k_2$};
\node at (0.75, -2.5) {$\vec k_1$};
\node[align=center] at (-4, 0) {$+$ \quad $2$ \quad};
\end{tikzpicture}
\caption{The ice-cream cone graph: the direction of the momenta are labelled for convenience in order to keep track of the signs in the loop integrals. The second and third diagrams follow the same labels as the first one. }\label{fig:icecream}
\end{figure}
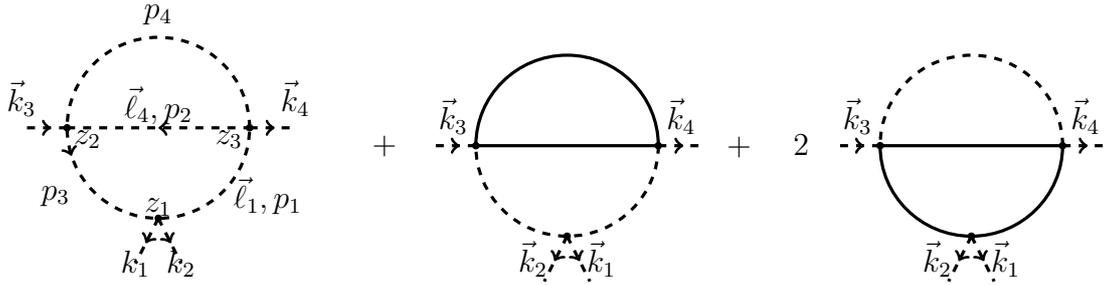
Since the rightmost diagram has a relative factor of 2 with respect to the other diagrams, it is convenient to write it as a sum of two diagrams where the solid internal line appears at the top or in the middle. Adding the four diagrams, we find
\eqs{\label{eq:ice_in_in}
I^{(4)}_{\widehat{\vee}}\propto& \lambda^3\int \frac{d^{3}l_{1}d^{3}l_{2}d p _{1}d p _{2}d p _{3}d p _{4}}{( p _1^2+l_1^2)( p _2^2+l_2^2)( p _3^2+l_3^2)( p _4^2+l_4^2)}
\int_{0}^\infty dz_{1}dz_{2}dz_{3} e^{-k_{12}z_{1}-k_{3}z_{2}-k_4z_3}\\
 \times & \Big( \sin p_{1}z_{1}\sin p_{3}z_{1}\sin p_{3}z_{3}\sin p_{1}z_{2}\left(\cos p_{4}z_{3}\cos p_{4}z_{2}\sin p_{2}z_{3}\sin p_{2}z_{2}+\cos\leftrightarrow\sin\right) \\
 & + \cos p_{1}z_{1}\cos p_{3}z_{1}\cos p_{3}z_{3}\cos p_{1}z_{2}\left(\cos p_{2}z_{3}\cos p_{4}z_{3}\cos p_{2}z_{2}\cos p_{4}z_{2}+\cos\leftrightarrow\sin\right) \Big),
}
where $\vec{l}_3 = \vec{l}_2-\vec{k}_{12}$ and $\vec{l}_3 = \vec{l}_{12}+\vec{k}_{4}$. Integrating out $z$ and performing several simplifications using the $ p _i \to - p _i$ symmetry, 
\begin{comment}
the in-in correlator can be simplified to
\eqs{I^{(4)}_{\widehat{\vee}}\propto&\lambda^3\int d p _1 d p _2 d p _3 d p _4 \frac{4k_{12} p _3  p _4}{(k_3^2+ p _3^2)(k_4^2+ p _4^2)\big(k_{12}^2+( p _3+ p _4)^2\big)}\\
\times &\frac{1}{({l}_1^2+ p _1^2)({l}_2^2+ p _2^2)(({l_1}+\vec{k}_{34})^2+(- p _1+ p _3+ p _4)^2)(({l}_1+{l}_2+\vec{k}_4)^2+( p _4- p _1- p _2)^2)}.
}
Again, 
\end{comment}
we can once again write the loop integral in terms of a four-dimension flat space integral:
\begin{multline}\label{ice-allminus}
I^{(4)}_{\widehat{\vee}}\propto\iint_{-\infty}^{\infty} d p _3 d p _4\frac{4k_{12}k_3k_4}{(k_3^2+ p _3^2)(k_4^2+ p _4^2)(k_{12}^2+( p _3+ p _4)^2)}\cr
 \times \int \frac{d^4L_1 d^4L_2}{(2\pi)^8}\frac{1}{L_1^2 L_2^2 (L_1+P_3+P_4)^2 (L_1+L_2+P_4)^2},
\end{multline}
where $L_i=(\vec{l}_i, p _i)$ $P_i=(\vec{k_i}, p _i)$.
\begin{comment}
\blue{Like in the case of sunset diagram,  even in this case the all $\phi_+$ correlator differs from this but also yields a simple result 
\begin{multline}
\propto\iint_{-\infty}^{\infty} d p _3 d p _4\frac{4k_{12} p _3 p _4}{(k_3^2+ p _3^2)(k_4^2+ p _4^2)(k_{12}^2+( p _3+ p _4)^2)}\cr
 \times \int \frac{d^4L_1 d^4L_2}{(2\pi)^8}\frac{1}{L_1^2 L_2^2 (L_1+P_3+P_4)^2 (L_1+L_2+P_4)^2},
\end{multline}}
\end{comment}
In appendix \ref{app:ice} we evaluate the leading order singularity of these integrals using a cutoff.  The leading and subleading singularties of \eqref{ice-allminus} will be evaluated using a different regularization scheme in Section~\ref{sec:analytic-reg}.

%------------------------------------------------
\subsection{One-Loop Polygons}\label{sec:polygon}

The structure described above can be extended to any general polygon at one loop. Similar to the scenario with Feynman diagrams in flat space, it can be demonstrated that polygons with three or more vertices do not exhibit ultraviolet divergences and do not need be regularized. For the interactions we consider in equation~\eqref{eq:EAdS_action}, the $n$-gon diagram corresponds to a $(2n)$-point function. In the appendix~\ref{app:tria-int} we derive the integrand for the triangle diagram using the recursion relations in section \ref{sec:recursion} and show that it can be expressed as follows:
\begin{eqn}\label{tria4D1}
&\begin{tikzpicture}[baseline]
\draw[very thick] (0, 1) -- (-1, 0);
\draw[very thick] (0, 1) -- (1, 0);
\draw[very thick] (-1, 0) -- (1, 0);
\node at (1, -0.25) {$x_3$};
\node at (-1, -0.25) {$x_2$};
\node at (0, 1.25) {$x_1$};
\node at (-0.75, 0.75) {$y_{12}$};
\node at (0.75, 0.75) {$y_{31}$};
\node at (0, -0.25) {$y_{23}$};

\draw[very thick, dashed] (0, 1) -- (-0.5, 1+0.5);
\draw[very thick, dashed] (0, 1) -- (0.5, 1+0.5);

\draw[very thick, dashed] (1, 0) -- (1+0.5, -0.5);
\draw[very thick, dashed] (1, 0) -- (1+0.5, +0.5);

\draw[very thick, dashed] (-1, 0) -- (-1-0.5, -0.5);
\draw[very thick, dashed] (-1, 0) -- (-1-0.5, +0.5);

\end{tikzpicture} 
+ 
\begin{tikzpicture}[baseline]
\draw[dashed, very thick] (0, 1) -- (-1, 0);
\draw[dashed, very thick] (0, 1) -- (1, 0);
\draw[dashed, very thick] (-1, 0) -- (1, 0);
\node at (1, -0.25) {$x_3$};
\node at (-1, -0.25) {$x_2$};
\node at (0, 1.25) {$x_1$};
\node at (-0.75, 0.75) {$y_{12}$};
\node at (0.75, 0.75) {$y_{31}$};
\node at (0, -0.25) {$y_{23}$};

\draw[very thick, dashed] (0, 1) -- (-0.5, 1+0.5);
\draw[very thick, dashed] (0, 1) -- (0.5, 1+0.5);

\draw[very thick, dashed] (1, 0) -- (1+0.5, -0.5);
\draw[very thick, dashed] (1, 0) -- (1+0.5, +0.5);

\draw[very thick, dashed] (-1, 0) -- (-1-0.5, -0.5);
\draw[very thick, dashed] (-1, 0) -- (-1-0.5, +0.5);

\end{tikzpicture} \\
&\propto \intinf \frac{x_1 x_2 x_3 dp dp'}{(p^2 + x_1^2) (p'^2 + x_2^2) \big( (p + p')^2 + x_3^2 \big)} \int \frac{d^4 L}{L^2  (L + P_1)^2 (L + P_2)^2},
\end{eqn}
(where we have suppressed the overall factors of $2\pi$ and $\lambda$) with $p = p_1 - p_3$, $p' = p_2 - p_1$ and  
\begin{eqn}
L= (p_3,  \vec l), \qquad P_1^\mu = (p_1 - p_3, \vec y_1), \qquad P_2= (p_2 - p_3,  \vec y_2)~
\end{eqn}
with $\vec y_1 = \vec  k_1 + \vec k_2$ and $\vec y_2 = \vec k_1 + \vec k_2 + \vec k_3 + \vec k_4$~. The $L$ integral is now equivalent to a four-dimensional Feynman integral of the triangle diagram. Since this integral is ultraviolet finite (as evident from power counting) and does not have infrared divergence (as $P_i^2 \neq 0$), we can integrate this without the need for any regulator. For a discussion on the wavefunction coefficient see Section~3.2.2 of~\cite{Chowdhury:2023khl}. This exemplifies the simplicity of the cosmological correlator compared to the corresponding wave function coefficient.

The value of the four-dimensional Feynman loop integral is given in terms of the Bloch-Wigner dilogarithm $\mathcal P_2(z):=\mbox{Im}\big(\mbox{Li}_2(z)\big) + \mbox{arg}(1 -z) \log|z|$ (see equation 4.13 of~\cite{Chavez:2012kn}):
\begin{eqn}
\int \frac{d^4 L}{L^2 (L + P_1)^2 (L + P_2)^2} = \frac{1}{(P_1 + P_2)^2}  \frac{2\mathcal P_2(z)}{z - \bar z},  
\end{eqn}
with $z, \bar z$ satisfying the following equations,
\begin{eqn}
z \bar z &= \frac{P_1^2}{(P_1 + P_2)^2}, \qquad
(1 - z) (1 - \bar z) = \frac{P_2^2}{(P_1 + P_2)^2}~.
\end{eqn}
Therefore the triangle diagram in equation~\eqref{tria4D1} is given as 
\begin{eqn}
\eqref{tria4D1}\propto\intinf  \frac{x_1 x_2 x_3 dp dp'}{(p^2 + x_1^2) (p'^2 + x_2^2) \big( (p + p')^2 + x_3^2 \big)} \frac{2}{(P_1 + P_2)^2} \frac{\mathcal P_2(z)}{z - \bar z}~.
\end{eqn}

It is interesting to note that the structure of the integrand in~\eqref{tria4D1} generalizes to an $n$-gon (corresponding to a $2n$-point correlator),
\begin{eqn}
&\begin{tikzpicture}[baseline]
\draw[very thick, dashed] (0, 1) -- (-0.5, 1+0.5);
\draw[very thick, dashed] (0, 1) -- (0.5, 1+0.5);

\draw[very thick, dashed] (0, -1) -- (-0.5, -1-0.5);
\draw[very thick, dashed] (0, -1) -- (0.5, -1-0.5);

\draw[very thick, dashed] (-1, 0) -- (-1-0.5,+0.5);
\draw[very thick, dashed] (-1, 0) -- (-1-0.5,-0.5);

\draw[very thick, dashed] (0.75, 0.75) -- (0.75+0, 0.75+0.5);
\draw[very thick, dashed] (0.75, 0.75) -- (0.75+0.5, 0.75+0);

\draw[very thick, dashed] (-0.75, 0.75) -- (-0.75+0, 0.75+0.5);
\draw[very thick, dashed] (-0.75, 0.75) -- (-0.75-0.5, 0.75+0);

\draw[very thick, dashed] (0.75, -0.75) -- (0.75+0, -0.75-0.5);
\draw[very thick, dashed] (0.75, -0.75) -- (0.75+0.5, -0.75+0);

\draw[very thick, dashed] (-0.75, 0.75) -- (-0.75+0, 0.75+0.5);
\draw[very thick, dashed] (-0.75, 0.75) -- (-0.75-0.5, 0.75+0);

\draw[very thick, dashed] (-0.75, -0.75) -- (-0.75+0, -0.75-0.5);
\draw[very thick, dashed] (-0.75, -0.75) -- (-0.75-0.5, -0.75+0);

\draw[very thick] (-0.75, 0.75) -- (0,1);
\draw[very thick] (0.75, 0.75) -- (0,1);
\draw[very thick] (-1, 0) -- (-0.75, 0.75);
\draw[very thick] (-1, 0) -- (-0.75, -0.75);
\draw[very thick]  (-0.75, -0.75) -- (0, -1);
\draw[very thick] (0, -1) -- (0.75, -0.75);
\draw[very thick] (0.75, 0.75) -- (0.9, 0.4);
\draw[very thick] (0.75, -0.75) -- (0.9, -0.4);
\node at (1, 0.1) {$\vdots$};

\node at (-1, -1) {$x_1$};
\node at (-1.25, 0) {$x_2$};
\node at (-1, 1) {$x_3$};
\node at (0, 1.25) {$x_4$};
\node at (1, 1) {$x_5$};
\node at (0, -1.25) {$x_n$};
\node at (1.25, -1) {$x_{n-1}$};
\end{tikzpicture}
+ 
\begin{tikzpicture}[baseline]
\draw[very thick, dashed] (0, 1) -- (-0.5, 1+0.5);
\draw[very thick, dashed] (0, 1) -- (0.5, 1+0.5);

\draw[very thick, dashed] (0, -1) -- (-0.5, -1-0.5);
\draw[very thick, dashed] (0, -1) -- (0.5, -1-0.5);

\draw[very thick, dashed] (-1, 0) -- (-1-0.5,+0.5);
\draw[very thick, dashed] (-1, 0) -- (-1-0.5,-0.5);

\draw[very thick, dashed] (0.75, 0.75) -- (0.75+0, 0.75+0.5);
\draw[very thick, dashed] (0.75, 0.75) -- (0.75+0.5, 0.75+0);

\draw[very thick, dashed] (-0.75, 0.75) -- (-0.75+0, 0.75+0.5);
\draw[very thick, dashed] (-0.75, 0.75) -- (-0.75-0.5, 0.75+0);

\draw[very thick, dashed] (0.75, -0.75) -- (0.75+0, -0.75-0.5);
\draw[very thick, dashed] (0.75, -0.75) -- (0.75+0.5, -0.75+0);

\draw[very thick, dashed] (-0.75, 0.75) -- (-0.75+0, 0.75+0.5);
\draw[very thick, dashed] (-0.75, 0.75) -- (-0.75-0.5, 0.75+0);

\draw[very thick, dashed] (-0.75, -0.75) -- (-0.75+0, -0.75-0.5);
\draw[very thick, dashed] (-0.75, -0.75) -- (-0.75-0.5, -0.75+0);

\draw[dashed, very thick] (-0.75, 0.75) -- (0,1);
\draw[dashed, very thick] (0.75, 0.75) -- (0,1);
\draw[dashed, very thick] (-1, 0) -- (-0.75, 0.75);
\draw[dashed, very thick] (-1, 0) -- (-0.75, -0.75);
\draw[dashed, very thick]  (-0.75, -0.75) -- (0, -1);
\draw[dashed, very thick] (0, -1) -- (0.75, -0.75);
\draw[dashed, very thick] (0.75, 0.75) -- (0.9, 0.4);
\draw[dashed, very thick] (0.75, -0.75) -- (0.9, -0.4);
\node at (1, 0.1) {$\vdots$};

\node at (-1, -1) {$x_1$};
\node at (-1.25, 0) {$x_2$};
\node at (-1, 1) {$x_3$};
\node at (0, 1.25) {$x_4$};
\node at (1, 1) {$x_5$};
\node at (0, -1.25) {$x_n$};
\node at (1.25, -1) {$x_{n-1}$};
\end{tikzpicture}\\
&\propto  \int \frac{x_1 x_2 \cdots x_n dp_1 dp_2 \cdots p_{n-1}}{(p_1^2 + x_1^2)  \cdots \big( p_{n-1}^2+ x_{n-1}^2 \big) \big( (p_1 + p_2 +  \cdots p_{n-1})^2 + x_n^2 \big)} \int \frac{d^4 L}{L^2  (L + P_1)^2 \cdots (L + P_{n-1})^2}~,
\end{eqn}
where $L= (p_n, \vec l)$ and
\begin{equation}
P_1 = (p_1, \vec y_1), \quad P_2 = (p_1 + p_2, \vec y_1 + \vec y_2), \dots,\quad 
P_{n-1} = (p_1 + \cdots + p_{n-1}, \vec y_1  + \cdots + \vec y_{n-1})~
\end{equation}
and 
\begin{eqn*}
\vec y_1 &= \vec k_1 + \vec k_2, \quad 
\vec y_2 = \vec k_1 + \vec k_2 + \vec k_3 + \vec k_4, \quad
\dots,\quad
\vec y_{n-1} = \vec k_1 + \vec k_2 + \cdots+ \vec k_{2n-2}~,\\
x_1 &= k_1 + k_2, \quad x_2 = k_3 + k_4, \quad \dots, \quad x_n = k_{2n-1} + k_{2n}~.
\end{eqn*}

Since the integrals for the \textit{in-in} correlators are now expressed in terms of standard Feynman integrals in flat space, it is also possible to use the cutting rules in flat space to study the higher point functions. We leave this problem for future work. It would also be an interesting mathematical problem to find if the integrand above can be obtained from a combinatorial geometry~\cite{Arkani-Hamed:2017fdk} along the lines of the cosmological polytope (see \cite{Benincasa:2024leu}).

\subsection{Renormalisation}\label{sec:renormalization}

In this subsection, we will renormalise the four-point correlator up to one-loop and the two-point correlator up to two loops. We will follow the standard approach used in flat space, by introducing a dimensionful renormalisation scale $\mu$. The resulting correlators will manifestly have the correct flat space limit but will not satisfy the conformal Ward identities. On the other hand we find that conformal symmetry can be restored by setting $\mu$ proportional to the energy, although this obscures the flat space limit. For the loop corrected 2-point function setting $\mu$ proportional to the energy removes branch cuts in the energy in agreement with the conclusion reached in~\cite{Senatore:2009cf}. For the four point function on the other hand, doing so introduces a branch cut in the energy and the resulting cosmological correlator contradicts the tree theorem in~\cite{Agui-Salcedo:2023wlq}.

In order to renormalise the correlators, we must include counterterms to cancel divergences. 
At two-points, this can be accomplished by adding the following counterterm Lagrangian to~\eqref{eq:action-conf1}:
\begin{equation}\label{2ptcounter}
\mathcal{L}_{ct}=\frac{1}{2}\left(-\left(Z_{-}-1\right)\left(\partial\phi_{-}\right)^{2}+\delta m^{2}\phi_{-}^{2}\right).
\end{equation}
Note that these terms correspond to mass and wavefunction renormalisations. We find that these counterterms are sufficient to cancel all the divergences that at arise up 2-loops using a cut-off so it appears that the action in \eqref{eq:EAdS_action_gen_pot} is renormalisable using this scheme. In section \ref{sec:analytic-reg} we will show that it is also renormalisable using an alternative regularisation. On the other hand, \cite{Bzowski:2023nef} recently pointed out that the action in \eqref{eq:EAdS_action_gen_pot} is not renormalisable for generic masses and polynomial interactions so it would be interesting to investigate this further.

Using \eqref{2ptcounter}, the two-point counterterm depicted below
\begin{eqn}
\begin{tikzpicture}
\draw[very thick] (-1, 0) -- (1, 0);
\node at (0, 0) {$\oplus$};
\end{tikzpicture}
\end{eqn}
is given by 
\begin{equation}
\delta I^{(2)}=\left(-2\left(Z_{-}-1\right)k^{2}+\delta m^{2}\right)\int_{0}^{\infty}dze^{-2kz}=-\left(Z_{-}-1\right)k+\frac{\delta m^{2}}{2k}.
\end{equation}
We may then set the one-loop correction to two-point correlator computed in~\eqref{tadpole1} to zero by choosing 
\begin{equation}
\delta m^{2}=-\frac{\frac{\lambda}{2}\Lambda^{2}}{8\pi^{2}}\,.
\end{equation}
At two loops, we can also choose the coefficients $Z_-$ and $\delta m^2$ to cancel all the divergent terms in~\eqref{2cactus} and~\eqref{sunsetintegrated} yielding the 2-point correlator
\begin{equation}\label{eq:I2log}
I^{(2)}=-\lambda^{2}\frac{i}{16}\frac{k\pi^{2}}{12}\left[30\log\left(\frac{2\mu}{3k}\right)+5\right],
\end{equation}
where $\mu$ is a dimensionful renormalisation scale. Note
that~\eqref{eq:I2log} features a growing logarithm in the sense
of~\cite{Senatore:2009cf}. On the other hand, the conformal Ward
identities imply that a two-point function of operators with scaling
dimension $\Delta$ should scale like $k^{2 \Delta-d}$, where $d$ is
the dimension of the conformal field
theory~\cite{Bzowski:2013sza,Bzowski:2015pba}. In the present case,
the logarithm spoils conformal invariance but this can be remedied by setting $\mu = \bm \delta k$, where $\bm \delta$ is a dimensionless renormalisation scale. After doing so, we find that the two-point function is indeed proportional to $k$, as expected for $d=3$ and $\Delta=2$. In particular, the growing log is absent. We will reach the same conclusion using an alternative regulator in section \ref{sec:analytic-reg}. Also, we will see a similar mechanism for restoring conformal symmetry at four points shortly.

Let us now renormalise the four-point correlator at one loop. For this purpose, we need to add the following counterterm Lagrangian:
\begin{equation}
\mathcal{L}_{ct}=-\frac{1}{2}(Z_{\lambda_R}-1)\frac{\lambda_R}{4!}(\phi_{-})^{4}~.
\end{equation}
Now we compute four-point counterterm: 
\begin{equation}
\delta I^{(4)}=\begin{gathered}\begin{tikzpicture}
\draw[very thick, dashed] (-1, 1) -- (1, -1);
\draw[very thick, dashed] (1, 1) -- (-1, -1);
\node at (0, 0) {$\otimes$};
\end{tikzpicture}\end{gathered}=-\frac{\lambda_R}{2}\left(Z_{\lambda_R}-1\right)\int_0^\infty dze^{-Ez}=\frac{\lambda_R}{2E}\left(Z_{\lambda_R}-1\right)\,,
\label{1loop4ptct}
\end{equation}
where $E=k_{12} + k_{34}$. The renormalised one-loop four-point correlator is then obtained by combining the bubble diagrams in~\eqref{1loop4ptcutoff} with the counterterm in~\eqref{1loop4ptct} and is given by 
\begin{multline}
I^{(4)}_{\circ}=\frac{\lambda_R}{2 E }+\frac{\lambda_R^2}{32 E(2\pi)^{2}}\Bigg[\ln\left(\frac{(k_{12} + |\vec k_{12}|)(k_{34} + \vec|k_{12}|)}{\mu^{2}}\right)
+\frac{E}{k_{12}-k_{34}}\ln\left(\frac{k_{34} + |\vec k_{12}|}{k_{12} + |\vec k_{12}|}\right)\cr+t-\textrm{channel}+u-\textrm{channel}\Bigg],\label{eq:4pt1loopa}
\end{multline}
where $\mu$ is the renormalisation scale, and we have indicated that one needs to add the 
$t$ and $u$ channel contributions obtained by exchanging $2\leftrightarrow4$ and $2\leftrightarrow3$ respectively.
We have made the choice 
\begin{equation}
Z_{\lambda_R}=1+\frac{3\frac{\lambda_R}{2}}{4(2\pi)^{2}}\ln\left(\frac{2\lambda_R}{\mu}\right)
\end{equation}
to absorb the divergences. It is straightforward to see that~\eqref{eq:4pt1loopa} has the
correct flat space limit:
\begin{equation}
\lim_{E\rightarrow0}EI^{(4)}_{\circ} =\frac{\lambda_R}{2}+\frac{\lambda_R^{2}}{32(2\pi)^{2}}\ln\left(\frac{stu}{\mu^{6}}\right).    
\end{equation}

Note that~\eqref{eq:4pt1loopa} breaks three-dimensional conformal symmetry since $\mu$
is a dimensionful scale. On the other hand, this symmetry must be
preserved since it corresponds to the isometry of the fixed background
spacetime, much like Poincar\'e invariance should be preserved when
regulating Feynman diagrams in flat background. This can be
remedied ad-hoc by introducing a dimensionless renormalisation scale $\bm \delta=\mu/E$.
After doing so, the renormalised correlator becomes 
\begin{multline}
I^{(4)}_{\circ}=\frac{\lambda_R}{2E}+\frac{(\frac{\lambda_R}{2})^{2}}{8(k_{12} + k_{34})(2\pi)^{2}}\Bigg[\ln\left(\frac{(k_{12} + |\vec k_{12}|)(k_{34} + |\vec k_{12}|)}{\bm \delta^{2}(k_{12} + k_{34})^{2}}\right) 
+\frac{k_{12} + k_{34}}{k_{12}-k_{34}}\ln\left(\frac{k_{34} + |\vec k_{12}|}{k_{12} + |\vec k_{12}|}\right)\cr+t-\textrm{channel}+u-\textrm{channel}\Bigg].\label{eq:4ptloop_2}
\end{multline}
While this expression is now manifestly scale invariant, this does imply conformal (i.e. de Sitter-)  invariance. Remarkably, it is indeed a solution
to the four-point conformal Ward identities, which can be seen by recasting
it in terms of the conformal cross ratios
\begin{equation}\label{e:uvdef}
u= \frac{|\vec{k}_{12}|}{k_{12}},\,\,\,v=\frac{|\vec{k}_{12}|}{k_{34}}.
\end{equation}
In particular the $s$-channel contribution to the renormalised correlator at $O(\lambda^2)$ is given by 
\begin{equation}
I^{(4)}_{\lambda^2}\propto\frac{\hat{F}(u,v)}{|\vec k_{12}|}
\end{equation}
with
\begin{equation}\label{hatF1}
\hat{F}(\hat u, \hat v)=\frac{uv}{u+v}\ln\left(\frac{uv(1+u)(1+v)}{\bm \delta^{2}(u+v)^{2}}\right)+\frac{uv}{u-v}\ln\left(\frac{u(1+v)}{v(1+u)}\right).
\end{equation}
Conformal symmetry is then implied by the Ward identity~\cite{Arkani-Hamed:2018kmz}
\begin{equation}\label{e:Delta2}
\left(\Delta_{u}-\Delta_{v}\right)\hat{F}=0,\qquad \mbox{where:} \quad \Delta_{u}=u^{2}(1-u^{2})\partial_{u}^{2}-2u^{3}\partial_{u},
\end{equation}
which is indeed satisfied. We should note that this in itself does not imply that~\eqref{eq:4ptloop_2} is the correct result because individual contributions are annihilated by~\eqref{e:Delta2}: 
\begin{align}
(\Delta_u- \Delta_v)\,{u v\over u+v}&=0,\cr
(\Delta_u-\Delta_v)\, {u v\over u+v}\log\left(uv(1+u)(1+v)\over (u+v)^2\right)&=0,\\
\nonumber
(\Delta_u-\Delta_v) {u v\over u-v}\log\left(u(1+v)\over v(1+u)\right)&=0.
\end{align}
In other words conformal invariance by itself does not imply a unique result, in particular it allows for the free parameter $\bm \delta$. 
In the next section we will confirm the correctness of~\eqref{eq:4ptloop_2}  using a de Sitter-invariant regularization. 

When taking the flat space limit of (\ref{eq:4ptloop_2}), we must take
$\bm \delta\rightarrow\infty$ while taking $E\rightarrow0$ holding $\mu = \bm \delta E$ fixed in order to avoid singular terms of the form $\ln E$, which would
spoil the flat space limit. Hence, the flat space limit is most easily
seen by restoring the dimensionful renormalisation scale $\mu=\bm \delta E$
which is held fixed. Moreover, we can compute the $\beta$-function
by demanding that the renormalised correlator is either independent
of the dimensionful renormalisation scale $\mu$ or the dimensionless
scale $\bm\delta$. Indeed, the calculation is almost identical to that
of flat space: 
\begin{eqn}
0=E \frac{dI^{(4)}}{d\log\bm\delta}=\frac{d\lambda_R}{d\ln\bm \delta}-\frac{3\lambda_R^{2}}{64\pi^{2}}+\mathcal{O}\left(\lambda_R^{3}\right)\,.
\end{eqn}

%%%%%%%%%%%%%%%%%%%%%%%%%%%%%%%%%%%%%%%%%%%%%%%%%%%%%%%%%%%%%%%%%%%%
\section{Analytic regularisation}\label{sec:analytic-reg}
Regularization in de Sitter space-time requires some consideration in order to take care of de Sitter-invariance of correlation functions. The cut-off regularization  described in the last section does not preserve de Sitter-invariance, although, as we showed there, at one-loop it can be restored by making a non-minimal and  non-local subtraction. Since this prescription is not unique, and it is not clear how generalize it to higher order we will describe in this section a manifestly dS-invariant regularization scheme leading to unambiguous results for all loop-corrected correlation functions. 

In position space the latter is guaranteed by expressing the regulated amplitudes in terms of regularized Green functions as a function of the geodesic distance only. Concretely, in~\cite{Bertan:2018khc} the regularized Green function in AdS for a conformally coupled scalar is chosen as
	\begin{equation}\label{eq:adsGG}
		G(\mathbf X,\mathbf Y;\Delta,\delta)=\left(\frac{1}{\ell_{AdS}2\pi}\right)^2\frac12
		\Bigg(\frac{K(\mathbf X,\mathbf Y)}{1+\delta-K(\mathbf X,\mathbf Y)}
		+(-1)^\Delta\frac{K(\mathbf X,\mathbf Y)}{1+\delta+K(\mathbf X,\mathbf Y)}\Bigg)\,.
	\end{equation}
Here $\delta>0$ regulates the short-distance singularity at $K=1$, where $K$, expressed in terms of embedding space coordinates $\mathbf X$ and $ \mathbf Y$ as 
	\begin{equation}
		\label{eq:Kdef}
		K(\mathbf X,\mathbf Y):=-\frac{\ell_{AdS}^2}{ \mathbf X
			\cdot \mathbf Y} =  \frac{2zw}{(\vec x - \vec y)^2 + z^2 + w^2}\,,
	\end{equation}
and is related to the geodesic distance as 
	\begin{align}
		d(\mathbf X,\mathbf Y)&=\ell_{AdS}\mathrm{arccosh}\left(-{\mathbf{X}\cdot\mathbf{Y}\over \ell_{AdS}^2}\right)\,.
	\end{align}
Since we focus on the momentum space representation at present we need to transform them accordingly. While this can be done (see Appendix~\ref{sec:delta}) the resulting expressions lead to complicated loop integrals in momentum space. 

An alternative regularization consists of replacing~\eqref{eq:adsGG} by 
\begin{equation}\label{eq:adsGdim}
		G(\mathbf X,\mathbf Y;\Delta;\kappa)=\frac{\ell_{AdS}^{2\kappa-2}}{(2\pi)^2}\frac12
		\Bigg(\frac{K(\mathbf X,\mathbf Y)^{1-\kappa}}{(1-K(\mathbf X,\mathbf Y))^{1-\kappa}}
		+(-1)^\Delta\frac{K(\mathbf X,\mathbf Y)^{1-\kappa}}{(1+K(\mathbf X,\mathbf Y))^{1-\kappa}}\Bigg)\,,
	\end{equation}
which unlike~\eqref{eq:adsGG} is still singular for $\mathbf X\to\mathbf Y$, but this singularity is integrable for $\kappa>0$, so that the loop integrals are well-defined.\footnote{The power of $\ell_{AdS}$ is fixed by the flat space limit using $K\sim 1-\frac{1}{2}(r/\ell_{AdS})^2+O(\ell_{AdS}^{-4})$  (eqn. (2.8) in~\cite{Bertan:2018afl}) where $a=\ell_{AdS}^{-1}$.} For $\Delta=2$~\eqref{eq:adsGdim} has the momentum representation 
\begin{equation}\label{eq:ds_bulk_e}
G_D(\vec{x},z, \vec{x}',z';\kappa)=\frac{ (z z'/\ell_{AdS}^2)^{1-\kappa}}{\pi}\int \frac{d^{3}\vec\ell}{(2\pi)^3} \int\limits_{-\infty}^\infty d p\;\frac{\sin(p z)\sin(p z')}{(p^2+\vec{\ell^2})^{1+\kappa}}e^{i\vec{\ell}\cdot(\vec{x}-\vec{x}')}\,,
\end{equation}
while for $\Delta=1$ the sine function is replaced by a cosine: 
\begin{equation}\label{eq:ds_bulk_en}
G_N(\vec{x},z, \vec{x}',z';\kappa)=-\frac{ ( z z'/\ell_{AdS}^2)^{1-\kappa}}{\pi}\int \frac{d^{3}\vec\ell}{(2\pi)^3} \int\limits_{-\infty}^\infty d p\;\frac{\cos(p z)\cos(p z')}{(p^2+\vec{
\ell^2})^{1+\kappa}}e^{i\vec{\ell}\cdot(\vec{x}-\vec{x}')}\,,
\end{equation}
where we changed the overall normalization with respect to~\eqref{eq:adsGdim}. Note that~\eqref{eq:ds_bulk_e} has a natural interpretation as the analogue of the analytic regularization in flat space with the extra feature of an $\kappa$-deformation of the $ z  z '$ prefactor of the momentum integral. In fact this is just the necessary modification to ensure de Sitter-invariance. For instance, the invariance under rescaling $\vec{x}\to \lambda\vec{x}$ is manifest when complementing it with $\vec{\ell}\to \frac{1}{\lambda}\vec{\ell}$, $p\to \frac{1}{\lambda}p $, $ z \to \lambda z $ thanks to the $ z ^{-\kappa}$-factor. 
On the other hand, this regularization is different from dimensional regularization of a canonical kinetic term in position space. This regularization rather corresponds to a non-canonical kinetic term\footnote{This is a characteristic of analytic regularization rather than working in de Sitter space-time.} in four dimensions corresponding to a quadratic action 
\begin{equation}
   \frac{1}{2} \int\frac{d  z  d^3x}{ (z/\ell_{AdS}) ^{-2\kappa}} \phi\left(\frac{\ell_{AdS}}{  z } 
    \Box^{1+\kappa}\; \frac{\ell_{AdS}}{ z }\right)\;\phi\equiv  \frac{1}{2}
    \int\frac{d  z  d^3x}{ (z/\ell_{AdS}) ^{4}} \phi\;\mathcal{D}^2_\kappa \;\phi\,,
\end{equation}
where $\Box$ is the d'Alembertian in flat space and $\Box^{1+\kappa}$ is defined in momentum space. For $\kappa=0$ this reduces to the conformally coupled free scalar in de Sitter. For  $ z < z '$ and $ z \to 0$, the bulk-to-bulk propagator asymptotes to 
\begin{equation}
    \frac{( z  z ')^{\Delta-\kappa}}{((\vec{x}-\vec{y})^2+z '^2)^{\Delta-\kappa}}\,,
\end{equation}
from which we read off the boundary conformal dimensions $\Delta-\kappa$ with $\Delta=1,2$. As a consequence, the effective action~\eqref{eq:EAdS_action_gen_pot} receives a $\kappa$-dependent modification. More precisely, from~\eqref{eq:EAdS_action_gen_pot}
we get for $d=3$, after substitution, to first non-trivial order in $\kappa$, 
\begin{multline} \label{eq:EAdS_action_phi4_k2}
    iS_c
	=-\frac{1}{2}\int\limits_0^{\infty}\frac{d z d^3x}{z^4}\Bigg[\mathcal{C}_\kappa \;\phi^+ \;\mathcal{D}^2_\kappa \;  \phi^+ - \mathcal{C}_\kappa\;\phi^- \;\mathcal{D}^2_\kappa \;  \phi^-\cr
	+\frac{\lambda}{4!}\left(\mathcal{C}_{2\kappa} \;{\phi^+}^4-6\;\mathcal{C}_{2\kappa}{\phi^+}^2{\phi^-}^2+\mathcal{C}_{2\kappa} \;{\phi^-}^4-4(2\pi\kappa) {\phi^+}^3\phi^-+4(2\pi\kappa) {\phi^-}^3\phi^+\right)\Bigg]\,,
\end{multline}
where 
\begin{equation}\label{e:Ckdef}
\mathcal{C}_\kappa=1-\frac{(\pi\kappa)^2}{2}.
\end{equation}
Note that $\lambda$ is now dimensionful since, with the $\kappa$-modified kinetic term, the fields $\phi_\pm$ have mass dimension $1-\kappa$.

It is also possible to see the correspondence between the bulk and the boundary fields by directly comparing the two-point functions in momentum space. From the bulk side, we take the simultaneous $z_i \to 0$ limit of the bulk-bulk propagator in~\eqref{eq:adsGdim} and obtain,
\begin{comment}
\begin{eqn}
 \frac{(z_1 z_2)^{1 - \kappa}}{\pi} \intinf \frac{dp}{(p^2 +\vec{\ell}^2)^{1+\kappa}} \sin(p z_1) \sin(p z_2) \sim  \frac{(z_1 z_2)^{2 - \kappa}}{\pi} \ell^{1 - 2 \kappa} \intinf \frac{y^2 dy}{(y^2 + 1)^{1+\kappa}}
\end{eqn}
\end{comment}
\begin{align}
     \frac{(z_1 z_2)^{1 - \kappa}}{\pi} \intinf \frac{dp}{(p^2 +\vec{\ell}^2)^{1+\kappa}} \cos(p z_1) \cos(p z_2) \sim \frac{(z_1 z_2)^{1 - \kappa}}{\pi} \ell^{-1 - 2 \kappa} \intinf \frac{ dy}{(y^2 + 1)^{1+\kappa}}
\end{align}
This can be compared with the form of the two-point function in the boundary~\cite{Bzowski:2013sza}~,
\begin{eqn}\label{eq:2ptCFT}
\braket{\braket{O(\vec \ell)O(- \vec \ell)}} = \frac{\pi ^{d/2} 2^{3-2 \Delta } \Gamma \left(\frac{1}{2} (d-2 \Delta )\right)}{\Gamma (\Delta )} \ell^{2\Delta - 3}
\end{eqn}
and allows us to identify $\Delta = 1-\kappa$. This relation also fixes the normalization between the boundary and the bulk operators.

\subsection{Feynman Rules in analytic regularization}
We summarize the Feynman rules in analytic regularization.\footnote{For notational symplicity we will suppress factors of $\mathcal{C}_{\kappa}$ and $\mathcal{C}_{2\kappa}$. They will then be restored in the calculations where necessary.}   Vertices are  represented by 
\begin{eqn}
\begin{tikzpicture}[baseline]

\draw[very thick] (-1, 0.5) -- (1, -0.5);
\draw[very thick] (1, 0.5) -- (-1, -0.5);
\end{tikzpicture} = \frac{\lambda}{2} 
, \qquad
\begin{tikzpicture}[baseline]
\draw[dashed, very thick] (-1, 0.5) -- (1, -0.5);
\draw[very thick] (1, 0.5) -- (-1, -0.5);
\end{tikzpicture} = -3\lambda
, \qquad
\begin{tikzpicture}[baseline]
\draw[dashed, very thick] (-1, 0.5) -- (1, -0.5);
\draw[dashed, very thick] (1, 0.5) -- (-1, -0.5);
\end{tikzpicture} = \frac{\lambda}{2}
, \qquad\cr\cr
\begin{tikzpicture}[baseline]
\draw[dashed, very thick] (-1, 0.5) -- (0, 0);
\draw[ very thick] (0, 0) -- (1, -0.5);
\draw[dashed, very thick] (1, 0.5) -- (-1, -0.5);
\end{tikzpicture} = 4\pi\kappa\lambda
, \qquad
\begin{tikzpicture}[baseline]
\draw[very thick] (-1, 0.5) -- (0, 0);
\draw[dashed, very thick] (0, 0) -- (1, -0.5);
\draw[very thick] (1, 0.5) -- (-1, -0.5);
\end{tikzpicture} = -4\pi\kappa\lambda
\,.
\end{eqn}
The bulk-to-bulk propagator for $\phi_+$ gives
\begin{eqn}\label{e:GDkkappa}
\begin{tikzpicture}[baseline]
\draw[very thick] (-1, 0) -- (1, 0);
\node at (-1, -0.25) {$z_1$};
\node at (1, -0.25) {$z_2$};
\node at (0, 0.25) {$\vec k$};    
\end{tikzpicture}
&\equiv G_D(\vec k,z_1,z_2;\kappa)= \frac{(z_1z_2/\ell_{AdS}^2)^{1-\kappa}}{\pi}\int\limits_{-\infty}^\infty dp\;\frac{\sin(p z_1)\sin(p z_2)}{(p^2+k^2)^{1+\kappa}},
\end{eqn}
while the bulk-to-bulk propagator for $\phi_-$ is given as, 
\begin{eqn}\label{e:GNkkappa}
\begin{tikzpicture}[baseline]
\draw[very thick, dashed] (-1, 0) -- (1, 0);
\node at (-1, -0.25) {$z_1$};
\node at (1, -0.25) {$z_2$};
\node at (0, 0.25) {$\vec k$};    
\end{tikzpicture}
&\equiv G_N(\vec k,z_1,z_2;\kappa)=-\frac{(z_1z_2/\ell_{AdS}^2)^{1-\kappa}}{\pi}\int\limits_{-\infty}^\infty dp\;\frac{\cos(p z_1)\cos(p z_2)}{(p^2+k^2)^{1+\kappa}},
\end{eqn}
and the bulk-to-boundary propagator for $\phi^+$  and $\phi^-$ are,
\begin{equation}
\begin{gathered}\begin{tikzpicture}[baseline]
\draw[very thick] (0, 1) -- (0.25, 0);
\node at (0, 1.25) {$\vec k$};
\end{tikzpicture}\end{gathered}
=\frac{ (z/\ell_{AdS})^{1-\kappa}}{\pi} \int\limits_{-\infty}^\infty dp\;\frac{p\; \sin(p z)}{(p^2+k^2)^{1+\kappa}}.
\end{equation}
Similarly, the bulk-boundary propagator for $\phi^-$ is,
\begin{equation}
\begin{gathered}\begin{tikzpicture}[baseline]
\draw[very thick, dashed] (0, 1) -- (0.25, 0);
\node at (0, 1.25) {$\vec k$};
\end{tikzpicture}\end{gathered}
\equiv \bar G(k,z;\kappa) = - \frac{ (z/\ell_{AdS})^{1-\kappa}}{\pi} \int\limits_{-\infty}^\infty dp\;\frac{\; \cos(p z)}{(p^2+k^2)^{1+\kappa}}.
\end{equation}

After integrating over $p$ the modified propagator takes the form
\begin{equation}
  \bar G(k,z;\kappa)=- (z/\ell_{AdS})^{1-\kappa}\frac{ z^{\frac{1}{2}+\kappa }
   K_{-\frac{1}{2}-\kappa }(k z)}{\sqrt{\pi } (2k)^{\kappa +\frac{1}{2}} \Gamma (\kappa +1)}  \,,
\end{equation}
with the small $\kappa$ expansion 
\begin{equation}\label{e:LambdaBdser}
  \bar G(k,z;\kappa)= - \frac{z}{2k\ell_{AdS}} e^{-kz}+O(\kappa)\,.
\end{equation}
In the following we will sometimes work in units where $a=1$ to simplify the notation \footnote{The $\frac1k$ factor in equation \eqref{e:LambdaBdser} shall appear in every diagram and it did not arise in the previous sections as mapping to the conformally flat metric had taken this into account.}.

%--------------------------------------------
\subsection{Mass renormalisation }
In this section we consider the graphs that contribute to the mass renormalisation up to two loops. 
%---------------------------------------------
\subsubsection{One-loop tadpole }\label{sec:1looptad}

At one-loop order we have the combination of the tadpole with the $\Delta=2$ and the $\Delta=1$ field running in the loop,  in  the $\kappa$-regularisation (with $a=1$) is given by
\begin{equation}\label{eq:tad-1-a}
I^{(2)}_\circ(k):= \begin{gathered}
    \begin{tikzpicture}[scale=0.6]
  \filldraw [color = black, fill=none, very thick] (0,1) circle (1cm);
  \draw [black,very thick, dashed] (0.5,-1) to (0,0);
    \draw [black,very thick, dashed] (-0.5,-1) to (0,0);
  \end{tikzpicture}\end{gathered}
  + \begin{gathered}\begin{tikzpicture}[scale=0.6]
  \filldraw [color = black, fill=none, very thick,dashed] (0,1) circle (1cm);
  \draw [black,very thick, dashed] (0.5,-1) to (0,0);
    \draw [black,very thick, dashed] (-0.5,-1) to (0,0);
\end{tikzpicture}
\end{gathered}
= \frac{\lambda}{2}\int_0^\infty {dz\over z^{4}} \bar G(k,z;\kappa)^2 \left(G_D(\vec k,z,z;\kappa)-G_N(\vec k,z,z;\kappa)\right).
\end{equation}
It is easy to see how the expression~\eqref{eq:tad-1-a} reduces to~\eqref{eq:tad_cut} for $\kappa\to 0$, where it then diverges. So this is the de Sitter-invariant modification of the latter.   Furthermore, this sum of two diagrams can be expressed in terms of the flat space tadpole in four dimensions by combining the integration over the 3d loop momentum $\vec \ell$
 and the energy $\omega$ into $L:=(\omega,\vec \ell)$ to give
 \begin{equation}
I^{(2)}_\circ(k)= \frac{\lambda}{(2\pi)^4}\int\limits_0^\infty dz \frac{\bar G(k,z;\kappa)^2}{z^{2+2\kappa}}\int  \frac{{\rm d}^4 L}{(L^2)^{1+\kappa}}.
\end{equation}
The $L$-integral vanishes in analytic regularization. In particular, there is no pole at $\kappa=0$ and by consequence no logarithmic correction to the spectral function ($\propto k$) of the two-point function.

%----------------------------------------------
\subsubsection{Two-loop tadpole }
The two-loop tadpole in the $\kappa$-regularisation is the first diagram that is sensitive to the $\kappa$-deformation of the action. The modified interaction term results in  \begin{align}
I_{\circ\circ}^{(2)}(k)=\frac{1}{4}\left(\frac{\lambda}{2}\right)^2\left(I_{\circ\circ}^{(2)+}+I_{\circ\circ}^{(2)-}\right)
\end{align}
where
\begin{equation}\label{eq:2-loop-tad}
I_{\circ\circ}^{(2)+}(k)=
\frac{\mathcal{C}^2_{2\kappa}}{\mathcal{C}^3_{\kappa}}\left(
\begin{gathered}
\begin{tikzpicture}[scale=0.6]
  \filldraw [color = black, fill=none, very thick] (0,3) circle (1cm);
  \filldraw [color = black, fill=none, very thick] (0,1) circle (1cm);
  \draw [black,very thick, dashed] (0.5,-1) to (0,0);
    \draw [black,very thick, dashed] (-0.5,-1) to (0,0);
\end{tikzpicture}    
\end{gathered}+\begin{gathered}\begin{tikzpicture}[scale=0.6]
  \filldraw [color = black, fill=none, very thick] (0,3) circle (1cm);
  \filldraw [color = black, fill=none, very thick,dashed] (0,1) circle (1cm);
  \draw [black,very thick, dashed] (0.5,-1) to (0,0);
    \draw [black,very thick, dashed] (-0.5,-1) to (0,0);
  \end{tikzpicture}\end{gathered}
  +\begin{gathered}\begin{tikzpicture}[scale=0.6]
  \filldraw [color = black, fill=none, very thick,dashed] (0,3) circle (1cm);
  \filldraw [color = black, fill=none, very thick] (0,1) circle (1cm);
  \draw [black,very thick, dashed] (0.5,-1) to (0,0);
    \draw [black,very thick, dashed] (-0.5,-1) to (0,0);
  \end{tikzpicture}
 \end{gathered}+\begin{gathered}\begin{tikzpicture}[scale=0.6]
   \filldraw [color = black, fill=none, very thick,dashed] (0,3) circle (1cm);
  \filldraw [color = black, fill=none, very thick,dashed] (0,1) circle (1cm);
  \draw [black,very thick, dashed] (0.5,-1) to (0,0);
    \draw [black,very thick, dashed] (-0.5,-1) to (0,0);
\end{tikzpicture} \end{gathered}\right)
\end{equation}
and 
\begin{equation}
I_{\circ\circ}^{(2)-}=
-(2\pi\kappa)^2\left(
\begin{gathered}\begin{tikzpicture}[scale=0.6]
  \draw [color = black, very thick,dashed] (0,3) circle (1cm);
  \draw [color = black, very thick] (0,2) arc (90:270:1cm);
  \draw [color = black, very thick,dashed] (0,2) arc (90:-90:1cm);
  \draw [black,very thick, dashed] (0.5,-1) to (0,0);
    \draw [black,very thick, dashed] (-0.5,-1) to (0,0);
\end{tikzpicture}\end{gathered}+\begin{gathered}\begin{tikzpicture}[scale=0.6]
  \draw [color = black,  very thick] (0,3) circle (1cm);
   \draw [color = black, very thick] (0,2) arc (90:270:1cm);
  \draw [color = black, very thick,dashed] (0,2) arc (90:-90:1cm);
  \draw [black,very thick, dashed] (0.5,-1) to (0,0);
    \draw [black,very thick, dashed] (-0.5,-1) to (0,0);
  \end{tikzpicture}\end{gathered}\right)\,.
\end{equation}
The first graph has the integral representation 
\begin{align}\label{eq:2loopmh_kappa}
 I_{\circ\circ}^{(2)+}\propto 
 \int\limits_{-\infty}^\infty dp_1\int\limits_{-\infty}^\infty  dp_2 \int\limits_{-\infty}^\infty  dp_3 \int d^3 \vec\ell_1&\int d^3 \vec\ell_2\int\limits_{-\infty}^{\infty}\frac{d z_1}{z_1^{2+2\kappa}} \frac{dz_2}{z_2^{4\kappa}} \bar G(k,z_1;\kappa)^2\cdot\\
&\frac{\left(\cos(p_+ z_1)  \cos(p_+z_2) +  \cos(p_-z_1)\cos(p_-z_2)\right)}{(p_1^2+\vec{\ell}_1^2)^{1+\kappa}(p_3^2+\vec{\ell}_2^2)^{1+\kappa}(p_2^2+\vec{\ell}_1^2)^{1+\kappa}}\nonumber\\
= 2
\int\limits_{-\infty}^\infty dp_1\int\limits_{-\infty}^\infty  dp_2 \int\limits_{-\infty}^\infty  dp_3 \int d^3 \vec\ell_1&\int d^3 \vec\ell_2\int\limits_0^{\infty}\frac{d z_1}{z_1^{2+2\kappa}} \frac{dz_2}{z_2^{4\kappa}} \bar G(k,z_1;\kappa)^2\cdot\nonumber\\
&\qquad\qquad\frac{ \cos(p_-z_1)\cos(p_-z_2)}{(p_1^2+\vec{\ell}_1^2)^{1+\kappa}(p_3^2+\vec{\ell}_2^2)^{1+\kappa}(p_2^2+\vec{\ell}_1^2)^{1+\kappa}}\nonumber\\
= 2
\int\limits_{-\infty}^\infty  dp_- \int d^4 L_1\int d^4 L_2\int\limits_0^{\infty}&\frac{d z_1}{z_1^{2+2\kappa}} \frac{dz_2}{z_2^{4\kappa}} \frac{ \bar G(k,z_1;\kappa)^2\cos(p_-z_1)\cos(p_-z_2)}{(L_1^2)^{1+\kappa}(L_2^2)^{1+\kappa}((Q+L_1)^2)^{1+\kappa}}=0\,.\nonumber
\end{align}
Here we suppressed an overall constant since the final result vanishes. Indeed, all integrals, including the $z_2$ integral,  are well-defined for $\frac{1}{4}>\kappa>0$. However, we won't need to compute the latter since the $L_2$ integral already gives a zero answer in analytic regularization. The contribution of the second set of graphs vanishes for $\kappa\to 0$ since, due to the relative minus sign, the leading $1/\kappa^2$ pole is cancelled. Again, there is no logarithmic correction to the spectral function.

\subsubsection{Sunset}

At two loops we have the sum of the two sunset diagrams: 
\begin{equation}
   I^{(2)}_{\circleddash}(k)=\frac{\mathcal{C}^2_{2\kappa}}{\mathcal{C}^3_{\kappa}}\left(\frac{\lambda}{2}\right)^2\left(\begin{gathered}
       \begin{tikzpicture}[scale=0.6]
  \filldraw [color = black, fill=none, very thick,dashed] (0,0) circle (1cm);
  \draw [dashed,black,very thick] (-2,0) to (2,0);
  \end{tikzpicture}
   \end{gathered} +3\; \begin{gathered}
       \begin{tikzpicture}[scale=0.6]
  \filldraw [color = black, fill=none, very thick] (0,0) circle (1cm);
  \draw [dashed,black,very thick] (-2,0) to (2,0);
  \end{tikzpicture}
\end{gathered}\right)+ O(\kappa^2)
\begin{gathered}
       \begin{tikzpicture}[scale=0.6]
  \filldraw [dashed,color = black, fill=none, very thick] (0,0) circle (1cm);
  \draw [dashed,black,very thick] (-2,0) to (-1,0);
    \draw [dashed,black,very thick] (1,0) to (2,0);
  \draw [black,very thick] (-1,0) to (1,0);
  \end{tikzpicture}
   \end{gathered}.
 \end{equation}
The  diagram  with  three dashed lines and one solid line comes with a
prefactor proportional to $\kappa^2$. This does not contribute to the
leading $1/\kappa$ divergence.

Using the symmetry between the vertices $z_1$, and $z_2$ this becomes
\begin{multline}
I_{\circleddash}^{(2)}(k)={\lambda^2\over 4}\frac{\mathcal{C}^2_{2\kappa}}{\mathcal{C}^3_{\kappa}}{1\over (2\pi)^9}\iint_0^\infty {dz_1dz_2\over z_1^{1+3\kappa}z_2^{1+3\kappa}}
 \bar G( k,z_1;\kappa) 
   \bar G( k,z_2;\kappa) \cr
   \times \iiint_{-\infty}^\infty dp_1dp_2dp_3 \cos((p_1+p_2+p_3)z_1)\cos((p_1+p_2+p_3)z_2)\cr
   \times \iint
   {d^3\ell_1d^3\ell_2d^3\ell_3\delta^3(\vec{\ell}_1+\vec{\ell}_2+\vec{\ell}_3-\vec
     k)\over
     (\ell_1^2+p_1^2)^{1+\kappa}(\ell_2+p_2^2)^{1+\kappa}(\ell_3^2+p_3^2)^{1+\kappa}}.
 \end{multline}
 Shifting $p_3$ we can rewrite this contribution as an integral of a four-dimensional flat space massless sunset 
 \begin{multline}
I_{\circleddash}^{(2)+}(k)={\lambda^2\over 4}\frac{\mathcal{C}^2_{2\kappa}}{\mathcal{C}^3_{\kappa}}{1\over (2\pi)^9}\iint_0^\infty {dz_1dz_2\over z_1^{1+3\kappa}z_2^{1+3\kappa}}
 \bar G( k,z_1;\kappa) 
   \bar G( k,z_2;\kappa)  \int_{-\infty}^\infty dp_3 \cos(p_3 z_1)\sin(p_3z_2)\cr
   \times \iint
   {d^4L_1d^4L_2\over
     (L_1^2)^{1+\kappa}(L_2^2)^{1+\kappa}((L_1+L_2+Q)^2)^{1+\kappa}},
 \end{multline}
 where $Q=(p_3,\vec k)$ and  
\begin{equation}
    \iint
   {d^4L_1d^4L_2\over
     (L_1^2)^{1+\kappa}(L_2^2)^{1+\kappa}((L_1+L_2+Q)^2)^{1+\kappa}}=\frac{\pi ^4 \Gamma (1-\kappa )^3 \Gamma (3 \kappa -1) \left(k^2+p_3^2\right)^{1-3 \kappa }}{\Gamma (3-3 \kappa ) \Gamma
   (\kappa +1)^3}\,. 
\end{equation}
Performing the $p_3$ integral for large enough $\kappa$ gives
\begin{equation}
 \int\limits_{-\infty}^\infty dp_3 {\cos(p_3 z_1)\cos(p_3z_2)\over 
 (\vec k^2+p_3^2)^{-1+3\kappa}}=
 \frac{\sqrt{\pi } 2^{\frac{3}{2}-3 \kappa } |\vec k|^{\frac{3}{2}-3 \kappa }}{\Gamma (3
   \kappa -1)}
   \left(\frac{K_{\frac{3}{2}-3 \kappa }\left(k \left| z_1-z_2\right|
   \right)}{\left|z_1-z_2\right|^{\frac{3}{2}-3 \kappa
   }} + \frac{K_{\frac{3}{2}-3 \kappa }\left(k \left(z_1+z_2\right) \right)}{\left(z_1+z_2\right)^{\frac{3}{2}-3 \kappa
   }} \right).
\end{equation}
Upon absorbing the energy $k$ in $z$, the sunset integral then becomes
\begin{multline}
  I_{\circleddash}^{(2)+}(k)={\lambda^2\, k^{1-2\kappa}\over 4(2\pi)^9} {\mathcal C^2_{2\kappa}\over
    \mathcal C_\kappa^3}  \frac{\pi ^{9/2} 2^{\frac{3}{2}-3 \kappa } \Gamma (1-\kappa )^3 }{\Gamma
   (3-3 \kappa ) \Gamma (\kappa +1)^3}\cr
 \times\iint_0^\infty {dz_1dz_2\over
    z_1^{1+3\kappa} z_2^{1+3\kappa}} \bar G(1,z_1;\kappa) \bar
  G(1,z_2;\kappa)   \left(\frac{K_{\frac{3}{2}-3 \kappa }\left(\left| z_1-z_2\right|
   \right)}{\left|z_1-z_2\right|^{\frac{3}{2}-3 \kappa
   }} +\frac{K_{\frac{3}{2}-3 \kappa }\left(z_1+z_2 \right)}{\left(z_1+z_2\right)^{\frac{3}{2}-3 \kappa
   }} \right)\,.
\end{multline}
The  integration over $z_1$ and $z_2$ has a  simple pole when $\kappa\to0$, 
 \begin{equation}
I_{\circleddash}^{(2)+}(k)\propto\lambda^2\frac{1}{\kappa}\,,
 \end{equation}
 plus finite terms arising from the $\kappa$ expansion of the measure.
 The simple pole in $\kappa$ is cancelled by a  regularization of the  mass
 shift  diagram
 \begin{equation}
     \delta I^{(2)}=\delta m\int\frac{dz}{z^4}\bar G(k,z;\kappa)^2 \,.
 \end{equation}
In addition there may be finite contributions involving $\log(k \ell_{dS})$ whose coefficient depends on the choice of subtraction scheme. Note that since $\delta m$ receives a contribution from the leading divergence, it is fixed by the result from the hard-cut off as discussed in section \ref{sec:renormalization}.

\subsection{Four-point function}
While our two-point functions do not contain any non-trivial dependence on the kinematic variables this is not so for the four-point functions. This is why the regularization enters crucially here.\footnote{In  more general situations, two-point functions may contain important dynamical information as well. See eg.~\cite{Senatore:2009cf, Chakraborty:2023qbp}.  } 
\subsubsection{Cross diagram}
In analytic regularization the cross diagram is also $\kappa$-dependent (since the boundary conformal dimensions are $\kappa$-dependent). Of course, since it is finite we may set $\kappa=0$ for its evaluation. However, this  $\kappa$-dependence will play a role for renormalisation, when combined with poles in the bare coupling. We have 
\begin{align}
I^{(4)}_\times=\begin{gathered}
\begin{tikzpicture}[scale=0.6]
  \draw [black,very thick] (0.5,-1) to (-0.5,1);
    \draw [black,very thick] (-0.5,-1) to (0.5,1);
  \end{tikzpicture}    
\end{gathered}= \frac{\lambda}{{2}}
\int_0^\infty {dz\over z^4 }\prod_{i=1}^4\bar G(k_i,z;\kappa)\,,
\end{align}
which, for $\kappa\to 0$, reduces to 
\begin{align}
  I^{(4)}_\times=  -\frac{\lambda}{2E} \times \frac{1}{k_1 k_2 k_3 k_4}
\end{align}

\subsubsection{One-Loop }
Collecting all diagrams originating from the effective action~\eqref{eq:EAdS_action_phi4_k2} we have 
\begin{equation}
    I^{(4)}_\circ= \frac{\lambda^2}{8}\frac{\mathcal{C}^2_{2\kappa}}{\mathcal{C}^2_{\kappa}}\left(\begin{gathered}
    \begin{tikzpicture}[scale=0.6]
  \filldraw [color = black, fill=none, very thick] (0,1) circle (1cm);
  \draw [black,very thick, dashed] (0.5,-1) to (0,0);
    \draw [black,very thick, dashed] (-0.5,-1) to (0,0);
    \draw [black,very thick, dashed] (0.5,3) to (0,2);
    \draw [black,very thick, dashed] (-0.5,3) to (0,2);
  \end{tikzpicture}\end{gathered} + \begin{gathered}\begin{tikzpicture}[scale=0.6]
  \filldraw [color = black, fill=none, very thick,dashed] (0,1) circle (1cm);
  \draw [black,very thick, dashed] (0.5,-1) to (0,0);
    \draw [black,very thick, dashed] (-0.5,-1) to (0,0);
    \draw [black,very thick, dashed] (0.5,3) to (0,2);
    \draw [black,very thick, dashed] (-0.5,3) to (0,2);
\end{tikzpicture}
\end{gathered}\right)
+O(\kappa^2)\;\begin{gathered}\begin{tikzpicture}[scale=0.6]
  \draw [color = black, very thick] (0,2) arc (90:270:1cm);
  \draw [color = black, very thick,dashed] (0,2) arc (90:-90:1cm);
  \draw [black,very thick, dashed] (0.5,-1) to (0,0);
    \draw [black,very thick, dashed] (-0.5,-1) to (0,0);
    \draw [black,very thick, dashed] (0.5,3) to (0,2);
    \draw [black,very thick, dashed] (-0.5,3) to (0,2);
\end{tikzpicture}
\end{gathered}
\end{equation}
To continue we keep only the terms to zeroth order in $\kappa$ since the diagrams have at most simple poles in $\kappa$:
\begin{equation}
    I^{(4)}_\circ=\frac{\lambda^2}{8}\left( \begin{gathered}
    \begin{tikzpicture}[scale=0.6]
  \filldraw [color = black, fill=none, very thick] (0,1) circle (1cm);
  \draw [black,very thick, dashed] (0.5,-1) to (0,0);
    \draw [black,very thick, dashed] (-0.5,-1) to (0,0);
    \draw [black,very thick, dashed] (0.5,3) to (0,2);
    \draw [black,very thick, dashed] (-0.5,3) to (0,2);
  \end{tikzpicture}\end{gathered} + \begin{gathered}\begin{tikzpicture}[scale=0.6]
  \filldraw [color = black, fill=none, very thick,dashed] (0,1) circle (1cm);
  \draw [black,very thick, dashed] (0.5,-1) to (0,0);
    \draw [black,very thick, dashed] (-0.5,-1) to (0,0);
    \draw [black,very thick, dashed] (0.5,3) to (0,2);
    \draw [black,very thick, dashed] (-0.5,3) to (0,2);
\end{tikzpicture}
\end{gathered}\right)+O(\kappa)\,,
\end{equation}
which evaluates to\footnote{In this section we keep the dependence on the de Sitter-radius, $\ell_{dS}\to i\ell_{AdS}$ after Wick rotation,  manifest to clarify the renormalization procedure.} 

\begin{align}\label{eq:loopmh_kappa}
 I^{(4)}_\circ= \frac{\lambda^2}{4(2\pi)^5}\int\limits_{-\infty}^\infty dp&\int\limits_{-\infty}^\infty  dp' \int d^3 \vec\ell\int\limits_{0}^{\infty}\frac{d z_1}{(z_1\ell_{AdS}^{-1})^{2+2\kappa}} \frac{dz_2}{(z_2\ell_{AdS}^{-1})^{2+2\kappa}} \prod\limits_{i=1}^2\bar G(k_i,z_1;\kappa)\prod\limits_{i=3}^4\bar G(k_i,z_2;\kappa)\nonumber\\
&\frac{\cos(p_+ z_1)  \cos(p_+z_2) +  \cos(p_-z_1)\cos(p_-z_2)}{(p^2+\vec{\ell}^2)^{1+\kappa}(p'^2+(\vec{\ell}+\vec{k}_{12})^2)^{1+\kappa}}\,.
\end{align}
Combining the three-dimensional loop momentum $\vec \ell$ with the energy $p$
into $L=(p,\vec \ell)$, we have an expression in terms of the four-dimensional massless bubble evaluated in analytic regularization:
\begin{align}
 I^{(4)}_\circ= \frac{\lambda^2}{2(2\pi)^5}\int\limits_{-\infty}^\infty dp_-\int d^4L &\int\limits_0^{\infty}\frac{d z_1}{(z_1\ell_{AdS}^{-1})^{2+2\kappa}} \frac{dz_2}{(z_2\ell_{AdS}^{-1})^{2+2\kappa}} \prod\limits_{i=1}^2\bar G(k_i,z_1;\kappa)\prod\limits_{i=3}^4\bar G(k_i,z_2;\kappa)\nonumber\\
&\times\frac{ \cos(p_-z_1)\cos(p_-z_2)}{(L^2)^{1+\kappa}((L+Q)^2)^{1+\kappa}}.
\end{align}
Here $p_-=p'-p$, $L=(p,\vec{\ell})$, $Q=(p_-,\vec{k}_{12})$. The $L$-integral is identical to the one-loop massless
bubble in flat space in analytic regularization giving
\begin{multline}\label{eq:aloopkf}
 I^{(4)}_\circ= \frac{\lambda^2}{2(2\pi)^5}\int\limits_{-\infty}^\infty dp_-\int\limits_0^{\infty}\frac{d z_1}{(az_1)^{2+2\kappa}} \frac{dz_2}{(az_2)^{2+2\kappa}} \prod\limits_{i=1}^2\bar G(k_i,z_1;\kappa)\prod\limits_{i=3}^4\bar G(k_i,z_2;\kappa)\cr
\times \cos(p_-z_1)\cos(p_-z_2) \frac{\pi^2\Gamma(1 - \kappa)^2\Gamma(2\kappa)}{\Gamma(2 - 2\kappa)\Gamma(1 + \kappa)^2
                (\vec{k}_{12}^2 + p_-^2)^{2\kappa}},
\end{multline}
which has the $\kappa$-expansion
\begin{multline}\label{eq:aloopkfexp}
 I^{(4)}_\circ
              = \frac{ (\lambda/2)^2}{2(2\pi)^3}\int\limits_{-\infty}^\infty dp_-\int\limits_0^{\infty}\frac{d z_1}{(z_1\ell_{AdS}^{-1})^{2}} \frac{dz_2}{(z_2\ell_{AdS}^{-1})^{2}} \prod\limits_{i=1}^2\bar G(k_i,z_1;\kappa)\prod\limits_{i=3}^4\bar G(k_i,z_2;\kappa)\cr
 \times\cos(p_-z_1)\cos(p_-z_2)\left({1\over 2\kappa}+1-\log(az_1az_2(\vec{k}_{12}^2 + p_-^2))+O(\kappa^2)\right),
\end{multline}
reproducing the expected pole in $\kappa$. The $p_-$-integral then evaluates to
\begin{multline}
     I^{(4)}_\circ|_{\kappa^{-1}}= \frac{\lambda^2}{32(2\pi)^2\kappa} \int\limits_0^{\infty}\frac{d z_1}{(z_1\ell_{AdS}^{-1})^{2}} \frac{dz_2}{(z_2\ell_{AdS}^{-1})^{2}} \prod\limits_{i=1}^2\bar G(k_i,z_1;\kappa) \prod\limits_{i=3}^4\bar G(k_i,z_2;\kappa) \cr\times\left(\delta(z_1+z_2)+\delta(z_1-z_2)\right),
\end{multline}
where we used that
\begin{equation}\label{e:intcosdelta}
  \int\limits_{-\infty}^\infty dp_- \cos(p_-z_1)\cos(p_-z_2)=
    \pi \,\left(\delta(z_1+z_2)+\delta(z_1-z_2)\right).
\end{equation}
Since the integration over $z_1$ and $z_2$ is only on the positive real line, the singular part  is given by the cross diagram
\begin{equation}\label{eq:1-l-cout}
   I^{(4)}_\circ|_{\kappa^{-1}}= \frac{\lambda^2}{32(2\pi)^2\kappa}
    \int\limits_0^{\infty}\frac{d z}{(z\ell_{AdS}^{-1})^{4}}
    \prod\limits_{i=1}^4\bar G(k_i,z;\kappa)
    =\frac{(\lambda/2)}{32\pi^2 \kappa}  I_\times.
\end{equation}
To determine the one-loop $\beta$-function we then write $ \lambda=\lambda_R\,\frac{\mu^{4\kappa}}{\ell_{AdS}\mu}+\delta\lambda$ where $\mu$ has the dimension of mass. Adding the $t$ and $u$-channel to~\eqref{eq:1-l-cout} we have 
\begin{align}
    \delta\lambda =- \frac{3 \lambda_R^2\mu^{4\kappa}}{ 64\pi^2\kappa}\,,
\end{align}
and thus the Callan-Symanzik equation,
\begin{align}
 0=\mu\partial_\mu\lambda_R  -\frac{3\lambda_R^2}{16\pi^2}\,.
\end{align}
This reproduces the flat space $\beta$-function (see also~\cite{Bertan:2018khc, Heckelbacher:2022hbq}) which was also obtained using the hard cutoff in Section~\ref{sec:renormalization}. 

In order to determine the finite contribution in the $s$-channel we first need to specify the subtraction scheme.  We use the scale-invariant scheme in which we subtract all terms which do not come from the $\kappa$-expansion of $1/(z_1^{2+2\kappa}z_2^{2+2\kappa}(\vec{k}_{12}^2 + p_-^2)^{2\kappa})$. The finite part is then given by  
\begin{multline}\label{eq:aloopkffinite}
 I^{(4)}_\circ|_{\kappa^0}=\frac{ (\lambda_R/2)^2}{2(2\pi)^3}\int\limits_{-\infty}^\infty
  dp_-\int\limits_0^{\infty}\frac{d z_1}{(z_1\ell_{AdS}^{-1})^{2}}
        \frac{dz_2}{(z_2\ell_{AdS}^{-1})^{2}}
        \prod\limits_{i=1}^2\bar G(k_i,z_1,0)\prod\limits_{i=3}^4\bar G(k_i,z_2,0)\cr
    \times     \cos(p_-z_1)\cos(p_-z_2)\left(1-\log\left(\frac{z_1z_2}{\mu^2\ell_{AdS}^{2}}(\vec{k}_{12}^2 +
                       p_-^2)\right)\right).
                   \end{multline}
 Using the expression for the leading contribution of the
 bulk-to-boundary propagator in~\eqref{e:LambdaBdser},    we can first integrate over $z_1$ and $z_2$ as described in Appendix~\ref{sec:integrals} and then over $p_-$ to find (with $\mu=a$) 
 \begin{multline}\label{analyticbubble1}
I^{(4)}_\circ|_{\kappa^0}=\frac{ (\lambda_R/2)^2}{2(4\pi)^2 k_1 k_2 k_3 k_4}
{1+2\gamma_E\over k_{12}+k_{34}}\cr
+\frac{ (\lambda_R/2)^2}{4(2\pi)^2 k_1 k_2 k_3 k_4} \left({1\over
    k_{12}-k_{34}}\log\left(k_{12}+|\vec k_{12}|\over k_{34}+|\vec k_{12}|\right)-
  {1\over k_{12}+k_{34}}\log\left((k_{12}+|\vec k_{12}|)( k_{34}+|\vec k_{12}|)\over
  (k_{12}+k_{34})^2\right)\right)\,.
 \end{multline}
The structure of this equation is very similar to the answer obtained via the hard-cutoff~\eqref{1loop4ptcutoff} when $\Lambda$ is replaced by $k_{12} + k_{34}$. This was  also noted in Section~\ref{sec:renormalization} by demanding that the renormalised correlator satisfies the conformal ward identity. Apart from the log's appearing in the second line above, we also see the appearance of the cross term in the first line, that contributes to the residue at the total energy pole and hence contributes to the flat space limit\footnote{These terms are often absent in the renormalised correlator by choosing the  $\overline{MS}$-scheme.}. We also note that de Sitter-invariant regularization scheme gives the expected divergent piece~\eqref{eq:1-l-cout}, which is conformally invariant and was absent for the answer obtained in~\eqref{eq:4ptloop_2}. Hence this example for the bubble diagram shows how this regularization scheme ensures that the correlators automatically satisfy the conformal Ward identity at one-loop. This is different from previous computations of the bubble diagram~\cite{ Salcedo:2022aal} for the cosmological correlator where the final answers do not satisfy the conformal ward identity. Substituting  their cutoff  $\Lambda \to k_{12} + k_{34}$ leads to an answer that satisfies the conformal Ward identity at one-loop but misses various divergent factors and total energy poles.

 %\newpage

%-----------------------------------------------
\subsubsection{Two-loop Necklace}
Keeping all terms that contribute when $\kappa\to 0$, we find 
\begin{equation}\label{e:Icircirckappa0}
I^{(4)}_{\circ\circ}=\left(\frac{\lambda}{2}\right)^3\frac{1}{4}\frac{\mathcal{C}^3_{2\kappa}}{\mathcal{C}^4_{\kappa}}\left(
\begin{gathered}
    \begin{tikzpicture}[scale=0.6]
  \filldraw [color = black, fill=none, very thick] (0,3) circle (1cm);
  \filldraw [color = black, fill=none, very thick] (0,1) circle (1cm);
  \draw [black,very thick, dashed] (0.5,5) to (0,4);
    \draw [black,very thick, dashed] (-0.5,5) to (0,4);
  \draw [black,very thick, dashed] (0.5,-1) to (0,0);
    \draw [black,very thick, dashed] (-0.5,-1) to (0,0);
  \end{tikzpicture}
  \end{gathered}
+%\lambda^3\left(1+3{(\pi\kappa)^2\over2}\right) \left(
\begin{gathered}
\begin{tikzpicture}[scale=0.6]
  \filldraw [color = black, fill=none, very thick] (0,3) circle (1cm);
  \filldraw [color = black, fill=none, very thick,dashed] (0,1) circle (1cm);
   \draw [black,very thick, dashed] (0.5,5) to (0,4);
    \draw [black,very thick, dashed] (-0.5,5) to (0,4);
  \draw [black,very thick, dashed] (0.5,-1) to (0,0);
    \draw [black,very thick, dashed] (-0.5,-1) to (0,0);
  \end{tikzpicture}\end{gathered}
  +\begin{gathered}\begin{tikzpicture}[scale=0.6]
  \filldraw [color = black, fill=none, very thick,dashed] (0,3) circle (1cm);
  \filldraw [color = black, fill=none, very thick] (0,1) circle (1cm);
   \draw [black,very thick, dashed] (0.5,5) to (0,4);
    \draw [black,very thick, dashed] (-0.5,5) to (0,4);
'  \draw [black,very thick, dashed] (0.5,-1) to (0,0);
    \draw [black,very thick, dashed] (-0.5,-1) to (0,0);
  \end{tikzpicture}
  \end{gathered}+\begin{gathered}\begin{tikzpicture}[scale=0.6]
   \filldraw [color = black, fill=none, very thick,dashed] (0,3) circle (1cm);
  \filldraw [color = black, fill=none, very thick,dashed] (0,1) circle (1cm);
   \draw [black,very thick, dashed] (0.5,5) to (0,4);
    \draw [black,very thick, dashed] (-0.5,5) to (0,4);
  \draw [black,very thick, dashed] (0.5,-1) to (0,0);
    \draw [black,very thick, dashed] (-0.5,-1) to (0,0);
\end{tikzpicture} 
\end{gathered}
\right)
\end{equation}
Applying the Feynman rules this set of graphs gives the integral representation 
\begin{multline}
 \frac{4}{(2\pi)^{10}}\int d^3\vec\ell_1 d^3\vec\ell_2 \int\limits_{-\infty}^{\infty} d p_1 d p_2 d p_3 d p_4 \int\limits_{-\infty}^{\infty}\frac{d z_1}{z_1^{2+2\kappa}}\frac{d z_2}{z_2^{4\kappa}} \frac{dz_3}{z_3^{2+2\kappa}} \prod\limits_{i=1}^2\bar G(k_i,z_1;\kappa)\prod\limits_{i=3}^4\bar G(k_i,z_3;\kappa)\times \cr
\frac{ \left(\cos(p^+_{12} z_1)  \cos(p^+_{12}z_2) +
    \cos(p^-_{12}z_1)\cos(p^-_{12}z_2)\right)\left(\cos(p^+_{34} z_2)
    \cos(p^+_{34}z_3) +
    \cos(p^-_{34}z_2)\cos(p^-_{34}z_3)\right)}{(p_1^2+\vec{\ell_1}^2)^{1+\kappa}
  ((\vec{\ell_1} +\vec{k}_{12})^2+p_2^2)^{1+\kappa} (p_3^2+\vec{\ell_2}^2)^{1+\kappa}((\vec{\ell_2} +\vec{k}_{34})^2+p_4^2)^{1+\kappa}}\,,
\end{multline}
where $p_{12}^\pm= p_1\pm p_2$ and $p_{34}^\pm=p_3\pm p_4$. After collecting identical integrals this simplifies to 
\begin{multline}
\frac{16}{(2\pi)^{10}} \int d^3\vec\ell_1 d^3\vec\ell_2 \int\limits_{-\infty}^{\infty} d p_1 d p_2 d p_3 d p_4 \int\limits_0^{\infty}\frac{d z_1}{z_1^{2+2\kappa}}\frac{d z_2}{z_2^{4\kappa}} \frac{dz_3}{z_3^{2+2\kappa}} \prod\limits_{i=1}^2\bar G(k_i,z_1;\kappa)\prod\limits_{i=3}^4\bar G(k_i,z_3;\kappa)\cr
\times\frac{ \cos(p^+_{12} z_1)  \cos(p^+_{12}z_2) \cos(p^+_{34} z_2)  \cos(p^+_{34}z_3) }{(p_1^2+\vec{\ell_1}^2)^{1+\kappa}
  ((\vec{\ell_1} +\vec{k}_{12})^2+p_2^2)^{1+\kappa} (p_3^2+\vec{\ell_2}^2)^{1+\kappa}((\vec{\ell_2} +\vec{k}_{34})^2+p_4^2)^{1+\kappa}}.
\end{multline}
%%%
Shifting $p_2\to p_2+p_1$ and $p_4\to p_4+p_3$,
one can again combine the integration over the three-dimensional momenta
$\vec{\ell}_1$ and $\vec{\ell}_2$ with the integration over $p_1$ and $p_2$
respectively into a four dimensional integral over $L_1=(p_1,\vec{\ell}_1)$
and $L_2=(p_2,\vec{\ell}_2)$:
\begin{multline}
 I^{(4)}_{\circ\circ}
=\left(\frac{\lambda}{2}\right)^3\frac{16}{(2\pi)^{10} } \frac{\mathcal{C}^3_{2\kappa}}{\mathcal{C}^4_{\kappa}}\int\limits_{-\infty}^{\infty} d p_2 d p_4 \int\limits_0^{\infty}\frac{d z_1}{z_1^{2+2\kappa}}\frac{d z_2}{z_2^{4\kappa}} \frac{dz_3}{z_3^{2+2\kappa}} \prod\limits_{i=1}^2\bar G(k_i,z_1;\kappa)\prod\limits_{i=3}^4\bar G(k_i,z_3;\kappa)\cr
\times \cos(p_2z_1)\cos(p_2z_2)
\cos(p_4z_2)\cos(p_4z_3) \cr \int_{\mathbb R^8} \frac{ d^4 L_1  d^4 L_2 }{(L_1^2)^{1+\kappa}((L_1+Q_1)^2)^{1+\kappa}(L_2^2)^{1+\kappa}((L_2+Q_2)^2)^{1+\kappa}}.
\end{multline}
The integrals over $L_1$ and $L_2$ are massless four dimensional
bubble integrals with $Q_1=(p_2,\vec{k}_{12})$ and
$Q_2=(p_4,\vec{k}_{34})$. Then performing the (flat-space)
$L$-integrals and using that $\vec k_{12}+\vec k_{34}=0$, we get\footnote{\label{fot:ck}In what follows we will ignore the factor $\frac{\mathcal{C}^3_{2\kappa}}{\mathcal{C}^4_{\kappa}}$ since it can be absorbed in a non-minimal subtraction of the cross counter term at two-loops. } 
\begin{multline}
 I^{(4)}_{\circ\circ}=\left(\frac{\lambda}{2}\right)^3\frac{1}{(2\pi)^6 }\int\limits_{-\infty}^{\infty}d p_2d p_4 \int\limits_0^{\infty}\frac{d z_1}{z_1^{2+2\kappa}}\frac{d z_2}{z_2^{4\kappa}} \frac{dz_3}{z_3^{2+2\kappa}} \prod\limits_{i=1}^2\bar G(k_i,z_1;\kappa)\prod\limits_{i=3}^4\bar G(k_i,z_3;\kappa)\cr
\times\frac{ \cos(p_2 z_1)\cos(p_2 z_2)  \cos(p_4 z_2)\cos(p_4 z_3)(\Gamma(1 - \kappa)^2\Gamma(2\kappa))^2}{\Gamma(2 - 2\kappa)^2\Gamma(1 + \kappa)^4
  (\vec{k}_{12}^2 + p_2^2)^{2\kappa}(\vec{k}_{12}^2 + p_4^2)^{2\kappa}}.
\end{multline}
The double pole contribution is then
\begin{multline}
  I^{(4)}_{\circ\circ}|_{\kappa^{-2}}=\left(\frac{\lambda}{2}\right)^3\frac{1}{(2\pi)^6}\frac{1}{2\kappa^2}\int\limits_{-\infty}^{\infty} d p_2
  \int\limits_{-\infty}^{\infty}d p_4 \int\limits_0^{\infty}{d
    z_1\over z_1^2}
  d z_2 {dz_3\over z_3^2}\prod\limits_{i=1}^2\bar G(k_i,z_1;\kappa)\prod\limits_{i=3}^4\bar G(k_i,z_3;\kappa)\; \cr
 \times\cos(p_2z_1)\cos(p_2z_2)  \cos(p_4z_2)\cos(p_4z_3).
\end{multline}
The final integral over $p_-^{12}$ and $p_-^{34}$ is done
using~\eqref{e:intcosdelta} so that that pole of order $\kappa^{-2}$
is proportional to the cross diagram
\begin{equation}
  I^{(4)}_{\circ\circ}|_{\kappa^{-2}}=\left(\frac{\lambda}{2}\right)^3\frac{2}{(4\pi)^4}\frac{1}{\kappa^2}
 \int\limits_0^{\infty}\frac{d
    z}{z^{4}}\prod\limits_{i=1}^4\bar G(\vec{k}_i,z;\kappa)=
  \left(\frac{\lambda}{2}\right)^2\frac{2}{(4\pi)^4}\frac{1}{\kappa^2} I_\times .
\end{equation}
As mentioned above, the correction resulting form the prefactor   $\frac{\mathcal{C}^3_{2\kappa}}{\mathcal{C}^4_{\kappa}}$ being proportional to  $\kappa^2$ merely results in a non-minimal subtraction of this cross term. 

For the sub-leading first-order pole of order $\kappa^{-1}$ we find  
\begin{multline}\label{eq:neckk-1}
  I^{(4)}_{\circ\circ}|_{\kappa^{-1}}=  -\left(\frac{\lambda}{2}\right)^3\frac{8}{(2\pi)^6 }\frac{1}{\kappa}
 \int\limits_{-\infty}^{\infty} d p_2 \int\limits_{-\infty}^{\infty}d p_4 \int\limits_0^{\infty}\frac{d z_1}{z_1^{2}}dz_2 \frac{dz_3}{z_3^{2}} \prod\limits_{i=1}^2\bar G(k_i,z_1,0)\prod\limits_{i=3}^4\bar G(k_i,z_3,0)\cr
\cos(p_2z_1)\cos(p_2z_2)  \cos(p_4z_2)\cos(p_4z_3)
\left(2-\log(z_1z_2(\vec k_{12}^2+p_2^2)-\log(z_2z_3(\vec k_{12}^2+p_4^2)\right).
\end{multline}
There is one sub-one-loop divergence after performing
the integration over $p_4$ using~\eqref{e:intcosdelta}
\begin{multline}
 \left(\frac{\lambda}{2}\right)^3\frac{4}{(2\pi)^5 }\frac{1}{\kappa}\int\limits_{-\infty}^{\infty} d p_2 \int\limits_0^{\infty}\frac{d z_1}{z_1^{2}}\frac{dz_2}{z_2^{2}} \prod\limits_{i=1}^2\bar G(k_i,z_1,0)\prod\limits_{i=3}^4\bar G(k_i,z_2,0)\cr
\times\cos(p_2z_1)\cos(p_2z_2)  
(1-\log(z_1z_2(\vec k_{12}^2+p_2^2))
\end{multline}
and similarly a second sub-one-loop divergence after performing the integration over $p_-^{12}$ 
\begin{multline}
\left(\frac{\lambda}{2}\right)^3\frac{4}{(2\pi)^5}\frac{1}{\kappa} \int\limits_{-\infty}^{\infty} d p_4 \int\limits_0^{\infty}\frac{d z_2}{z_2^{2}}\frac{dz_3}{z_3^{2}} \prod\limits_{i=1}^2\bar G(k_i,z_2,0)\prod\limits_{i=3}^4\bar G(k_i,z_3,0)\cr
\times\cos(p_4z_2)\cos(p_4 z_3)  
(1-\log(z_2z_3(\vec k_{12}^2+p_4^2)).
\end{multline}
In our subtraction scheme, defined in the last subsection, the residues of the simple poles are then equal to the finite part of the one-loop bubble 
in~\eqref{eq:aloopkffinite}. Therefore the simple pole contribution reads
\begin{equation}
  I_{\circ\circ}^{(4)}|_{\kappa^{-1}}= \left(\frac{\lambda_R}{2}\right)\frac{8}{(2 \pi)^2}\frac{1}{\kappa}I_\circ|_{\kappa^0}\,.
\end{equation}

Finally, we turn to the fine part which, in our scheme, is then given by 
\begin{multline}
I^{(4)}_{\circ\circ}|_{\kappa^{0}}=  \left(\frac{\lambda}{2}\right)^3\frac{1/2}{(2\pi)^6}\int\limits_{-\infty}^{\infty} dp_2
  \int\limits_{-\infty}^{\infty}dp_4 \int\limits_0^{\infty}{d
    z_1\over z_1^2}
  d z_2 {dz_3\over z_3^2}\prod\limits_{i=1}^2\bar G(k_i,z_1,0)\prod\limits_{i=3}^4\bar G(k_i,z_3,0)\; \cr
 \cos(p_2z_1)\cos(p_2z_2)
 \cos(p_4z_2)\cos(p_4z_3)
 \Big(6+ \left(\log(z_1z_2(\vec
     k_{12}^2+p_2^2))+\log(z_2z_3(\vec
     k_{12}^2+p_4^2))\right)^2\cr-4 \left(\log(z_1z_2(\vec
     k_{12}^2+p_2^2))+\log(z_2z_3(\vec
     k_{12}^2+p_4^2))\right) \Big)\,.
\end{multline}
Using the same techniques as before we write the finite piece as
 $   I_{\circ\circ}|_{\kappa^{0}}=
I^{(4)}_{\circ\circ}|_{\kappa^{0}}^{(a)}+
I^{(4)}_{\circ\circ}|_{\kappa^{0}}^{(b)}$.
With a first piece given by
\begin{multline}
     I^{(4)}_{\circ\circ}|_{\kappa^{0}}^{(a)}=\frac{1}{2k_1 k_2 k_3 k_4}\left(\lambda\over 4\pi\right)^3
{3+8\gamma_E\over 4(k_{12}+k_{34})}\cr
+\frac{1}{2k_1 k_2 k_3 k_4}\left(\lambda\over 4\pi\right)^3 \left({1\over
    k_{12}-k_{34}}\log\left(k_{12}+|\vec k_{12}|\over k_{34}+|\vec k_{12}|\right)-
  {1\over k_{12}+k_{34}}\log\left((k_{12}+|\vec k_{12}|)( k_{34}+|\vec k_{12}|)\over
  (k_{12}+k_{34})^2\right)\right).
\end{multline}
This expression the sum of finite piece of the cross and one-loop
contribution. This piece satisfies the conformal ward identity.
The new contribution arising at two-loop 
\begin{multline}\label{eq:2-loop_kappa_2}
 I_{\circ\circ}|_{\kappa^{0}}^{(b)}= \left(\lambda\over4\pi\right)^3
  \int\limits_{-\infty}^\infty {dp_-  \over2\pi } \left\{-\left(\frac{\log(\vec{k}_{12}^2+p_-^2)}{k_{12}-ip_-}+\frac{\log(\vec{k}_{12}^2+p_-^2)}{k_{12}+ip_-}\right)\left(\frac{\log(\vec{k}_{12}^2+p_-^2)}{k_{34}-ip_-}+\frac{\log(\vec{k}_{12}^2+p_-^2)}{k_{34}+ip_-}\right)\right.\cr
     \left. +2\left(\frac{\log(\vec{k}_{12}^2+p_-^2)}{k_{12}-ip_-}+\frac{\log(\vec{k}_{12}^2+p_-^2)}{k_{12}+ip_-}\right)\left(\frac{\log(k_{34}-ip_-)}{k_{34}-ip_-}+\frac{\log(k_{34}+ip_-)}{k_{34}+ip_-}\right)\right.\cr
  \left.+2\left(\frac{\log(k_{12}-ip_-)}{k_{12}-ip_-}+\frac{\log(k_{12}+ip_-)}{k_{12}+ip_-}\right)\left(\frac{\log(\vec{k}_{12}^2+p_-^2)}{k_{34}-ip_-}+\frac{\log(\vec{k}_{12}^2+p_-^2)}{k_{34}+ip_-}\right)\right.\cr
  \left.-3\left(\frac{\log(k_{12}-ip_-)}{k_{12}-ip_-}+\frac{\log(k_{12}+ip_-)}{k_{12}+ip_-}\right)\left(\frac{\log(k_{34}-ip_-)}{k_{34}-ip_-}+\frac{\log(k_{34}+ip_-)}{k_{34}+ip_-}\right)\right.\cr
  \left. -\frac{1}{2}\left(\frac{\log^2(ik_{12}+p_-)+ \log^2(-ik_{12}-p_-)}{k_{12}-ip_-}+\frac{\log^2(ik_{12}-p_-)+ \log^2(-ik_{12}+p_-)}{k_{12}+ip_-}\right)\right.\cr
  \left.\times\left(\frac{1}{k_{34}-ip_-}+\frac{1}{k_{34}+ip_-}\right)
   \right\} \frac{1}{k_1 k_2 k_3 k_4}.
\end{multline}
This expression is easily integrated  using Panzer's
\texttt{HyperInt}~\cite{Panzer:2014caa}  and is given by a combination of weight 2 polylogarithms (see~\eqref{e:Li11def}
and~\eqref{e:Li2def} for a definition of the weight 2 polylogarithms
entering this expression)
\begin{multline}\label{e:necklace-analytic-finite1}
 I^{(4)}_{\circ\circ}|_{\kappa^{0}}^{(b)}=\left(\lambda\over 4\pi\right)^3
 {1\over 4|\vec k_{12}|}{uv\over u^2-v^2}\Big[-(u-v)   \left(\Mpl([1, 1],[-u/v , v])+\Mpl([1, 1],[-v/u   , u])+4   \Mpl([1, 1],[1, -v/u  
])\right)\cr-(u+v)   \left(\Mpl([1, 1],[u/v   , -v])-\Mpl([1, 1],[v/u  , -u])-2   \Mpl([2],[v/u  ])\right)\cr+
2   v   \left(\Mpl([2],[-v])+\Mpl([1, 1],[-1, v])-\Mpl([2],[v])\right)-2   u   \left(\Mpl([2],[-u])+\Mpl([1,
  1],[-1, u])-\Mpl([2],[u])\right)\cr
+3 u \ln \! \left(u \right)^{2}-u \ln \! \left(v \right)^{2}
+\Big(-2 u \ln \! \left(v \right)+\left(u +v \right) \ln \! \left(1-u
  \right)-2\left(u +v \right) \ln \! \left(u -v \right)\cr+\left(-u
    +v \right) \ln \! \left(1+u \right)-2\left( u -v \right) \ln \!
  \left(u +v \right)+2 \ln \! \left(1+v \right) v \Big) \ln \!
\left(u \right)\cr
+\Big(2\left( u - v \right) \ln \! \left(u +v \right)-\left(u+v
  \right) \ln \! \left(1-v \right)+2\left( u + v \right) \ln \!
  \left(u -v \right)+\left(-u +v \right) \ln \! \left(1+v \right)\cr
  -2
  \ln \! \left(1+u \right) u \Big) \ln \! \left(v \right)
+\Big(\left(u -v \right) \ln \! \left(1-u \right)+\left(u -v \right)
  \ln \! \left(1-v \right)\cr+\left(\ln \! \left(1+v \right)+\ln \!
    \left(1+u \right)\right) \left(u -v \right)\Big) \ln \! \left(u
  +v \right)
+2 \ln \! \left(1-u \right) u \ln \! \left(2\right)-2 \ln \! \left(2\right) \ln \! \left(1-v \right) v \cr-2 \ln \! \left(1+u \right) u \ln \! \left(2\right)+2 \ln \! \left(1+v \right) \ln \! \left(2\right) v -\frac{11 \pi^{2} \left(\frac{5 v}{11}+u \right)}{24}
\Big] \frac{1}{k_1 k_2 k_3 k_4} 
\end{multline}
where we have made use of the conformal cross ratios $u=|\vec
k_{12}|/k_{12}$ and $v=|\vec k_{12}|/k_{34}$.
One can check that~\eqref{e:necklace-analytic-finite1} satisfies the
conformal ward identities.
   
Note that at two-loops it is not enough to replace the renormalization
scale in the hard cut-off result~\ref{app:necklace-hard} by
$\bm{\delta} (k_{12} + k_{34})$ to recover the de Sitter-invariant result~\eqref{e:necklace-analytic-finite1}.

 \subsubsection{Ice-cream}
 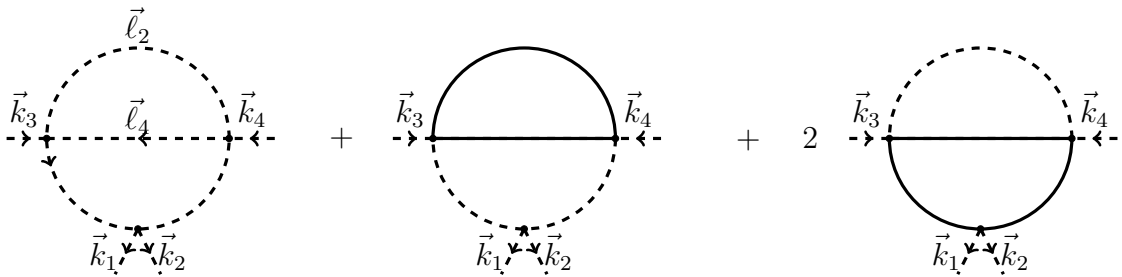
\begin{figure}[H]\centering
\begin{tikzpicture}[scale=0.6]
\node at (0, 2.5) {$\vec \ell_2$};
\node at (0, 0.5) {$\vec \ell_4$};
\filldraw [fermion, dashed, color = black, fill=none, very thick] (0,0) circle (2cm);
\draw [fermionbar, black,very thick, dashed] (-2,0) to (2,0);
\filldraw [black] (2,0) circle (2pt);
\filldraw [black] (0,-2) circle (2pt);
\filldraw [black] (-2,0) circle (2pt);
\draw [fermionbar, dashed, black,very thick] (-2,0) to (-3,0);
\draw [fermion, dashed, black,very thick]  (3,0) to (2,0);
\draw [fermion, dashed, black,very thick] (0,-2) to (0.5,-3);
\draw [fermion, dashed, black,very thick] (0,-2) to (-0.5,-3);
\node at (-2.5, 0.65) {$\vec k_3$};
\node at (2.5, 0.65) {$\vec k_4$};
\node at (-0.75, -2.5) {$\vec k_1$};
\node at (0.75, -2.5) {$\vec k_2$};
\end{tikzpicture}
\quad
\begin{tikzpicture}[scale=0.6]
\draw[color = black, fill=none, very thick]  (-2,0)--(2,0) arc(0:180:2);
\draw[color = black, fill=none, very thick,dashed] (2,0) arc(0:-180:2) --cycle;
\filldraw [black] (2,0) circle (2pt);
\filldraw [black] (0,-2) circle (2pt);
\filldraw [black] (-2,0) circle (2pt);
\draw [fermionbar, dashed, black,very thick] (-2,0) to (-3,0);
\draw [fermion, dashed, black,very thick] (3,0) to (2,0);
\draw [fermion, dashed, black,very thick] (0,-2) to (0.5,-3);
\draw [fermion, dashed, black,very thick] (0,-2) to (-0.5,-3);
\node at (-2.5, 0.65) {$\vec k_3$};
\node at (2.5, 0.65) {$\vec k_4$};
\node at (-0.75, -2.5) {$\vec k_1$};
\node at (0.75, -2.5) {$\vec k_2$};

  \node[align=center] at (-4, 0) {$+$};
\end{tikzpicture}\qquad
\begin{tikzpicture}[scale=0.6]
\draw[color = black, fill=none, very thick,dashed] (2,0) arc(0:180:2) --cycle;
\draw[color = black, fill=none, very thick] (-2,0) -- (2,0);
\draw[color = black, fill=none, very thick] (2,0) arc(0:-180:2);
\filldraw [black] (2,0) circle (2pt);
\filldraw [black] (0,-2) circle (2pt);
\filldraw [black] (-2,0) circle (2pt);
\draw [fermionbar, dashed, black,very thick] (-2,0) to (-3,0);
\draw [fermion, dashed, black,very thick] (3,0) to (2,0);
\draw [fermion, dashed, black,very thick] (0,-2) to (0.5,-3);
\draw [fermion, dashed, black,very thick] (0,-2) to (-0.5,-3);
\node at (-2.5, 0.65) {$\vec k_3$};
\node at (2.5, 0.65) {$\vec k_4$};
\node at (-0.75, -2.5) {$\vec k_1$};
\node at (0.75, -2.5) {$\vec k_2$};

  \node[align=center] at (-4, 0) {$+$ \quad $2$ \quad};
\end{tikzpicture}
\caption{Ice-cream diagrams: The configuration of the loop momenta for the second and third diagram are similar to the first. }\label{fig:icecream2}
\end{figure}
 We start from Figure \ref{fig:icecream2} which translates into~\eqref{eq:ice_in_in} with the $\kappa$ regulator added. After some trigonometric manipulations this becomes
 \begin{align}\label{eq:ice_ana}
I^{in-in}_{\widehat{\vee}}=&
\frac{16(\frac{\lambda}{2})^3}{(2\pi)^{10}}\int \frac{dz_1}{z_1^{2+2\kappa}}\frac{dz_2}{z_2^{1+3\kappa}}\frac{dz_3}{z_3^{1+3\kappa}}d^{3}\vec\ell_{1}d^{3}\vec\ell_{2}d p_{1}dp_{2}d p_{3}d p_{4}\\
&\frac{\bar G(k_1,z_1;\kappa)\bar G(k_2,z_1;\kappa)\bar G(k_3,z_2;\kappa)\bar G(k_4,z_3;\kappa)\cos(p_{234}^+z_2)\cos(p_{124}^+z_3)\cos(p_{13}^-z_1)}{(p_2^2+\vec{l}_2^2)^{1+\kappa}(p_4^2+\vec{l}_4^2)^{1+\kappa}(p_3^2+(\vec{k}_3+\vec{
\ell}_2+\vec{\ell}_4)^2)^{1+\kappa}(p_1^2+(-\vec{k}_4+\vec{
\ell}_2+\vec{\ell}_4)^2)^{1+\kappa} } \nonumber\\
=&
\frac{16(\frac{\lambda}{2})^3}{(2\pi)^{10}}\int \frac{dz_1}{z_1^{2+2\kappa}}\frac{dz_2}{z_2^{1+3\kappa}}\frac{dz_3}{z_3^{1+3\kappa}}d^{4}L_{4}d^{4}L_{2}d p_{234}^+dp_{124}^+\nonumber\\
&\frac{\bar G(k_1,z_1;\kappa)\bar G(k_2,z_1;\kappa)\bar G(k_3,z_2;\kappa)\bar G(k_4,z_3;\kappa)\cos(p_{234}^+z_2)\cos(p_{124}^+z_3)\cos(p_{13}^-z_1)}{(L_2^2)^{1+\kappa}(L_4^2)^{1+\kappa}((L_4+L_2+Q)^2)^{1+\kappa}((L_4+L_2+\tilde Q)^2))^{1+\kappa} } \nonumber
\end{align}
where $L_i=(\vec{\ell}_i,p_i)$, $i=2,4$, $p^+_{ijk}=p_i+p_j+p_k$, $Q=(\vec{k}_3,-p_{243}^+)$, $\tilde Q=(-\vec{k}_4,-p_{124}^+)$ and $p_{13}^-=p_1-p_3=p_{124}^+- p_{234}^+$. We also suppressed a prefactor as in footnote~\ref{fot:ck}. 
Performing a loop-by-loop integration we find 
\begin{align}\label{eq:flat_ice}
   &\int \frac{d^{4}L_{4}d^{4}L_{2}}{(L_2^2)^{1+\kappa}(L_4^2)^{1+\kappa}((L_4+L_2+Q)^2)^{1+\kappa}((L_4+L_2+\tilde Q)^2))^{1+\kappa} }   \\
&= \frac{i\pi^2\Gamma(1-\kappa)^2\Gamma(2\kappa)}{\Gamma(1+\kappa)^2\Gamma(2-2\kappa)}\int \frac{d^{4}L_{2}}{(L_2^2)^{2\kappa}((L_2+Q)^2)^{1+\kappa}((L_2+\tilde Q)^2))^{1+\kappa} }\nonumber \\
& =\frac{i\pi^4\Gamma(1-\kappa)^2\Gamma(2\kappa)}{2\Gamma(1+\kappa)^2\Gamma(2-2\kappa)}\left(\frac{1}{2 \kappa}   + 2- \log((Q - \tilde Q)^2(Q+\tilde Q)^2 )   + O(\kappa)\right)\nonumber\\
& =\frac{i\pi^4\Gamma(1-\kappa)^2\Gamma(2\kappa)}{2\Gamma(1+\kappa)^2\Gamma(2-2\kappa)}\left(\frac{1}{2 \kappa}   + 2- \log(\vec{k}_{34}^2+p_{13}^{-2})   -\log\big((\vec{k}_{3}-\vec{k}_{4})^2+(p_{234}^++p_{124}^+)^2\big) \right)\nonumber
\end{align}
up to $O(\kappa^0)$ terms. The first two terms, multiplying a double pole and a simple pole respectively, reduce to the cross diagram after integration over $p_{234}$ and $p_{124}$. This term is then cancelled in the usual way by the cross counter-term. Concerning the next term we change then integration variables to $p_{13}^-$ and $p_{124}^++ p_{234}^+$ and then integrate over $p_{124}^++ p_{234}^+$ using~\eqref{e:intcosdelta} which results in $\pi\delta(z_2-z_3)$, representing the collapsed loop. Together with the $\kappa$-expansion of the measure (with $4\kappa$ in the exponential absorbed in the collapsing loop), this combines to give the same logarithmic dependence as in~\eqref{eq:aloopkfexp}, as required by renormalisability at two-loops.  This leaves us with the last term which looks problematic at first sight since it differs from the one-loop bubble sub-diagram. To see that this term does not contribute we proceed as follows. Since we are focusing on the simple pole of~\eqref{eq:flat_ice} we may set $\kappa=0$ everywhere else. Then the integral over $z_i$ implement a  conservation of energy at each vertex. We can then process as in flat space (see e.g. Section~8.5 of~\cite{web:Schrans}) to show that the singular part of~\eqref{eq:flat_ice} depends only on $\vec{k}_{34}$. Thus the simple pole of~\eqref{eq:ice_ana} (together with the diagram with $\vec{k}_3,\vec{k}_4\leftrightarrow  \vec{k}_{12}$) combined with the divergent sub-diagram in the necklace~\eqref{eq:neckk-1} is cancelled by the one-loop bubble combined with the cross counter term.  This then proves renormalisability of the effective action~\eqref{eq:EAdS_action_phi4_k2} up to two-loops. The finite contribution can be computed as well. However, we will not display the lengthy result since we have no use for it here.

%-----------------------------------------------
\subsection{Dimensional regularization}

To summarize, in this section we have so far proposed a variant of analytic regularization that preserves (A)dS-invariance and demonstrated consistency and calculability of this scheme by computing various correlators up to two-loops. In doing so we also established the renormalisability of the effective action~\cite{DiPietro:2021sjt} given in eq.~\eqref{eq:EAdS_action_phi4_k2} up to this order. A natural extension for a higher number of loops and legs is to resort to the recursion relations formulated in Section~\ref{sec:recursion}. However, it turns out that the derivation of these recursions no longer applies with analytic regularizations. The reason for this is that the Cauchy residue-formula is no-longer applicable for the energy integrals for a non-integer power of the propagators which is a defining feature of analytic regularization.   In this subsection we will the briefly describe another dS-invariant regulator that preserves the nature of the propagator. However, we will see that explicit calculations in this scheme quickly become very complicated.  

We start with the following action (which is obtained from eq.~\eqref{eq:EAdS_action_gen_pot} for a general dimension):
\begin{equation}
    S_{D}[\phi_{+},\phi_{-}]=-\frac{1}{2}\int\frac{dzd^{D}x}{z^{D+1}}\left(\frac{(\p\phi_{+})^{2}}{z^{2}}-m^{2}\phi_{+}^{2}-\frac{(\p\phi_{-})^{2}}{z^{2}}+m^{2}\phi_{+}^{2}+\frac{\lambda}{4!}V(\phi_{+},\phi_{-},D)\right)
\end{equation}
where $D=3-\epsilon$, $m$ is the conformally coupled mass corresponding to $\Delta_+=2-\epsilon/2$ and $\Delta_-=1-\epsilon/2$, 
 and
\begin{equation}
V(\phi_+,\phi_-,D)=\cos\left(\frac{\pi \epsilon}{2}\right)\left(  \phi_+^4 + \phi_-^4 - 6 \phi_+^2 \phi_-^2 \right)+\sin\left(\frac{\pi \epsilon}{2}\right)\left(\phi_+\ph_-^3-\phi_-\ph_+^3\right)
\end{equation}
We then make the conformal mapping to half of $\mathbb{R}^{4}$ with a boundary at $z=0$ through 
\begin{equation}
    g_{\mu\nu}\to \frac{1}{z^2}g_{\mu\nu}\,,\quad\phi_\pm\to z^{D-1}\phi_\pm\,,\quad \lambda\to z^{-\epsilon}\lambda,
\end{equation}
and get 
\begin{equation}
     S_{D = 3 - \e}[\phi_+,\phi_-]=-\frac{1}{2}\int dz d^{D}x\left( (\p \phi_+)^2  -(\p \phi_-)^2 
    + \frac{z^{-\epsilon}\lambda}{4!}V(\phi_+,\phi_-,D)\right).
\label{dimregaction}
\end{equation}
When computing scattering amplitudes using dimensional regularisation, we would normally include a renormalisation scale by inserting $\mu^\epsilon$ in front of the interaction terms, which is required by dimensional analysis. Comparing to~\eqref{dimregaction}, we identify $\mu=z^{-1}$.

The \textit{in-in} correlator corresponding to the bubble diagram in this scheme is then given by 
\begin{align}\label{eq:loopmh_kappay}
 \frac{\lambda^2}{(4\pi)^{5-\e}}\int\limits_{0}^\infty dp& dp' \int d^D \ell\int\limits_0^{\infty}\frac{d z_1}{z_1^{\epsilon}} \frac{dz_2}{z_2^{\epsilon}} e^{-k_{12}z_1-k_{34}z_2}
 \frac{\left(\cos(p_+ z_1)  \cos(p_+z_2) +  \cos(p_-z_1)\cos(p_-z_2)\right)}{(p^2+\vec{\ell}^2)(p'^2+(\vec{\ell}+\vec{k})^2)}
\end{align}
By evaluating the $p, p'$ integrals we get
\begin{multline}
\int {d^D\vec\ell\over (p^2+\vec\ell^2)
({p'}^2+(\vec\ell+\vec k)^2)}=  \pi ^{\frac{3}{2}-\frac{\epsilon}{2} } \Gamma \left(\frac{\epsilon}{2} +\frac{1}{2}\right) 
   \left(-\frac{\vec k^2}{(z-1) (\bar z-1)}\right)^{-\frac{\epsilon}{2} -\frac{1}{2}} \cr\times\left(\frac{\, _2F_1\left(1,\frac{\epsilon}{2} +\frac{1}{2};
   \epsilon +1;\frac{z-\bar z}{z-1}\right)}{ (z-1) \epsilon}-\frac{z (z \bar z)^{-\frac{\epsilon}{2} -\frac{1}{2}} \,
   _2F_1\left(1,\frac{\epsilon}{2}+\frac{1}{2};\epsilon +1;\frac{z-\bar z}{(z-1) \bar z}\right)}{ (z-1) \epsilon}\right), 
\end{multline}
with 
\begin{equation}
   (1-z)(1-\bar z) =-{\vec k_{12}^2\over p^2}; \qquad z \bar  z={p'^2\over p^2} .
\end{equation}
The integral above is highly non-trivial to manipulate analytically. This illustrates the difficulty in working with this regularization scheme even at 1-loop.

\section{Conclusion}\label{sec:conclusion}
In this paper we have found evidence that \textit{in-in} correlators appear to be much simpler than wavefunction coefficients. In particular their loop integrands can be recast in terms of four-dimensional flat space Feynman integrals and after loop integration their analytic structure appears to be very closely related to that of scattering amplitudes. This fact is obscured by the standard definition of \textit{in-in} correlators in terms of the square the wavefunction, but becomes more manifest after mapping the calculation to Witten diagrams derived from an effective action in EAdS. From that point of view, the simplicity arises from a subtle interplay of boundary conditions, notably the Neumann boundary conditions of the shadow fields and the Dirichlet boundary conditions of the original fields in de Sitter space. 

We have demonstrated this explicitly in a number of examples up to two loops for the conformally coupled $\phi^4$ theory, but we believe that similar simplicity of \textit{in-in} correlators will exist in more general theories. As a simple example, let us consider the 1-loop tadpole contribution to the \textit{in-in} correlator in the case of a massless scalar. The effective potential in EAdS is the same as the conformally coupled case\footnote{This is more generally true for all integer $\Delta$.} so we find that two diagrams will contribute analogous to the ones in~\eqref{eq:tad_cut}\footnote{Since this is just for illustration, we consider the case with external $\phi_+$.}, and their sum is given by  $\mathcal{M}_{2}^{(1)}=A_{3/2}+A_{-3/2}$ where ($\nu = 3/2$) 
\begin{equation}
A_{\nu}=\int\frac{d^{3}ldp p}{p^{2}+l^{2}}\int_{0}^{\infty}\frac{dz}{z^{4}}\left(z^{3/2}J_{\nu}(p z)\right)^{2}\left(\left(kz\right)^{3/2}K_{\nu}(kz)\right)^{2}.
\end{equation}
While the integrand for each diagram is a rather complicated object
consisting of trigonometric functions, the integrand of the sum is
remarkably simple:
\begin{equation}
\mathcal{M}_{2}^{(1)}=\int\frac{d^{3}ldp}{p^{2}\left(p^{2}+l^{2}\right)}\int_{0}^{\infty}\frac{dz}{z^{4}}e^{-2kz}\left(1+kz\right)^{2}\left(1+p^{2}z^{2}\right),
\end{equation}
so once again we find a dramatic simplification of the \textit{in-in} correlator
compared to the wavefunction already at the integrand level. One important
difference compared to the conformally coupled case is that now there are infrared divergences when $z=0$, but this can be easily regulated by
taking lower limit of the $z$ integral to be $\epsilon\ll1$ \footnote{For recent work on infrared divergences of cosmological correlators, see \cite{Gorbenko:2019rza,Cespedes:2020xqq,Cespedes:2023aal}.}. 
After performing the $z$ integral, we are left with
\begin{equation}
\mathcal{M}_{2}^{(1)}=C_{1}\int\frac{d^{3}ldp}{\left(p^{2}+l^{2}\right)}+C_{2}\int\frac{d^{3}ldp}{p^{2}\left(p^{2}+l^{2}\right)},
\label{massles1looptadpole}
\end{equation}
where $C_{i}$ are prefactors that diverge as $\epsilon \rightarrow 0$.
Hence, we are left with a linear combination of two simple loop integrals. Recall that the integral over $\omega$ can be carried out by summing over residues in the upper half-plane. After doing so the second term in~\eqref{massles1looptadpole} vanishes, and we are just left with the first term, whose integrand has four-dimensional Lorentz invariance. Hence, we are once again left with a four-dimensional  flat space integral. This integral is identical to the conformally coupled case and can be set to zero after renormalisation, as expected from the flat space limit.

Another key result of this paper is the construction of a manifestly de Sitter-invariant regularization scheme that preserves much of the flat-space structure of standard Feynman integrals. In this way the regulated loop integrals become almost as simple as those of scattering amplitudes, allowing us in particular to identify the recursive renomalizeability of the Euclidean effective action and to systematically derive de Sitter-invariant expressions for the finite parts of loop-level correlators. On the other hand the loop integrands in this regularisation scheme  are not amenable to the recursion relations derived in Section~\ref{sec:recursion}. Alternative de Sitter-invariant regulators which are compatible with recursion exists but do not lead to calculable integrals even at one-loop. This leads to the question of whether a calculable, de Sitter-preserving regularization scheme exists that is compatible with recursion. 

Another subtlety of demanding conformal invariance at loop-level is the flat space limit. In particular, when taking the flat space limit we must break conformal symmetry to Poincar\'e symmetry since the isometry group of the background gets broken. In order to do so, we must introduce a dimensionful renormalisation scale. In the simple examples that we considered, this can be accomplished by setting $\bm\delta = \mu / E$ (where $\bm\delta$ is the dimensionless renormalisation scale in the de Sitter-invariant renormalisation scheme, $\mu$ is a dimensionful renormalisation scale that arises in the flat space limit, and $E$ is the energy),  and taking $E\rightarrow 0$ holding $\mu= \bm \delta E$ fixed. This subtlety when taking the flat space limit does not occur at tree-level and to our knowledge has not been previously pointed out at loop-level. It would be interesting to see if this simple prescription for breaking conformal symmetry in order to recover the flat space limit is valid more generally.

There are a number of future directions to consider. As mentioned
above, it would be important to understand how the simplicity of
\textit{in-in} correlators extends to more general masses and
interactions, in particular the case of massless scalar fields and
$\phi^3$ interactions. It would also be of interest to generalise the
Euclidean effective action to other conformally flat FLRW
backgrounds. For conformally coupled $\phi^4$ theory, the
generalisation is trivial since it can always be mapped to a massless
scalar theory in half of flat space via a conformal
transformation. More generally, mapping a scalar theory with a general
mass and polynomial interactions in a general FLRW background to half
of flat space will introduce time dependence in the masses and
interactions~\cite{Benincasa:2019vqr}. Nevertheless, we may still perform a Wick rotation to half of Euclidean space and introduce ghost fields to obtain a Euclidean effective action analogous to the one in~\eqref{eq:action-conf1} whose Feynman rules can then be used to compute \textit{in-in} correlators in the original FLRW background.

It would also be of interest to have a more systematic understanding of the singularity structure of \textit{in-in} correlators analogous to that of wavefunction coefficients. For wavefunction coefficients, this is partially encoded by the cosmological optical theorem (COT)~\cite{Goodhew:2020hob}, which relates unitarity cuts of individual Feynman diagrams in de Sitter background to products of shifted lower-point Feynman diagrams. While it is unclear how to extend this to \textit{in-in} correlators starting from their standard definition in terms the square of the wavefunction, our results suggest that the COT can be straightforwardly extended to \textit{in-in} correlators by taking into account the contributions from the ghost fields appearing in the Euclidean effective action. We leave a detailed investigation of these important questions for future work.

\begin{center}
\textbf{Acknowledgements}
\end{center}
We thank Paolo Benincasa, Martin Beneke, Gordon Lee, Ioannis Matthaiakakis,  Paul McFadden,  Shota Komatsu, Kajal Singh, and Kostas Skenderis for helpful discussions. AL is supported an STFC Consolidated Grant ST/T000708/1. JM is supported by a Durham-CSC Scholarship. CC is supported by the STFC consolidated grant (ST/X000583/1) “New Frontiers in Particle Physics, Cosmology and Gravity”. I.S. is supported by the Excellence Cluster Origins of the DFG under Germany’s Excellence Strategy EXC-2094 390783311. 
CC would like to thank ICTP-Trieste (and the organizers of Workshop on Scattering Amplitudes and Cosmology), CMI-Chennai (and Alok Laddha), IIT Kharagpur (and Jyotirmoy Bhattacharya) \& IIT Mandi (and the organizers of the ST4 conference) for hospitality where a part of the work was completed. I.S. would like to thank the CERN Theory group for hospitality while this work was completed.  The work of P.V. has received funding from the ANR grant ``SMAGP''
ANR-20-CE40-0026-01. P.V. acknowledges support of the Institut Henri Poincar\'e (UAR 839 CNRS-Sorbonne Universit\'e), and LabEx CARMIN (ANR-10-LABX-59-01).

\appendix

\section{\textit{in-in} correlators from wavefunctions}\label{sec:wave}

We start with the definition of cosmological correlator in terms of wavefunction
\begin{equation}
\left\langle \sigma_{1}\sigma_{2}...\sigma_{n}\right\rangle =\frac{\int\mathcal{D}\sigma\left|\psi(\sigma)\right|^{2}\prod_{i}\sigma_{i}}{\int\mathcal{D}\sigma\left|\psi(\sigma)\right|^{2}},
\label{psi2}
\end{equation}
where the wavefunction takes the form
\begin{equation}
\psi(\sigma)\propto\exp\left[-\frac{1}{2}\int\sigma_{1}\sigma_{2}\psi_{2}-\frac{1}{4!}\int\sigma_{1}\cdots\sigma_{4}\psi_{4}-\frac{1}{6!}\int\sigma_{1}\cdots\sigma_{6}\psi_{6}+\cdots\right],
\end{equation}
and we use the shorthand notation
\begin{equation}
\int\sigma_{1}\cdots\sigma_{n}=\int\prod_{i=1}^{n}\frac{d^{3}p_{i}}{(2\pi)^{3}}\sigma\left(\vec{p}_{i}\right)\delta^{3}\left(\sum_{i=1}^{n}\vec{p}_{i}\right),
\end{equation}
and 
\begin{equation}
\psi_{n}=\psi_{n}\left(\vec{p}_{1},\dots,\vec{p}_{n}\right).
\end{equation}
The wavefunction coefficients $\psi_n$ are computed from Witten diagrams and have a
perturbative expansion in the coupling of the bulk $\lambda\phi^{4}$
theory. At lowest order, $\psi_{4}$ comes from a tree-level contact
diagram and is $\mathcal{O}\left(\lambda\right)$, while $\psi_{6}$
comes from exchange diagrams and is $\mathcal{O}\left(\lambda^{2}\right)$.

We can rephrase the path integral in~\eqref{psi2} in terms of an effective action:
\begin{equation}
\left\langle \sigma_{1}\sigma_{2}\cdots\sigma_{n}\right\rangle =\frac{\int\mathcal{D}\sigma e^{-S(\sigma)}\prod_{i}\sigma_{i}}{\int\mathcal{D}\sigma e^{-S(\sigma)}},
\end{equation}
where
\begin{equation}
S(\sigma)=\frac{1}{2}\int\sigma_{1}\sigma_{2}\mathrm{Re}\psi_{2}+\frac{1}{4!}\int\sigma_{1}\cdots\sigma_{4}\mathrm{Re}\psi_{4}+\frac{1}{6!}\int\sigma_{1}\cdots\sigma_{6}\mathrm{Re}\psi_{6}+\cdots
\end{equation}

The Feynman rules of this action are easy to read off:
\begin{equation}
\begin{gathered}
\begin{tikzpicture}[baseline]

\draw (0, 0) -- (2, 0);
\node at (1, -0.2) {$\vec{k}$};

\end{tikzpicture}\end{gathered}
=\frac{1}{\mathrm{Re}\psi_2(\vec{k})}~, 
\quad 
\begin{gathered}
\begin{tikzpicture}[baseline]

\draw (-0.7, -0.7) -- (0.7, 0.7);
\draw (-0.7, 0.7) -- (0.7, -0.7);

\end{tikzpicture}\end{gathered}
=\mathrm{Re}\psi_4, \quad
\begin{gathered}
\begin{tikzpicture}[baseline]

\draw (-0.7, -0.7) -- (0.7, 0.7);
\draw (-0.7, 0.7) -- (0.7, -0.7);
\draw (0, 0.8) -- (0, -0.8);

\end{tikzpicture}\end{gathered}
=\mathrm{Re}\psi_6, \quad \dots
\end{equation}

While the underlying theory only has a $\phi^{4}$ interaction vertex, the effective action used to compute the \textit{in-in} correlators has an infinite number of vertices,
each of which has a pertrubative expansion in the bulk coupling. We
can then compute loop diagrams using the effective action, whose vertices
will contain different powers of the coupling and then keep all contributions
that contain a given order in the coupling. For example, at leading
order in the coupling, the four-point \textit{in-in} correlator comes from a four-point
contact diagram where we only keep the tree-level contribution to
the coefficient $\mathrm{Re}\psi_{4}^{(0)}$. We also dress the vertex with
external propagators, keeping only the tree-level contributions $\mathrm{Re}\psi_{2}^{(0)}$
\begin{equation}
\la \phi(\vec{k}_1) \phi(\vec{k}_2)\phi(\vec{k}_3)\phi(\vec{k}_4) \ra=
    \begin{tikzpicture}[baseline]

\draw (-0.7, -0.7) -- (0.7, 0.7);
\draw (-0.7, 0.7) -- (0.7, -0.7);

\end{tikzpicture}
=\frac{\mathrm{Re}\psi_4}{\prod\limits_{i=1}\limits^4 \mathrm{Re}\psi_2(\vec{k}_i)}
\end{equation}

At one-loop, the \textit{in-in} correlator receives contributions from three
types of diagrams: a tree-level diagram whose vertex is 1-loop coefficient
$\mathrm{Re}\psi_{4}^{(1)}$, a 1-loop bubble diagram whose vertices are the
tree-level coefficients $\mathrm{Re}\psi_{4}^{(0)}$ and a 1-loop diagram
whose 6-point vertex is the tree-level coefficient $\mathrm{Re}\psi_{6}^{(0)}$:
\begin{equation}
\la \phi(\vec{k}_1) \phi(\vec{k}_2)\phi(\vec{k}_3)\phi(\vec{k}_4)\ra^{(2)}=
    \begin{tikzpicture}[baseline]

\draw (-0.7, -0.7) -- (0.7, 0.7);
\draw (-0.7, 0.7) -- (0.7, -0.7);

\end{tikzpicture}
\quad 
+
\begin{tikzpicture}[baseline]

\draw (0, 0) -- (-0.7, -0.7);
\draw (0, 0) -- (-0.7, 0.7);
\draw  (0.7,0) circle (0.7cm);
\draw (2.1, -0.7) -- (1.4, 0);
\draw (2.1, 0.7) -- (1.4, 0);

\end{tikzpicture}
\quad
+ 
\begin{tikzpicture}[baseline]

\draw (-0.7, -0.7) -- (0, 0);
\draw (0.7, -0.7) -- (0, 0);
\draw (1, 0) -- (0, 0);
\draw (-1, 0) -- (0, 0);
\draw  (0,0.5) circle (0.5cm);

\end{tikzpicture}
+ \mathrm{perms} 
\quad
\end{equation}
Note
that all of these contributions are $\mathcal{O}\left(\lambda^{2}\right)$. 
This gives
\begin{multline}
    \langle \phi(\vec{k}_1) \phi(\vec{k}_2)\phi(\vec{k}_3)\phi(\vec{k}_4) \rangle^{(2)} = \\
    \frac{1}{\prod\limits_{i=1}^4 \mathrm{Re}\psi_2(\vec{k}_i)} \left( \mathrm{Re}\psi_4^{(2)} + \sum\limits_{\text{perms}} \frac{\mathrm{Re}\psi_4(\vec{k}_1,\vec{k}_2,\vec{\ell},\vec{\ell}_2)\mathrm{Re}\psi_4(\vec{k}_3,\vec{k}_4,\vec{\ell},\vec{\ell}_2)}{2\mathrm{Re}\psi_2(\vec{\ell})\mathrm{Re}\psi_2(\vec{\ell}_2)} \right. \\
    \left. + \frac{\mathrm{Re}\psi_6(\vec{k}_1,\vec{k}_2,\vec{\ell},\vec{k}_3,\vec{k}_4,\vec{\ell})}{2\mathrm{Re}\psi_2(\vec{\ell})}+\frac{\mathrm{Re}\psi_6(\vec{k}_1,\vec{k}_2,\vec{\ell}_2,\vec{k}_3,\vec{k}_4,\vec{\ell}_2)}{2\mathrm{Re}\psi_2(\vec{\ell}_2)} \right)
\end{multline}
where $\psi_4^{(2)}$ is the one-loop wavefunction coefficient
\eqs{
 \psi_4^{(2)}&=\lambda^2 \int d\eta_1 d\eta_2 \phi(\eta_1,k_1) \phi(\eta_1,k_2) G(\eta_1,\eta_2,\ell)G(\eta_1,\eta_2,\ell_2) \phi(\eta_2,k_3)  \phi(\eta_2,k_4)\\
&=2\frac{E+\ell+\ell_2}{E(E+\ell) (E+\ell_2 )(k_1+k_2+\ell+\ell_2)(k_3+k_4+\ell+\ell_2)}
}
then the tree level wavefunction coefficients are
\eqs{
\psi_2(\vec{k})&=k \\
\psi_4(\vec{k}_1,\vec{k}_2,\vec{k}_3,\vec{k}_4)&=\frac{\lambda}{k_1+k_2+k_3+k_4} \\
\psi_6(\vec{k}_1,\vec{k}_2,\vec{\ell},\vec{k}_3,\vec{k}_4,\vec{\ell})&=\lambda^2 \int d\eta_1 d\eta_2 \phi(\eta_1,k_1) \phi(\eta_1,k_2) \phi(\eta_1,\ell) G(\eta_1,\eta_2,\ell) \phi(\eta_2,k_3)  \phi(\eta_2,k_4) \phi(\eta_2,\ell) \\
&=\frac{\lambda^2}{(k_1+k_2+k_3+k_4+2\ell) (k_1+k_2+\ell+\ell_2)(k_3+k_4+\ell+\ell_2)}
}

After adding up all the diagrams and integrating out the energies
we land on,
\begin{equation}
    \langle \phi(\vec{k}_1) \phi(\vec{k}_2)\phi(\vec{k}_3)\phi(\vec{k}_4) \rangle^{(2)} =\frac{\lambda^2(E+\ell+\ell_2)}{E\ell \ell_2(k_1+k_2+\ell+\ell_2)(k_3+k_4+\ell+\ell_2)}
\end{equation}
which reproduces the result we obtained using shadow
fields in~\eqref{bubble-rec} after integrating out $\omega$.

\section{Integrand of triangle diagram using recursion}\label{app:tria-int}
We demonstrate the usage of the recursion relation developed in Section~\ref{sec:recursion} for the triangle diagram. This is a six-point function in the $\phi^4$ theory. 

For the six-point function with external legs composed of $\phi_+$ we have the following two diagrams (for this appendix we suppress the external legs and represent them with $\bullet$ in order to avoid a clutter of notations\footnote{This is similar to the notation introduced in~\cite{Arkani-Hamed:2017fdk}.})
\begin{eqn}
\begin{tikzpicture}[baseline]
\draw (0, 1) -- (-1, 0);
\draw (0, 1) -- (1, 0);
\draw (-1, 0) -- (1, 0);
\node at (1, -0.25) {$x_3$};
\node at (-1, -0.25) {$x_2$};
\node at (0, 1.25) {$x_1$};
\node at (-0.75, 0.75) {$y_{12}$};
\node at (0.75, 0.75) {$y_{13}$};
\node at (0, -0.25) {$y_{23}$};
\node at (1, 0) {\textbullet};
\node at (-1, 0) {\textbullet};
\node at (0, 1) {\textbullet};
\end{tikzpicture}
+ 
\begin{tikzpicture}[baseline]
\draw[dashed] (0, 1) -- (-1, 0);
\draw[dashed] (0, 1) -- (1, 0);
\draw[dashed] (-1, 0) -- (1, 0);
\node at (1, -0.25) {$x_3$};
\node at (-1, -0.25) {$x_2$};
\node at (0, 1.25) {$x_1$};
\node at (-0.75, 0.75) {$y_{12}$};
\node at (0.75, 0.75) {$y_{31}$};
\node at (0, -0.25) {$y_{23}$};
\node at (1, 0) {\textbullet};
\node at (-1, 0) {\textbullet};
\node at (0, 1) {\textbullet};
\end{tikzpicture}
\end{eqn}
Here $x_1 = k_1 + k_2$, $x_2 = k_3 + k_4$, $x_3 = k_5 + k_6$ and $y_{12} = |\vec l + \vec k_1 + \vec k_2|$, $y_{23} = |\vec l + \vec k_1 + \vec k_2 + \vec k_3 + \vec k_4| = |\vec l - \vec k_5 - \vec k_6|$ and $y_{31} = |\vec l|$~.

Consider the first diagram, 
\begin{eqn}
\begin{tikzpicture}[baseline]
\draw (0, 1) -- (-1, 0);
\draw (0, 1) -- (1, 0);
\draw (-1, 0) -- (1, 0);
\node at (1, -0.25) {$x_3$};
\node at (-1, -0.25) {$x_2$};
\node at (0, 1.25) {$x_1$};
\node at (-0.75, 0.75) {$y_{12}$};
\node at (0.75, 0.75) {$y_{31}$};
\node at (0, -0.25) {$y_{23}$};
\node at (1, 0) {\textbullet};
\node at (-1, 0) {\textbullet};
\node at (0, 1) {\textbullet};
\end{tikzpicture}.
\end{eqn}
By expanding this in terms of the propagators we get 
\begin{eqn}
\begin{tikzpicture}[baseline]
\draw (0, 1) -- (-1, 0);
\draw (0, 1) -- (1, 0);
\draw (-1, 0) -- (1, 0);
\node at (1, -0.25) {$x_3$};
\node at (-1, -0.25) {$x_2$};
\node at (0, 1.25) {$x_1$};
\node at (-0.75, 0.75) {$y_{12}$};
\node at (0.75, 0.75) {$y_{31}$};
\node at (0, -0.25) {$y_{23}$};
\node at (1, 0) {\textbullet};
\node at (-1, 0) {\textbullet};
\node at (0, 1) {\textbullet};
\end{tikzpicture} = \intsinf dz_1 dz_2 dz_3 e^{- x_1 z_1} e^{- x_2 z_2} e^{- x_3 z_3} G_D(z_1, z_2, y_{12}) G_D(z_2, z_3, y_{23}) G_D(z_3, z_1, y_{13}).
\end{eqn}
Following the procedure described above we can insert the $z-$translation operator inside the integral and obtain the following 
\begin{eqn}\label{triangleDDD}
\begin{tikzpicture}[baseline]
\draw (0, 1) -- (-1, 0);
\draw (0, 1) -- (1, 0);
\draw (-1, 0) -- (1, 0);
\node at (1, -0.25) {$x_3$};
\node at (-1, -0.25) {$x_2$};
\node at (0, 1.25) {$x_1$};
\node at (-0.75, 0.75) {$y_{12}$};
\node at (0.75, 0.75) {$y_{31}$};
\node at (0, -0.25) {$y_{23}$};
\node at (1, 0) {\textbullet};
\node at (-1, 0) {\textbullet};
\node at (0, 1) {\textbullet};
\end{tikzpicture}  = \frac{1}{x_1 + x_2 + x_3} \qquad &\Bigg[ \HlineDD{x_1 + y_{12}}{x_3}{x_2 + y_{12}}{y_{31}}{y_{23}} + \HlineDD{x_1 + y_{13}}{x_2}{x_3 + y_{13}}{y_{12}}{y_{23}} \\
&\qquad + \HlineDD{x_2 + y_{23}}{x_1}{x_3 + y_{23}}{y_{12}}{y_{13}}  \Bigg].
\end{eqn}
Now consider the second diagram 
\begin{eqn}\label{triangleNNN1}
\begin{tikzpicture}[baseline]
\draw[dashed] (0, 1) -- (-1, 0);
\draw[dashed] (0, 1) -- (1, 0);
\draw[dashed] (-1, 0) -- (1, 0);
\node at (1, -0.25) {$x_3$};
\node at (-1, -0.25) {$x_2$};
\node at (0, 1.25) {$x_1$};
\node at (-0.75, 0.75) {$y_{12}$};
\node at (0.75, 0.75) {$y_{31}$};
\node at (0, -0.25) {$y_{23}$};
\node at (1, 0) {\textbullet};
\node at (-1, 0) {\textbullet};
\node at (0, 1) {\textbullet};
\end{tikzpicture} = \intsinf dz_1 dz_2 dz_3 e^{- x_1 z_1} e^{- x_2 z_2} e^{- x_3 z_3} G_N(z_1, z_2, y_{12}) G_N(z_2, z_3, y_{23}) G_N(z_3, z_1, y_{13}).
\end{eqn}
By inserting the $z-$translation operator we get the following 
\begin{eqn}
 \intsinf dz_1 dz_2 dz_3 \Big( \pd{}{z_1} + \pd{}{z_2} + \pd{}{z_3}  \Big) e^{- x_1 z_1} e^{- x_2 z_2} e^{- x_3 z_3} G_N(z_1, z_2, y_{12}) G_N(z_2, z_3, y_{23}) G_N(z_3, z_1, y_{13}) 
\end{eqn}
The terms that come from the derivative operators hitting the integrand are exactly the same as in equation~\eqref{triangleDDD} with $G_D$ replaced by $G_N$. However, in addition to them we also have terms that appear from the boundary $z_i \to 0$ , for $i = 1, 2, 3$~. This gives the following set of boundary terms,
\begin{eqn}
&\intsinf dz_1 dz_2 e^{- x_1 z_1} e^{- x_2 z_2} G_N(z_1, z_2, y_{12}) G_N(z_2, 0, y_{23}) G_N(0, z_1, y_{13}) + \mbox{permutations(1, 2, 3)}  \\
&=\intsinf dz_1 dz_2 e^{- x_1 z_1} e^{- x_2 z_2} G_N(z_1, z_2, y_{12}) \frac{e^{- y_{23} z_2}}{y_{23}} \frac{e^{- y_{13} z_1}}{y_{13}} + \mbox{permutations(1, 2, 3)}  \\
&= \vertex{y_{31}}\vertex{y_{23}} \HlineN{x_1 + y_{31}}{x_2 + y_{23}}{y_{12}} + \mbox{permutations(1, 2, 3)} ~. 
\end{eqn}
Therefore the integrand in diagram~\eqref{triangleNNN1} becomes 
\begin{eqn}
&\begin{tikzpicture}[baseline]
\draw[dashed] (0, 1) -- (-1, 0);
\draw[dashed] (0, 1) -- (1, 0);
\draw[dashed] (-1, 0) -- (1, 0);
\node at (1, -0.25) {$x_3$};
\node at (-1, -0.25) {$x_2$};
\node at (0, 1.25) {$x_1$};
\node at (-0.75, 0.75) {$y_{12}$};
\node at (0.75, 0.75) {$y_{31}$};
\node at (0, -0.25) {$y_{23}$};
\node at (1, 0) {\textbullet};
\node at (-1, 0) {\textbullet};
\node at (0, 1) {\textbullet};
\end{tikzpicture} \\
&=\frac{1}{x_1 + x_2 + x_3} \Bigg[ -\HlineNN{x_1 + y_{12}}{x_3}{x_2 + y_{12}}{y_{31}}{y_{23}} - \HlineNN{x_1 + y_{13}}{x_2}{x_3 + y_{13}}{y_{12}}{y_{23}} - \HlineNN{x_2 + y_{23}}{x_1}{x_3 + y_{23}}{y_{12}}{y_{13}} \\
&+\vertex{y_{31}}\vertex{y_{23}} \HlineN{x_1 + y_{31}}{x_2 + y_{23}}{y_{12}} + \vertex{y_{12}}\vertex{y_{23}} \HlineN{x_1 + y_{12}}{x_3 + y_{23}}{y_{13}} + \vertex{y_{12}}\vertex{y_{13}} \HlineN{x_2 + y_{12}}{x_3 + y_{13}}{y_{23}} \Bigg]
\end{eqn}
By adding the two we obtain
\begin{eqn}
&\frac{1}{\left(x_1+x_2+x_3\right) y_{12} y_{23} y_{31}} \frac{1}{\left(x_2+y_{12}+y_{23}\right) \left(x_1+x_3+y_{12}+y_{23}\right) \left(x_1+y_{12}+y_{31}\right) } \\
&\times \frac{1}{ \left(x_2+x_3+y_{12}+y_{31}\right) \left(x_1+x_2+y_{23}+y_{31}\right) \left(x_3+y_{23}+y_{31}\right)} \\
&\times \Bigg[ x_1^3 \left(x_2+x_3+y_{12}+y_{23}+y_{31}\right)+2 x_1^2 \left(x_2+x_3+y_{12}+y_{23}+y_{31}\right){}^2 \\
&+x_1 \Big(4 x_2^2 \left(x_3+y_{12}+y_{23}+y_{31}\right) \\
&+x_2 \left(9 x_3 \left(y_{12}+y_{23}+y_{31}\right)+4 x_3^2+4 y_{12}^2+4 \left(y_{23}+y_{31}\right){}^2+9 y_{12} \left(y_{23}+y_{31}\right)\right)\\
&+4 x_3^2 \left(y_{12}+y_{23}+y_{31}\right)+x_3 \left(4 \left(y_{12}+y_{23}\right){}^2+9 y_{31} \left(y_{12}+y_{23}\right)+4 y_{31}^2\right)+x_2^3+x_3^3\\
&+\left(y_{12}+y_{23}+y_{31}\right) \left(y_{12}^2+3 \left(y_{23}+y_{31}\right) y_{12}+y_{23}^2+y_{31}^2+3 y_{23} y_{31}\right)\Big)\\
&+\left(x_2+x_3+y_{12}+y_{31}\right) \Big(x_2^2 \left(x_3+y_{12}+y_{23}+y_{31}\right)+x_2 \left(x_3+y_{12}+y_{23}+y_{31}\right) \left(x_3+y_{12}+2 y_{23}+y_{31}\right)\\
&+\left(y_{12}+y_{23}+y_{31}\right) \left(x_3+y_{12}+y_{23}\right) \left(x_3+y_{23}+y_{31}\right)\Big) \Bigg].
\end{eqn}

%%%%%%%%%%%%%%%%%%%%%%%%%%%%%%%%%%%%%%%%%%%%%%%%%%%%%%
\subsection*{Integrals in terms of four-dimensional flat space Feynman  integrals}
Although the expression above is quite messy, it is interesting to note that this can also be written as a four-dimensional Feynman  integral. To see this, we note that the triangle with all solid legs is given as 
\begin{eqn}
\begin{tikzpicture}[baseline]
\draw (0, 1) -- (-1, 0);
\draw (0, 1) -- (1, 0);
\draw (-1, 0) -- (1, 0);
\node at (1, -0.25) {$x_3$};
\node at (-1, -0.25) {$x_2$};
\node at (0, 1.25) {$x_1$};
\node at (-0.75, 0.75) {$y_{12}$};
\node at (0.75, 0.75) {$y_{31}$};
\node at (0, -0.25) {$y_{23}$};
\node at (1, 0) {\textbullet};
\node at (-1, 0) {\textbullet};
\node at (0, 1) {\textbullet};
\end{tikzpicture} &=  \intsinf dz_1 dz_2 dz_3 \intinf dp_1 dp_2 dp_3 \int d^3 l \ e^{- x_1 z_1} e^{- x_2 z_2} e^{- x_3 z_3}  \\
&\qquad\times \frac{\sin(p_1 z_1) \sin(p_1 z_2) \sin(p_2 z_2) \sin(p_2 z_3) \sin(p_3 z_3) \sin(p_3 z_1)}{(p_1^2 + y_{12}^2)(p_2^2 + y_{23}^2)(p_3^2 + y_{31}^2)}.
\end{eqn}
Similarly the diagram with the dashed lines becomes
\begin{eqn}
\begin{tikzpicture}[baseline]
\draw[dashed] (0, 1) -- (-1, 0);
\draw[dashed] (0, 1) -- (1, 0);
\draw[dashed] (-1, 0) -- (1, 0);
\node at (1, -0.25) {$x_3$};
\node at (-1, -0.25) {$x_2$};
\node at (0, 1.25) {$x_1$};
\node at (-0.75, 0.75) {$y_{12}$};
\node at (0.75, 0.75) {$y_{31}$};
\node at (0, -0.25) {$y_{23}$};
\node at (1, 0) {\textbullet};
\node at (-1, 0) {\textbullet};
\node at (0, 1) {\textbullet};
\end{tikzpicture} &=  \intsinf dz_1 dz_2 dz_3 \intinf dp_1 dp_2 dp_3 \int d^3 l \ e^{- x_1 z_1} e^{- x_2 z_2} e^{- x_3 z_3}  \\
&\qquad\times \frac{\cos(p_1 z_1) \cos(p_1 z_2) \cos(p_2 z_2) \cos(p_2 z_3) \cos(p_3 z_3) \cos(p_3 z_1)}{(p_1^2 + y_{12}^2)(p_2^2 + y_{23}^2)(p_3^2 + y_{31}^2)}.
\end{eqn}

By first performing the $z$ integrals and adding the two diagrams we obtain\footnote{Since the $p_i$ integrals range from $(-\infty, \infty)$, any $p$-integral with an odd power does not contribute. This is similar to the other integrals we encountered before.} 
\begin{eqn}
&\begin{tikzpicture}[baseline]
\draw (0, 1) -- (-1, 0);
\draw (0, 1) -- (1, 0);
\draw (-1, 0) -- (1, 0);
\node at (1, -0.25) {$x_3$};
\node at (-1, -0.25) {$x_2$};
\node at (0, 1.25) {$x_1$};
\node at (-0.75, 0.75) {$y_{12}$};
\node at (0.75, 0.75) {$y_{31}$};
\node at (0, -0.25) {$y_{23}$};
\node at (1, 0) {\textbullet};
\node at (-1, 0) {\textbullet};
\node at (0, 1) {\textbullet};
\end{tikzpicture} 
+ 
\begin{tikzpicture}[baseline]
\draw[dashed] (0, 1) -- (-1, 0);
\draw[dashed] (0, 1) -- (1, 0);
\draw[dashed] (-1, 0) -- (1, 0);
\node at (1, -0.25) {$x_3$};
\node at (-1, -0.25) {$x_2$};
\node at (0, 1.25) {$x_1$};
\node at (-0.75, 0.75) {$y_{12}$};
\node at (0.75, 0.75) {$y_{31}$};
\node at (0, -0.25) {$y_{23}$};
\node at (1, 0) {\textbullet};
\node at (-1, 0) {\textbullet};
\node at (0, 1) {\textbullet};
\end{tikzpicture} \\
&= \intinf dp_1 d p_2 dp_3 d^3 l \\
&\times\frac{x_1 x_2 x_3}{\left(\left(p_1-p_3\right){}^2+x_1^2\right) \left(\left(p_1-p_2\right){}^2+x_2^2\right) \left(\left(p_2-p_3\right){}^2+x_3^2\right) \left(p_1^2+y_{12}^2\right) \left(p_2^2+y_{23}^2\right) \left(p_3^2+y_{31}^2\right)}.
\end{eqn}
These three-vectors can be embedded in a four-vector in the Euclidean signature,
\begin{eqn}
L = (p_3, \vec l), \qquad P_1 = (p_1 - p_3, \vec x_1), \qquad P_2 = (p_2 - p_3, \vec x_2)
\end{eqn}
with $\vec x_1 = \vec k_1 + \vec k_2$ and $\vec x_2 = \vec k_1 + \vec k_2 
+ \vec k_3 + \vec k_4$, the integral above can be expressed as 
\begin{eqn}\label{tria4D}
&\begin{tikzpicture}[baseline]
\draw (0, 1) -- (-1, 0);
\draw (0, 1) -- (1, 0);
\draw (-1, 0) -- (1, 0);
\node at (1, -0.25) {$x_3$};
\node at (-1, -0.25) {$x_2$};
\node at (0, 1.25) {$x_1$};
\node at (-0.75, 0.75) {$y_{12}$};
\node at (0.75, 0.75) {$y_{31}$};
\node at (0, -0.25) {$y_{23}$};
\node at (1, 0) {\textbullet};
\node at (-1, 0) {\textbullet};
\node at (0, 1) {\textbullet};
\end{tikzpicture} 
+ 
\begin{tikzpicture}[baseline]
\draw[dashed] (0, 1) -- (-1, 0);
\draw[dashed] (0, 1) -- (1, 0);
\draw[dashed] (-1, 0) -- (1, 0);
\node at (1, -0.25) {$x_3$};
\node at (-1, -0.25) {$x_2$};
\node at (0, 1.25) {$x_1$};
\node at (-0.75, 0.75) {$y_{12}$};
\node at (0.75, 0.75) {$y_{31}$};
\node at (0, -0.25) {$y_{23}$};
\node at (1, 0) {\textbullet};
\node at (-1, 0) {\textbullet};
\node at (0, 1) {\textbullet};
\end{tikzpicture} \\
&= \intinf \frac{x_1 x_2 x_3 dp dp'}{(p^2 + x_1^2) (p'^2 + x_2^2) \big( (p + p')^2 + x_3^2 \big)} \int \frac{d^4 L}{L^2  (L + P_1)^2 (L + P_2)^2}.
\end{eqn}
where $p = p_1 - p_3$, $p' = p_2 - p_1$ $\implies p + p' = p_2 - p_3$~. As noted in Section~\ref{sec:polygon} this structure can be generalized to any $n$-gon diagram at one-loop.

%-----------------------------------------------
\section{$\delta$-regularization}\label{sec:delta}
The $\delta$-regulated Green function~\eqref{eq:adsGG} has the momentum representation (here $z,z'>0$ and $a=1$)
\begin{equation}
G_D(\vec{k},\omega,z,z')=\frac{zz'}{\pi}\int   d\omega\;\frac{\sin(\frac{\omega}{2}(\Sigma_z-\Delta_\delta)) \sin(\frac{\omega}{2}(\Sigma_z+\Delta_\delta))}{\omega^2+k^2} ,
\end{equation}
and
\begin{equation}
G_N(\vec{k},\omega,z,z')=-\frac{zz'}{\pi}\int   d\omega\;\frac{\cos(\frac{\omega}{2}(\Sigma_z-\Delta_\delta)) \cos(\frac{\omega}{2}(\Sigma_z+\Delta_\delta))}{\omega^2+k^2} ,
\end{equation}
where $\Delta_\delta=\sqrt{\Delta_z^2+2\delta zz'}$, $\Delta_z=z-z'$ and $\Sigma_z=z+z'$. Adding the two contribution as before, we get for the one-loop, four-point function 
\begin{align}\label{eq:loopmh}
I^{(4)}_\circ= \frac{(\lambda/2)^2}{(2\pi)^5}\int\limits_{-\infty}^\infty dp\int\limits_{-\infty}^\infty  dp'&\int d^3 \ell\int\limits_0^{\infty} dz dz'e^{-k_{12}z-k_{34}z'}\\
&\times\frac{\left(\cos(p\Sigma_z)  \cos(p'\Sigma_z) +  \cos(p\Delta_\delta)\cos(p'\Delta_\delta)\right)}{(p^2+\vec{\ell}^2)(p'^2+(\vec{\ell}-\vec{k}_{12})^2)}\nonumber
\end{align}
Integration over $p$ and $p'$ gives 
\begin{equation}
I^{(4)}_\circ=\frac{(\lambda/2)^2}{4(2\pi)^3}\int d^3 \ell\int\limits_0^{\infty} d z d z' e^{-k_{12}z-k_{34}z'}\frac{\left(e^{-(|\vec{\ell}|+|\vec{\ell}-\vec{k}_{12}|)\Sigma_z}+e^{-(|\vec{\ell}|+|\vec{\ell}-\vec{k}_{12}|)\Delta_\delta}\right)}{|\vec{\ell}||\vec{\ell}-\vec{k}_{12}|}.
\end{equation}
Integration over $\ell$ then gives 
\begin{equation}
 I^{(4)}_\circ= \frac{(\lambda/2)^2}{4(2\pi)^2} \int\limits_0^{\infty} dz dz' e^{-k_{12}z-k_{34}z'}\left( \frac{e^{-|\vec{k}_{12}|\Sigma_z}}{\Sigma_z}+\frac{e^{-(|\vec{k}_{12}|)\Delta_\delta}}{\Delta_\delta}\right).
\end{equation}
The first term is easily integrated over $z$ and $z'$ to give
\begin{align}
\frac{(\lambda/2)^2}{4(2\pi)^2}\frac{1}{(k_{34}-k_{12})}\log\left(\frac{k_{34}+|\vec{k_{12}}|}{k_{12}+|\vec{k_{12}}|}\right)
\end{align}
For the second term we write
\begin{align}
   &\int\limits_0^{\infty} dz dz' e^{-k_{12}z-k_{34}z'}\frac{e^{-|\vec{k}_{12}|\Delta_\delta}}{\Delta_\delta}
   =\int\limits_0^{\infty} dz_+\int\limits_{-z_+}^{z_+} dz_- e^{-k_{+}z_+-k_{-}z_-}\frac{e^{-|\vec{k}_{12}|\Delta_\delta}}{\Delta_\delta}\nonumber\\
&=\int\limits_{0}^{|\vec{k_{12}}|}\frac{d}{d x} \int\limits_0^{\infty} dz_+\int\limits_{-z_+}^{z_+} dz_- e^{-k_{+}z_+-k_{-}z_-}\frac{e^{-|\vec{k}_{12}|\Delta_\delta}}{\Delta_\delta}+\int\limits_0^{\infty} dz_+\int\limits_{-z_+}^{z_+} dz_- \frac{e^{-k_{+}z_+-k_{-}z_-}}{\Delta_\delta},
\end{align}
where $z_\pm=\frac{1}{\sqrt{2}}(z\pm z')$ and $k_\pm=\frac{1}{\sqrt{2}}(k_{12}\pm k_{34})$. For the first term we can set $\delta=0$, since it is finite, resulting in 
\begin{equation}
  \int\limits_{0}^{|\vec{k_{12}}|} dx\int\limits_0^{\infty} dz_+\int\limits_{-z_+}^{z_+} dz_- e^{-k_{+}z_+-k_{-}z_-}e^{-x|\Delta_z|}=\frac{1}{(k_{12}+k_{34})}\log\left(\frac{(k_{34}+|\vec{k_{12}}|)(k_{12}+|\vec{k_{12}}|)}{k_{12}k_{34}}\right).
\end{equation}
The second is not a standard integral but, after some effort can be evaluated to 
\begin{equation}
- \frac{1}{(k_{12}+k_{34})}\log\left( \frac{-\delta\; k_{12}k_{34}}{8(k_{12}+k_{34})^2}\right) .
\end{equation}
Putting all terms together we end up with
\begin{equation}
  I^{(4)}_\circ=  \frac{(\lambda/2)^2}{4(2\pi)^2}\left(\frac{1}{(k_{34}-k_{12})}\log\left(\frac{k_{34}+|\vec{k_{12}}|}{k_{12}+|\vec{k_{12}}|}\right) -\frac{1}{(k_{34}+k_{12})}\log\left(\frac{-\delta}{8}\frac{(k_{34}+|\vec{k_{12}}|)(k_{12}+|\vec{k_{12}}|)}{(k_{34}+k_{12})^2}\right)\right)
\end{equation}
We see that (up to an overall normalization) the finite part agrees with~\eqref{analyticbubble1} from analytic regularization and also with~\eqref{1loop4ptcutoff} of the cut-off regularization if the cut-off is replaced by the total energy.

\section{Some integrals}\label{sec:integrals}

In this appendix we list some useful integrals required for the computation of the bubble, necklace and ice-cream diagrams,
\begin{enumerate}
\item \begin{equation}
\int_{-\infty}^\infty {dp\over (a^2+p^2)(b^2+p^2)}= {\pi\over ab(a + b)}
\end{equation}

\item 
\begin{equation}
  \int_0^\infty {\log(a^2+x^2)\over a^2+x^2} dx={\pi\over a}\log(a(1+a))
\end{equation}

\item 
\begin{equation}
  \int_0^\infty {\log(a^2+x^2)\over b^2+x^2} dx={\pi\over b}\log(a+b)
\end{equation}

\item Using Panzer's {\it HyperInt} program~\cite{Panzer:2014caa} we get 
\begin{multline}
  \int_0^\infty {\log(a^2+x^2)^2\over b^2+x^2} dx= {2  \pi\over b}  (\Mpl([1,
  1],[1, -a/b])+\ln\left(2 ab)\right) \log\left(a+b\over b\right)\cr
  +\Mpl([2],[-a/b])+\Mpl([1, 1],[-1, a/b])-\ln\left(2a\over b\right) \log\left(b-a\over b\right)+{\pi^2\over4}+\ln(b)^2-\Mpl
([2],[a/b])) 
\end{multline}

\item
\begin{eqn}
&\int_0^\infty {\log(a^2+x^2)\log(b^2+x^2)\over c^2+x^2} dx \\
&=  (\Mpl([1, 1],[c/b, -a/c])-\Mpl([1, 1],[-c/b, a/c])\\
&+(\Mpl([1],[-a/c])-\ln(b)-\ln(c)+
\Mpl([1],[a/c]))  \Mpl([1],[-b/c])+(-\ln(b)-\ln(c))  \Mpl([1],[-a/c])\\
&+\Mpl([2],[-b/c])+
(\ln(b)-\ln(c))  \Mpl([1],[a/c])+(\ln(b)-\ln(c))  \Mpl([1],[b/c])\\
&+1/2  \pi^2+2  \ln(c)^2-
\Mpl([2],[b/c]))  \pi/c~.
\end{eqn}

\end{enumerate}
One can express these integrals in terms of the standard dilogarithms using the identity
\begin{equation}\label{e:Li11def}
\Mpl([1,1],[x,y])=\Mpl([2],[y(x-1)\over (1-y)])-\Mpl([2],[y\over (y-1)])-\Mpl([2],[xy])
\end{equation}
with
\begin{equation}\label{e:Li2def}
    \Mpl([2],[x])=\sum_{n\geq1} {x^n\over n^2} \qquad \textrm{for}\quad |x|<1
  \end{equation}
  or
  \begin{equation}
    \Mpl([2],[z]):= -\int_0^z \log(1-u){du\over u}\qquad \textrm{for}\quad
  z\in\mathbb C\backslash [1,\infty[.
    \end{equation}

\section{Necklace Integral using Hard Cutoff}\label{app:necklace-hard}
We evaluate the value of the integral~\eqref{2loopnecklacerecursion} in this appendix using the hard-cutoff approach. We find that the transcedentality of the integral evaluated using the hard cutoff is the same as the answer obtained using the ads-invariant regulator. 

\begin{eqn}
I_{Necklace}^\Lambda&= \int d^3l_1 d^3 l_2
\frac{1}{E_T y_1 y_2 y_3 y_4 (x_1 + \sigma_1)(x_1 + \sigma_2) (x_2 + \sigma_1) (x_2 + \sigma_2) } \\ &\quad \times \Big[  (E_T + \sigma_1)  (E_T + \sigma_2) + \frac{E_T x_1 x_2}{\sigma_1 + \sigma_2} \Big] \\
&\equiv I_{Bubble^2} + I_{non-fact},
\end{eqn}
where
\begin{eqn}
I_{Bubble^2}&= \int d^3l_1 d^3 l_2
\frac{1}{E_T y_1 y_2 y_3 y_4 (x_1 + \sigma_1)(x_1 + \sigma_2) (x_2 + \sigma_1) (x_2 + \sigma_2) }  (E_T + \sigma_1)  (E_T + \sigma_2) \\
&= \frac{1}{E_T}\Bigg[ \int d^3 l_1 \frac{E_T + \s_1}{y_1 y_2 (x_1 + \s_1)(x_2 + \s_2)} \Bigg]^2.
\end{eqn}
The integral above is the square of the one-loop bubble integral given in~\eqref{1loop4ptcutoff}. The ultraviolet divergent piece of the necklace is captured by this term. The other term in $I^\Lambda_{Necklace}$ contributes to the finite part and is evaluated below. To simplify things we shall express the factor $\frac{1}{\s_1 + \s_2}$ of the integrand by an auxiliary integral with over a variable $\a$ as shown below,
\begin{eqn}\label{non-fact1}
I_{non-fact} &= \int d^3l_1 d^3 l_2\frac{1}{y_1 y_2 y_3 y_4 (x_1 + \sigma_1)(x_1 + \sigma_2) (x_2 + \sigma_1) (x_2 + \sigma_2) } \frac{x_1 x_2}{\sigma_1 + \sigma_2} \\
&= \frac{x_1 x_3}{\pi} \intinf d\a \Bigg[ \frac{d^3 l_1}{(x_1 + \s_1)(x_3 + \s_1)(\a^2 + \s_1^2)} \frac{\s_1}{y_1 y_2} \Bigg]^2
\end{eqn}  
Where we have used the identity 
\begin{equation}
\frac{1}{\s_1 + \s_2} = \frac{\s_1 \s_2}{\pi} \intinf \frac{d\a}{(\a^2 + \s_1^2)(\a^2 + \s_2^2)}.
\end{equation}
Using this representation we can compute the integral~\eqref{non-fact1}. 
\begin{eqn}
&I_{non-fact}\\
&=\frac{\pi ^4}{x_1+x_2} +\frac{8 \pi^2 x_2 x_1 \log \left(\frac{x_1}{k}+1\right) \log \left(\frac{x_2}{k}+1\right)}{\left(x_1-x_2\right){}^2 \left(x_1+x_2\right)} \\
&+ \frac{4 \pi^2 \log (2) \left(x_2 \log \left(1-\frac{x_1}{k}\right)-x_1 \log \left(1-\frac{x_2}{k}\right)\right)}{x_1^2-x_2^2} \\
&+\frac{4 \pi ^2 x_1 \big[-(x_1+x_2)+ (x_1-x_2) \log 2 \big] \log \left(\frac{x_2}{k}+1\right)}{\left(x_1-x_2\right){}^2 \left(x_1+x_2\right)} + (x_1 \leftrightarrow x_2) \\
&+ \frac{4 \pi ^2 x_1 x_2 \left(\frac{\log \left(\frac{x_1}{k}\right)}{k+x_1}+\frac{\log \left(\frac{x_2}{k}\right)}{k+x_2}\right)}{\left(x_1-x_2\right)^2} \\
&+ \frac{\pi^2 \log \left(\frac{x_2}{k}\right)}{(x_1 - x_2)}  \left(-\frac{8 x_2 x_1^2 \log \left(\frac{x_1}{k}+1\right)}{\left(x_1-x_2\right)^2 \left(x_1+x_2\right)}+\frac{2 \left(x_1+x_2\right) x_1 \log \left(\frac{x_2}{k}+1\right)}{\left(x_1-x_2\right)^2}-\frac{2 x_1 \log \left(1-\frac{x_2}{k}\right)}{\left(x_1+x_2\right)}\right) + (x_1 \leftrightarrow x_2) \\
&+\frac{2\pi^2}{\left(x_1-x_2\right)^3 \left(x_1+x_2\right)}  \Bigg[ x_1 \Bigg\{\left(x_1^2+6 x_2 x_1+x_2^2\right) \text{Li}_2\left(-\frac{x_2}{k}\right) + 2 \left(x_1^2 - x_2^2\right) \text{Li}_2\left(\frac{x_2}{k+x_2}\right)\\
&\qquad -\left(x_1^2-x_2^2\right) \Big(-\text{Li}_2\left(\frac{x_2}{k}\right)-2 \text{Li}_2\left(-\frac{2 x_2}{k-x_2}\right)-\log ^2\left(1-\frac{x_2}{k}\right)\Big)\Bigg\}\Bigg] +  (x_1 \leftrightarrow x_2)
\end{eqn}
where $k = |\vec k_1 + \vec k_2|$. Therefore we have the full expression for the necklace graph using hard cut-off. By comparing with the expression obtained using analytic regularization~\eqref{e:necklace-analytic-finite1} we see that a simple replacement of $\Lambda \to k_{12} + k_{34}$ does not ensure that the conformal ward identity is satisfied.

%%%%%%%%%%%%%%%%%%%%%%%%%%%%%%%%%%%%%%%%%%%%%%%%%%%%%%%
\section{Leading Singularity of the Ice-Cream using Hard-Cutoff}\label{app:ice}
By using the recusion relations we can evaluate the integrand for the ice-cream diagram. The full expresion is submitted with the arxiv submission in a mathematica notebook titled \reference{recursions.nb}. For our purpose we only need to extract the leading singularity from the loop integral and that is given as 
\begin{equation}
\frac{1}{k_1 + k_2 + k_3 + k_4}\int \frac{d^3 l_1 d^3 l_2}{y_1^2 y_2 y_3^2 y_4} 
= \frac{(4\pi)^2}{k_1 + k_2 + k_3 + k_4} \log \frac{k}{\Lambda} \log \frac{|\vec k + \vec k_3|}{\Lambda},
\end{equation}
with $y_1 = |\vec l_1|,$ $y_2 = |\vec l_1 + \vec k|,$ $y_3 = |\vec l_2|$, $y_4 = |\vec k_3 + \vec l_1 + \vec l_2| $  and the integral is regularized in hard-cutoff $\Lambda$ and we omit the terms that go as a power law in $\Lambda$ as they are absent in any other scale-invariant regularization. By comparing with the leading singularities from Section~\ref{sec:analytic-reg} we see that it is not easy to convert between the two regularization schemes at 2-loops.

\providecommand{\href}[2]{#2}\begingroup\raggedright\endgroup

\end{document}